\documentclass[11pt,oneside,english,onesided]{amsart}
\usepackage[T1]{fontenc}
\usepackage[latin1]{inputenc}

\makeatletter
 \theoremstyle{plain}
 \newtheorem{thm}{Theorem}[section]
 \numberwithin{equation}{section} %% Comment out for sequentially-numbered
 \numberwithin{figure}{section} %% Comment out for sequentially-numbered
 \theoremstyle{plain}
 \theoremstyle{definition}
  
 \theoremstyle{definition}
 
 \theoremstyle{remark}
 \newtheorem{rem}[thm]{Remark}
 \theoremstyle{plain}
 \newtheorem{prop}[thm]{Proposition} %%Delete [thm] to re-start numbering
 \theoremstyle{plain}
 \newtheorem{cor}[thm]{Corollary} %%Delete [thm] to re-start numbering
 \theoremstyle{plain}
\usepackage{amsmath, amsfonts, amssymb, amsthm, graphicx, stmaryrd, fancyhdr}
\usepackage{cite}

\newcommand{\cA}{\mathcal{A}}

\newcommand{\cF}{\mathcal{F}}
\newcommand{\cG}{\mathcal{G}}

\newcommand{\cM}{\mathcal{M}}

\newcommand{\carl}{\curvearrowleft}
\newcommand{\carr}{\curvearrowright}

\newcommand{\IP}{\mathbb{P}}

\newcommand{\IE}{\mathbb{E}}

\def\ineps#1#2{\includegraphics[scale=#1]{#2}}

\usepackage{babel}

\makeatother
\begin{document}

\title{Multicast Capacity of Optical WDM Packet Ring for Hotspot Traffic}

\thanks{Supported by the DFG Research Center \textsc{Matheon}
   ``Mathematics for key technologies'' in Berlin.}

\thanks{M.\ an der Heiden, M.\ Sortais,
and M. Scheutzow are with the
Department of Mathematics, Technical University Berlin, 10623
Berlin, Germany
(e-mail: \texttt{Matthias.an.der.Heiden@alumni.TU-Berlin.DE}, 
   \texttt{sortais@math-info.univ-paris5.fr}, 
   \texttt{ms@math.tu-berlin.de}).}

\thanks{M.\ Reisslein
is with the Dept. of Electrical
Eng., Arizona State Univ., Tempe, AZ 85287--5706, USA
(e-mail: \texttt{reisslein@asu.edu}).}

\thanks{M.\ Maier is with the Institut
National de la Recherche Scientifique (INRS), Montr\'eal, QC, H5A
1K6, CANADA (e-mail: \texttt{maier@ieee.org}).}

\author{Matthias an der Heiden, Michel Sortais,
Michael Scheutzow, Martin Reisslein, and Martin Maier
}

\begin{abstract}
Packet-switching WDM ring networks with a hotspot transporting
unicast, multicast, and broadcast traffic
are important components of high-speed metropolitan area networks.
For an arbitrary
multicast fanout traffic model with uniform, hotspot destination, and
hotspot source packet traffic,
we analyze the maximum achievable long-run average
packet throughput, which we refer to as
\textit{multicast capacity}, of
bi-directional shortest-path routed
WDM rings.
We identify three segments that can experience the maximum
utilization, and thus, limit the multicast capacity.
We characterize the segment utilization probabilities
through bounds and approximations, which we verify through
simulations.
We discover that shortest-path routing can lead to
utilization probabilities above one half for moderate to large portions
of hotspot source multi- and broadcast traffic, and consequently
multicast capacities of less than two simultaneous packet transmissions.
We outline a one-copy routing strategy that guarantees
a multicast capacity of at least two simultaneous
packet transmissions for arbitrary hotspot source traffic.

\end{abstract}

\maketitle

\section{Introduction}
\label{intro:sec}
\baselineskip  0.275in
Optical packet-switched ring wavelength division multiplexing (WDM)
networks have emerged as a promising
solution to alleviate the capacity shortage in the metropolitan
area, which is commonly referred to as metro gap.
Packet-switched ring networks, such as the Resilient
Packet Ring (RPR)~\cite{DYGU04, SSCW04, GYK04}, overcome many
of the shortcomings of circuit-switched ring networks, such as
low provisioning flexibility for packet data
traffic~\cite{GPC02}.
In addition, the use of multiple wavelength channels in WDM ring
networks, see e.g.,~\cite{FJRA+98, MBLM+96j,
MBLM+99, MLMN00, BA02,HeMR04, JeEl02, SS00, WhHH03}, overcomes
a key limitation of RPR, which was originally designed for a
single-wavelength channel in each ring direction.
In optical packet-switched ring
networks, the destination nodes typically remove (strip) the
packets destined to them from the ring. This \textit{destination
stripping} allows the destination node as well as other nodes
downstream to utilize the wavelength channel for their own
transmissions. With this so-called
\textit{spatial wavelength reuse},
multiple simultaneous transmissions can take place on any given
wavelength channel.
Spatial wavelength reuse is maximized through shortest path routing,
whereby the source node sends a packet in the ring direction that
reaches the destination with the smallest hop distance, i.e.,
traversing the smallest number of intermediate network nodes.

Multicast traffic is widely expected to account for a
large portion of the metro area traffic due to multi-party
communication applications, such as tele-conferences~\cite{FeSY02},
virtual private network interconnections,
interactive distance learning,
distributed games, and
content distribution.
These multi-party applications are expected to demand substantial
bandwidths due to the trend to deliver the video component of
multimedia content
in the High-Definition Television (HDTV) format or in
video formats with even higher resolutions, e.g., for digital cinema
and tele-immersion applications.
While there is at present
scant quantitative information about the multicast traffic volume,
there is ample anecdotal evidence of the emerging significance of
this traffic type~\cite{BeCl03,SaAl05}.
As a result, multicasting has been
identified as an important service in optical
networks~\cite{Rouskas03,SaMu00} and has begun to attract significant
attention in optical networking research as outlined in
Section~\ref{lit:sec}.

Metropolitan area networks consist typically of edge rings that interconnect
several access networks (e.g., Ethernet Passive Optical Networks)
and connect to a metro core ring~\cite{GPC02}.
The metro core ring interconnects
several metro edge rings and connects to the wide area network. The
node connecting a metro edge ring to the metro core ring is typically
a traffic hotspot as it collects/distributes traffic destined to/originating
from other metro edge rings or the wide area network. Similarly, the
node connecting the metro core ring to the wide area network is typically
a traffic hotspot. Examining the capacity of optical packet-switched
ring networks for hotspot traffic is therefore very
important.

In this paper we examine the
multicast capacity (maximum achievable long run average multicast
packet throughput) of bidirectional WDM optical ring networks
with a single hotspot for a general fanout traffic model comprising
unicast, multicast, and broadcast traffic. We consider an arbitrary
traffic mix composed of uniform traffic, hotspot destination traffic
(from regular nodes to the hotspot), and hotspot source traffic (from
the hotspot to regular nodes).
We study the widely considered node architecture that allows nodes
to transmit on all wavelength channels, but to receive only on one
channel.
We initially examine shortest path routing by
deriving bounds and approximations for the ring segment
utilization probabilities due to uniform, hotspot destination, and
hotspot source packet traffic.
We prove that there are three ring segments (in a given ring direction)
that govern the maximum segment utilization probability.
For the clockwise direction in a network with nodes $1, 2, \ldots, N$ and
wavelengths $1, 2, \ldots, \Lambda$ (with $N/\Lambda \geq 1$),
whereby node 1 receives on wavelength 1, node
2 on wavelength 2, $\ldots$, node $\Lambda$ on wavelength $\Lambda$,
node $\Lambda+ 1$ on wavelength 1, and so on, and with node $N$ denoting the
index of the hotspot node, the three critical segments are
identified as:
\begin{itemize}
\item[$(i)$] the segment connecting the hotspot,
node $N$, to node 1 on wavelength 1,
\item[$(ii)$] the segment connecting node $\Lambda - 1$ to node $\Lambda$ on
 wavelength $\Lambda$, and
\item[$(iii)$] the segment connecting node $N-1$ to node $N$ on wavelength
$\Lambda$.
\end{itemize}
The utilization on these three segments limits the maximum
achievable multicast packet throughput. We observe from the derived
utilization probability expressions that the utilizations of the
first two identified segments exceed 1/2 (and approach 1)
for large fractions of
hotspot source multi- and broadcast traffic,
whereas the utilization of the third
identified segment is always less than or equal to 1/2. Thus,
shortest path routing achieves a long run average multicast
throughput of less than two simultaneous packet transmissions
(and approaching one simultaneous packet transmission) for
large portions of hotspot source multi- and broadcast traffic.

We specify one-copy routing which sends only one packet copy
for hotspot source traffic, while uniform and hotspot
destination packet traffic is still served using shortest path
routing. One-copy routing ensures a capacity of at
least two simultaneous packet transmissions for arbitrary hotspot source
traffic, and at least approximately two simultaneous packet transmissions
for arbitrary overall traffic.
We verify the accuracy of our bounds and approximations
for the segment utilization probabilities, which are exact in the
limit $N/\Lambda \rightarrow \infty$, through comparisons with
utilization probabilities obtained from discrete event simulations.
We also quantify the gains in maximum achievable multicast
throughput achieved by the one-copy routing strategy over shortest
path routing through simulations.

This paper is structured as follows. In the following subsection, we
review related work. In Section~\ref{model:sec}, we introduce the
detailed network and traffic models and formally define the
multicast capacity. In Section~\ref{gen_prop:sec}, we establish
fundamental properties of the ring segment utilization in WDM packet
rings with shortest path routing. In Section~\ref{util_laneqLa:sec},
we derive bounds and approximations for the ring segment utilization
due to uniform, hotspot destination, and hotspot source packet
traffic on the wavelengths that the hotspot is not receiving on,
i.e., wavelengths $1, 2, \ldots, \Lambda-1$ in the model outlined
above. In Section~\ref{util_laeqLa:sec}, we derive similar
utilization probability bounds and approximations for wavelength
$\Lambda$ that the hotspot receives on. In
Section~\ref{max_segment:sec}, we prove that the three specific
segments identified above govern the maximum segment utilization and
multicast capacity in the network, and discuss implications for
packet routing. In Section~\ref{num:sec}, we present numerical
results obtained with the derived utilization bounds and
approximations and compare with verifying simulations. We conclude
in Section~\ref{concl:sec}.

\subsection{Related Work}
\label{lit:sec}
There has been increasing
research interest in recent years for the wide range of aspects of multicast
in general mesh circuit-switched WDM
networks, including
lightpath design, see for instance~\cite{SiSM06},
traffic grooming, see e.g.,~\cite{UlKa06},
routing and wavelength assignment,
see e.g.,~\cite{PoZh06,SaSu03,WaQC06},
and connection carrying capacity~\cite{QiYa04}.
Similarly, multicasting in
packet-switched single-hop star
WDM networks has been intensely investigated, see
for instance~\cite{HaKa02,HsLH04,LiWa01,NaWK04}.
In contrast to these studies, we focus on
packet-switched WDM ring networks in this paper.

Multicasting in circuit-switched WDM rings,
which are
fundamentally different from the packet-switched networks
considered in this paper, has been
extensively examined in the literature.
The scheduling of connections and cost-effective design of bidirectional
WDM rings was addressed, for instance in \cite{ZhQi99}.
Cost-effective traffic
grooming approaches in WDM rings have been studied for instance in
\cite{GRS00,WCVM01}. The routing and wavelength assignment in
reconfigurable bidirectional WDM rings with wavelength converters was
examined in \cite{ChMo05}.
The wavelength assignment for multicasting
in circuit-switched WDM ring networks has been studied
in~\cite{Din04,FeMM06,JiHR02,LiHD02,WaCU03,ZY02}.
For unicast traffic, the throughputs achieved by different
circuit-switched and packet-switched optical ring network
architectures are compared in~\cite{VeLA02}.

Optical \textit{packet-switched}
WDM ring networks have been experimentally demonstrated,
see for instance~\cite{WhHH03,CFFG04},
and studied for unicast traffic, see for instance~\cite{FJRA+98,
MBLM+96,MBLM+96j,
MBLM+99, MLMN00,BA02,HeMR04, JeEl02, SS00, WhHH03}.
Multicasting in packet-switched WDM ring networks has received
increasing interest in recent years~\cite{HCT02,HeMR04}.
The photonics level issues involved in multicasting over ring
WDM networks are explored in~\cite{BoLW03}, while
a node architecture suitable for multicasting is studied in~\cite{AlBe00}.
The general network architecture and MAC protocol
issues arising from multicasting in packet-switched WDM ring networks
are addressed in~\cite{CFFG04,ShFO03}.
The fairness issues arising when transmitting a
mix of unicast and multicast traffic in a ring WDM network
are examined in~\cite{PiRN04}.
The multicast capacity of packet-switched
WDM ring networks has been examined for uniform
packet traffic in~\cite{ChHC07,ScSMR07,ScSMR07it,ZaSRM07}.
In contrast, we consider non-uniform traffic with a hotspot node,
as it commonly arises in metro edge rings~\cite{SaSi99}.

Studies of non-uniform traffic in optical networks have generally
focused on issues arising in circuit-switched optical networks, see for
instance~\cite{BeMo05,HuTo05,LiWL00,SuAS98,WaCVM01,XuYa04,ZhQi00}.
A comparison of circuit-switching to optical burst switching network
technologies, including a brief comparison for non-uniform traffic,
was conducted in~\cite{ZaDS04}.
The throughput
characteristics of a mesh network interconnecting routers on an
optical ring through fiber shortcuts for non-uniform unicast traffic were
examined in~\cite{RuHu97}.
The study~\cite{ChHC05} considered the throughput characteristics of a
ring network with uniform unicast traffic, where the nodes may
adjust their send probabilities in a non-uniform manner.
The multicast capacity of a single-wavelength packet-switched
ring with non-uniform traffic was examined in~\cite{HeSSR07}.
In contrast to these works, we consider
non-uniform traffic with an
arbitrary fanout, which accommodates a wide range of unicast, multicast,
and broadcast traffic mixes, in a WDM ring network.

\section{System Model and Notations}
\label{model:sec}
We let $N$ denote
the number of network nodes,
which we index sequentially by $i,\,\, i=1,\ldots,N$,
in the clockwise direction and let $\cM:=\left\{ 1,\ldots,N\right\} $
denote the set of network nodes.
For convenience, we label the nodes modulo $N$, e.g.,
node $N$ is also denoted by $0$ or $-N$.
We consider the family of node structures where each node can
transmit on any wavelength using either one or multiple
tunable transmitters $\left(TTs\right)$ or an array of $\Lambda$
fixed-tuned transmitters $\left(FT^{\Lambda}\right)$,
and receive on one wavelength using a single fixed-tuned receiver
$\left(FR\right)$.

For $N=\Lambda$, each node has its own home channel for
reception. For $N>\Lambda$, each wavelength is shared by $\eta:=N/\Lambda$
nodes, which we assume to be an integer.
For $1\leq i\leq N$,
we let $\overset{\carr}{u}_{i}$ denote the clockwise oriented ring
segment connecting node $i-1$ to node $i$.
Analogously, we let $\overset{\carl}{u}_{i}$
denote the counter clockwise oriented ring segment connecting node
$i$ to node $i-1$. Each ring deploys the same set of wavelength
channels $\left\{ 1,\ldots,\Lambda\right\} $, one set on the clockwise
ring and another set on the counterclockwise ring.
The nodes $n=\lambda+k\Lambda$ with $k\in\left\{ 0,1,\ldots,\eta-1\right\} $
share the drop wavelength $\lambda$.
We refer to the incoming edges
of these nodes, i.e., the edges $\overset{\carr}{u}_{\lambda+k\Lambda}$
and $\overset{\carl}{u}_{\lambda+1+k\Lambda}$, as
\emph{critical edges} on $\lambda$.
\begin{figure}
\begin{center}
\ineps{.45}{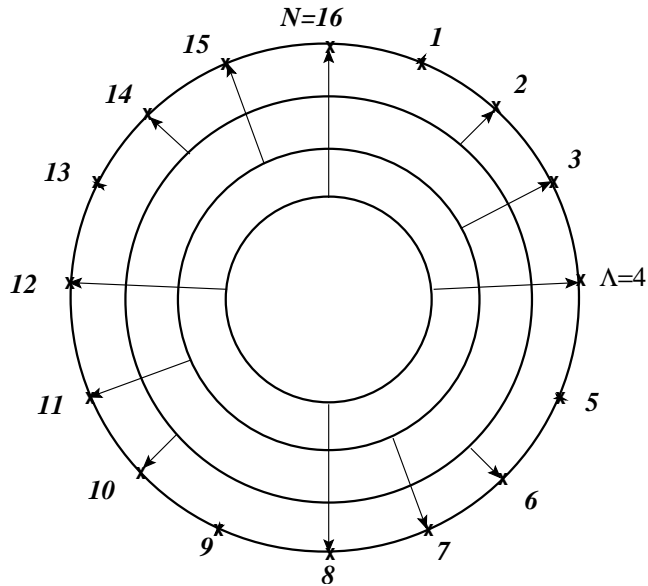} \centering
\end{center}
\caption{Illustration of the clockwise wavelength channels
of a WDM ring network with $N=16$ nodes and $\Lambda=4$ wavelength channels.}
\label{fig:RPRmodel}
\end{figure}

For multicast traffic, the sending node generates a copy of the multicast
packet for each wavelength that is drop wavelength for at least one
destination node.
Denote by $S$ the node that is the sender. We introduce the random
set of destinations (fanout set)
$\mathcal{F}\subset(\{1,2,\ldots,N\}\setminus\{ S\}).$
Moreover, we define the set of active nodes $\cA$ as the union of
the sender and all destinations, i.e., $\cA:=\cF\cup\left\{ S\right\} $.

We consider a traffic model combining a portion $\alpha$ of uniform
traffic, a portion $\beta$ of hotspot destination traffic, and a
portion $\gamma$ of hotspot source traffic with $\alpha,\beta,\gamma\geq0$
and $\alpha+\beta+\gamma=1$:

\begin{description}
\item [Uniform$\,$traffic]A given generated packet is a \emph{uniform
traffic} packet with probability $\alpha$. For such a packet, the
sending node is chosen uniformly at random amongst all network nodes
$\{1,2,\ldots,N\}$. Once the sender $S$ is chosen, the number of
receivers (fanout) $l\in\{1,2,\ldots,N-1\}$ is chosen at random according
to a discrete probability distribution $(\mu_{l})_{l=1}^{N-1}$. Once
the fanout $l$ is chosen, the random set of destinations (fanout
set) $\mathcal{F}\subset(\{1,2,\ldots,N\}\setminus\{ S\})$ is chosen
uniformly at random amongst all subsets of $\{1,2,\ldots,N\}\setminus\{ S\}$
having cardinality $l$. We denote by $P_{\alpha}$ the probability
measure associated with uniform traffic.
\item [Hotspot$\,$destination$\,$traffic]A given packet is a \emph{hotspot
destination traffic} packet with probability $\beta$. For a hotspot
destination traffic packet, node $N$ is always a destination. The
sending node is chosen uniformly at random amongst the other nodes
$\{1,2,\ldots,N-1\}$. Once the sender $S$ is chosen, the fanout
$l\in\{1,2,\ldots,N-1\}$ is chosen at random according to a discrete
probability distribution $(\nu_{l})_{l=1}^{N-1}$. Once the fanout
$l$ is chosen, a random fanout subset $\mathcal{F}'\subset(\{1,2,\ldots,N-1\}\setminus\{ S\})$
is chosen uniformly at random amongst all subsets of $\{1,2,\ldots,N-1\}\setminus\{ S\}$
having cardinality $(l-1)$, and the fanout set is $\mathcal{F}=\mathcal{F}'\cup\{ N\}$.
We denote by $Q_{\beta}$ the probability measure associated with
hotspot destination traffic.
\item [Hotspot$\,$source$\,$traffic]A given packet is a \emph{hotspot
source traffic} packet with probability $\gamma$. For such a packet,
the sending node is chosen to be node $N$. The fanout $1\leq l\leq(N-1)$
is chosen at random according to a discrete prob. distribution $(\kappa_{l})_{l=1}^{N-1}$.
Once the fanout $l$ is chosen, a random fanout set $\mathcal{F}\subset\{1,2,\ldots,N-1\}$
is chosen uniformly at random amongst all subsets of $\{1,2,\ldots,N-1\}$
having cardinality $l$. We denote by $Q_{\gamma}$ the probability
measures associated with hotspot source traffic.
\end{description}
While our analysis assumes that the traffic type, the source node,
the fanout, and the fanout set are drawn independently at random,
this independence assumption is not critical for the analysis. Our
results hold also for traffic patterns with correlations, as long
as the long run average segment utilizations are equivalent to the
utilizations with the independence assumption. For instance, our results
hold for a correlated traffic model where a given source node transmits
with a probability $p<1$ to exactly the same set of destinations
as the previous packet sent by the node, and with probability $1-p$
to an independently randomly drawn number and set of destination nodes.

We denote by $P_{\alpha}^{l}$ the probability measure $P_{\alpha}$
conditioned upon $\left|\cF\right|=l$, and define $Q_{\beta}^{l}$
and $Q_{\gamma}^{l}$ analogously.
We denote the set of nodes with drop wavelength $\lambda$ by
\begin{equation}
\cM_{\lambda}:=\left\{ \lambda+k\Lambda\,|\,
k\in\left\{ 0,\ldots,\eta-1\right\} \right\}.
\end{equation}
The set of all destinations with drop wavelength $\lambda$ is then
\begin{equation}
\cF_{\lambda}:=\cF\cap \cM_{\lambda}.
\end{equation}

Moreover, we use the following notation:
For $\ell\in\left\{ 1,\ldots,N-1\right\}$ we denote the probability of $\ell$
destinations on wavelength $\lambda$
by $\mu_{\lambda,\ell}$, $\nu_{\lambda,\ell}$, and $\kappa_{\lambda,\ell}$
for uniform, hotspot destination, and hotspot source traffic, respectively.
Since the fanout set is chosen
uniformly at random among all subsets of $\{1,2,\ldots,N\}\setminus\{ S\}$
having cardinality $l$,
these usage-probabilities can be expressed by
$\mu_{l}$, $\nu_{l}$, and $\kappa_{l}$. Depending on whether
the sender is on the drop-wavelength or not, we obtain slightly
different expressions.
As will become evident shortly, it suffices to focus on the
case where the sender is on the considered
drop wavelength $\lambda$, i.e., $S\in \cM_{\lambda}$, since
the relevant probabilities are estimated through comparisons
with transformations (enlarged, reduced or right/left-shifted ring
introduced in Appendix~\ref{AppendixA})
that put the sender in $\cM_{\lambda}$.

Through elementary combinatorial considerations we obtain the
following probability distributions:
For uniform traffic, the probability for having
$\ell\in\left\{ 0,\ldots,l\wedge \eta\right\}$
destinations on wavelength $\lambda$ is
\begin{equation} \label{mu_laell:eqn}
\mu_{\lambda,\ell}:=\sum_{l=\max(1, \ell)}^{N-1}
   \frac{{{\eta \choose \ell}}{{N-\eta \choose l-\ell}}}
    {{{N \choose l}}}\mu_{l}.
\end{equation}
For hotspot destination traffic, we obtain for
wavelengths $\lambda\neq\Lambda$ and
$\ell\in\left\{ 0,\ldots,(l-1)\wedge \eta\right\}$
\begin{equation}
\nu_{\lambda,\ell}:=\sum_{l=\max(1, \ell)}^{N-1}
\frac{{{\eta \choose \ell}}{{N-\eta-1 \choose l-\ell-1}}}
{{{N-1 \choose l-1}}}\nu_{l},
\end{equation}
as well as for wavelength $\Lambda$ homing the hotspot and
$\ell\in\left\{ 1,\ldots,l\wedge \eta\right\}$
\begin{equation}
\nu_{\Lambda,\ell}:=\sum_{l=1}^{N-1}
\frac{{{\eta-1 \choose \ell-1}}{{N-\eta \choose l-\ell}}}
{{{N-1 \choose l-1}}}\nu_{l}.
\end{equation}
Finally, for hotspot source traffic, we obtain for
$\lambda\neq\Lambda$ and $\ell\in\left\{ 0,\ldots,l\wedge \eta\right\}$
\begin{equation}
\kappa_{\lambda,\ell}:=\sum_{l=\max(1, \ell)}^{N-1}
   \frac{{{\eta \choose \ell}}
            {{N-1-\eta \choose l-\ell}}}{{{N-1 \choose l}}} \kappa_{l}
\end{equation}
and for $\lambda = \Lambda$ and
$\ell\in\left\{ 0,\ldots,l\wedge (\eta-1)\right\}$
\begin{equation}
\kappa_{\Lambda,\ell}:=\sum_{l=\max(1, \ell)}^{N-1}
 \frac{{{\eta-1 \choose \ell}}
  {{N-\eta \choose l-\ell}}}{{{N-1 \choose l}}} \kappa_{l}.
\end{equation}

For a given wavelength $\lambda$,
we denote by $p_{\alpha,\lambda}^{\ell}$
the probability measure $P_{\alpha}$ conditioned
upon $\left|\cF_{\lambda}\right|=\ell$,
and define $q_{\beta,\lambda}^{\ell}$ and $q_{\gamma,\lambda}^{\ell}$
analogously.
\begin{rem}
Whenever it is clear which wavelength $\lambda$ is considered we omit
the subscript $\lambda$ and write $p_{\alpha}^{\ell},\, q_{\beta}^{\ell}$,
or $q_{\gamma}^{\ell}$.
\end{rem}

We introduce the set of active nodes $\cA_{\lambda}$ on a given
drop wavelength $\lambda$ as
\begin{equation}
\cA_{\lambda}:=\cF_{\lambda}\cup\left\{ S\right\}.
\end{equation}
We order the nodes in this set in increasing order of their node
indices, i.e.,
\begin{equation}
\cA_{\lambda}=\{ X_{\lambda,1},X_{\lambda,2},\ldots,X_{\lambda,\ell+1}\},\quad1\leq X_{\lambda,1}<X_{\lambda,2}<\ldots<X_{\lambda,\ell+1}\leq N,\label{indices}
\end{equation}
and consider the ``gaps''
\begin{equation}
X_{\lambda,1}+(N-X_{\lambda,\ell+1}),\,(X_{\lambda,2}-X_{\lambda,1}),\ldots,
\,(X_{\lambda,\ell+1}-X_{\lambda,\ell}),
\end{equation}
between successive nodes in the set $\cA_{\lambda}.$ We have denoted
here again by $\ell\equiv \ell_{\lambda}$ the random number of
destinations with
drop wavelength $\lambda$.

For shortest path routing, i.e., to maximize spatial wavelength reuse,
we determine the largest of these gaps. Since there may be a tie among
the largest gaps (in which case one of the largest gaps is chosen
uniformly at random),
we denote the selected largest gap as {}``$CLG_{\lambda}$''\"{}
(for {}``Chosen Largest Gap''). Suppose the $CLG_{\lambda}$ is
between nodes $X_{\lambda,i-1}$ and $X_{\lambda,i}$. With shortest
path routing, the packet is then sent from the sender $S$ to node
$X_{\lambda,i-1}$, and from the sender $S$ to node $X_{\lambda,i}$
in the opposite direction. Thus, the largest gap is not traversed
by the packet transmission.

Note that by symmetry, $\IP\{\overset{\carr}u_{1}\text{ is used}\}=\IP\{\overset{\carl}u_{N}\text{ is used}\}$,
and $\IP\{\overset{\carr}u_{N}\text{ is used}\}=\IP\{\overset{\carl}u_{1}\text{ is used}\}$.
More generally, for reasons of symmetry, it suffices to compute the
utilization probabilities for the clockwise oriented edges. For $n\in\left\{ 1,\ldots,N\right\} $,
we abbreviate
\begin{equation}
\overset{\carr}{n}_{\lambda}:=\overset{\carr}{u_{n}}\textrm{ is used on wavelength }\lambda.
\end{equation}
It will be convenient to call node $N$ also node $0$.
We let $\cG_{\lambda},\ \cG_{\lambda} = 0, \ldots, N-1$,
be a random variable denoting the first
node bordering the chosen largest gap on wavelength
$\lambda$, when this gap is considered clockwise.

The utilization probability for the clockwise segment $n$ on wavelength
$\lambda$ is given by
\begin{equation}
\IP\left( \overset{\carr}{n}_{\lambda} \right)  =
\sum_{\ell=0}^{\eta} \left( \alpha \cdot p_{\alpha}^{\ell}
\left(\overset{\carr}{n}_{\lambda}\right)
\cdot \mu_{\lambda,\ell} + \beta \cdot q_{\beta}^{\ell}
\left(\overset{\carr}{n}_{\lambda}\right)
\cdot \nu_{\lambda,\ell}
+ \gamma \cdot q_{\gamma}^{\ell}
\left(\overset{\carr}{n}_{\lambda}\right)
\cdot \kappa_{\lambda,\ell} \right).
\label{segment_util:eqn}
\end{equation}
Our primary performance metric is the maximum packet throughout (stability
limit). More specifically, we define the (effective) multicast capacity
$C_{M}$ as the maximum number of packets (with a given traffic pattern)
that can be sent simultaneously in the long run, and note that $C_{M}$
is given as the reciprocal of the largest ring segment utilization
probability, i.e.,
\begin{eqnarray}
C_{M} & := & \frac{1}{\max_{n\in\left\{ 1,\ldots,N\right\} }
\max_{\lambda\in\left\{ 1,\ldots,\Lambda\right\} }\IP\left(\overset{\carr}{n}_{\lambda}\right)}.
\label{CM:eqn}
\end{eqnarray}

\section{General Properties of Segment Utilization}
\label{gen_prop:sec}
First, we prove a general recursion formula for shortest path routing.

\begin{prop}
\label{Prop31}
Let $\lambda\in\left\{ 1,\ldots,\Lambda\right\}$ be a fixed wavelength.
For all nodes $n\in\left\{ 0,\ldots,N-1\right\}$,
\begin{eqnarray}
\IP\left(\overset{\carr}{(n+1)}_{\lambda}\right) & = &
\IP\left(\overset{\carr}{n}_{\lambda}\right)+\IP\left(S=n\right)-
\IP\left(\cG_{\lambda}=n\right).
\label{recursion for p beta}
\end{eqnarray}
\end{prop}

\begin{proof}
There are two complementary events leading to
$\overset{\carr}{(n+1)}_{\lambda}$:
(A) the packet traverses (on wavelength $\lambda$) both the clockwise
segment $\overset{\carr}u_{n+1}$ and the preceding clockwise segment
$\overset{\carr}u_{n}$, i.e., the sender is a node $S\neq n$, and
(B) node $n$ is the sender ($S=n$) and transmits the packet in the
clockwise direction, so that it traverses segment $\overset{\carr}{u}_{n+1}$
following node $n$ (in the clockwise direction).
Formally,
\begin{eqnarray}
\IP\left(\overset{\carr}{(n+1)}_{\lambda}\right) & = &
\IP\left(\overset{\carr}{n}_{\lambda}
\text{ and }\overset{\carr}{(n+1)}_{\lambda}\right)+
\IP\left(S=n\textrm{ and }\overset{\carr}{(n+1)}_{\lambda}\right).
\label{equation for beta}
\end{eqnarray}
Next, note that the event that the clockwise segment $\overset{\carr}u_{n}$
is traversed can be decomposed into two complementary events, namely
(a) segments $\overset{\carr}u_{n}$ and $\overset{\carr}u_{n+1}$
are traversed, and (b) segment $\overset{\carr}u_{n}$ is traversed,
but not segment $\overset{\carr}u_{n+1}$, i.e.,
\begin{eqnarray}
\IP\left(\overset{\carr}{n}_{\lambda}\right) & = & \IP\left(\overset{\carr}{n}_{\lambda}\text{ and }\overset{\carr}{(n+1)}_{\lambda}\right)+\IP\left(\overset{\carr}{n}_{\lambda}\textrm{ and not }\overset{\carr}{(n+1)}_{\lambda}\right).
\label{segndecompose}
\end{eqnarray}
Similarly, we can decompose the event of node $n$ being the sender
as
\begin{equation}
\IP\left(S=n\right)=\IP\left(S=n\textrm{ and }\overset{\carr}{(n+1)}_{\lambda}\right)+\IP\left(S=n\textrm{ and not }\overset{\carr}{(n+1)}_{\lambda}\right).
\end{equation}
Hence, we can express $\IP\left(\overset{\carr}{(n+1)}_{\lambda}\right)$
as
\begin{eqnarray}
\IP\left(\overset{\carr}{(n+1)}_{\lambda}\right) & = & \IP\left(\overset{\carr}{n}_{\lambda}\right)-\IP\left(\overset{\carr}{n}_{\lambda}\textrm{ and not }\overset{\carr}{(n+1)}_{\lambda}\right)\nonumber \\
 &  & +\IP\left(S=n\right)-\IP\left(S=n\textrm{ and not }\overset{\carr}{(n+1)}_{\lambda}\right).
\end{eqnarray}
Now, note that there are two complementary events that result in
the CLG to start at node $n$, such that clockwise segment $n+1$
is inside the CLG: $(i)$ node $n$ is the last destination node reached
by the clockwise transmission, i.e., segment $n$ is used, but segment
$n+1$ is not used, and $(ii)$ node $n$ is the sender and transmits
only a packet copy in the counter clockwise direction. Hence,
\begin{eqnarray}
\IP\left(\cG_{\lambda}=n\right) & = & \IP\left(\overset{\carr}{n}_{\lambda}\textrm{ and not }\overset{\carr}{(n+1)}_{\lambda}\right)+\IP\left(S=n\textrm{ and not }\overset{\carr}{(n+1)}_{\lambda}\right).
\end{eqnarray}
Therefore, we obtain the general recursion
\begin{equation}
\IP\left(\overset{\carr}{(n+1)}_{\lambda}\right)=\IP\left(\overset{\carr}{n}_{\lambda}\right)+\IP\left(S=n\right)-\IP\left(\cG_{\lambda}=n\right).
\end{equation}

\end{proof}

We introduce the left (counter clockwise) shift and the right
(clockwise) shift of node $n$ to be $\left\lfloor n\right\rfloor
_{\lambda}$ and $\left\lceil n\right\rceil _{\lambda}$ given by
\begin{equation}
\left\lfloor n\right\rfloor_{\lambda}:=
\left\lfloor \frac{n-\lambda}{\Lambda}\right\rfloor\Lambda+\lambda
\textrm{ and }
\left\lceil n\right\rceil_{\lambda}
:=\left\lceil \frac{n-\lambda}{\Lambda}\right\rceil\Lambda+\lambda.
\end{equation}
The counter clockwise shift maps a
node $n$ not homed on $\lambda$ onto the nearest node in the counter
clockwise direction that is homed on $\lambda$. Similarly, the
clockwise shift maps a node $n$ not homed on $\lambda$ onto the
closest node in the clockwise direction that is homed on $\lambda$.

For the traffic on wavelength $\lambda$,
we obtain by repeated application of Proposition \ref{Prop31}
\begin{eqnarray}
\IP\left(\overset{\carr}
{\left(\left\lceil n\right\rceil _{\lambda}\right)}_{\lambda}\right)
& = &
\IP\left(\overset{\carr}n_{\lambda}\right)+
  \sum_{i = n}^{\lceil n \rceil _{\lambda}-1}
    \IP\left(S = i \right)
-   \sum_{i = n}^{\lceil n \rceil _{\lambda}-1}
  \IP\left(\cG_{\lambda} = i \right)
 \\
&=& \IP\left(\overset{\carr}n_{\lambda}\right)+
\IP\left( S \in \left\{ n,\ldots,
      \lceil n \rceil _{\lambda} -1 \right\} \right)
-\IP\left(\cG_{\lambda}\in
          \left\{ n,\ldots,\lceil n \rceil _{\lambda}-1\right\} \right).
\label{Pnlambda1}
\end{eqnarray}
Note that the CLG on $\lambda$ can only start $(i)$ at the source
node, irrespective of whether it is on $\lambda$, or $(ii)$ at a
destination node on $\lambda$. Consider a given node $n$ that is not
on $\lambda$, then the nodes in $\{n,\ n+1, \ldots, \lceil n \rceil
_{\lambda}-1 \}$ are not on $\lambda$. (If node $n$ is on $\lambda$,
i.e., $n = \lceil n \rceil _{\lambda}$, then trivially the set
$\{n,\ n+1, \ldots, \lceil n \rceil _{\lambda}-1 \}$ is empty and
$\IP\left(\overset{\carr} {\left(\left\lceil n\right\rceil
_{\lambda}\right)}_{\lambda}\right) =
\IP\left(\overset{\carr}n_{\lambda}\right)$.) Hence, the CLG on
$\lambda$ can only start at a node in $\{n,\ n+1, \ldots, \lceil n
\rceil _{\lambda}-1 \}$ if that node is the source node, i.e.,
\begin{equation}
\IP\left(\cG_{\lambda}\in
          \left\{ n,\ldots,\lceil n \rceil _{\lambda}-1\right\} \right)
 = \IP\left(\cG_{\lambda}=S\in
    \left\{ n,\ldots,  \lceil n \rceil _{\lambda}-1\right\} \right).
\label{GlambdaequalS}
\end{equation}
Next, note that the event that a node in
$\{n,\ n+1, \ldots, \lceil n \rceil _{\lambda}-1 \}$ is the source node
can be decomposed into the two complementary events
$(i)$ a node in $\{n,\ n+1, \ldots, \lceil n \rceil _{\lambda}-1\}$
is the source node and the CLG on $\lambda$ starts at that node, and
$(ii)$ a node in $\{n,\ n+1, \ldots, \lceil n \rceil _{\lambda}-1\}$
is the source node and the CLG does not start at that node.
Hence,
\begin{equation}
\IP\left(S\in\left\{ n,\ldots,
      \lceil n \rceil _{\lambda} -1\right\}\right) =
 \IP\left(\cG_{\lambda}=S\in
    \left\{ n,\ldots,  \lceil n \rceil _{\lambda}-1\right\} \right) +
  \IP\left(S\in\left\{ n,\ldots,m-1\right\} ,\cG_{\lambda}\neq S\right).
 \label{PScomplement}
\end{equation}
Inserting (\ref{GlambdaequalS}) and (\ref{PScomplement}) in
(\ref{Pnlambda1}) we obtain
\begin{equation}
\IP\left(\overset{\carr}
{\left(\left\lceil n\right\rceil _{\lambda}\right)}_{\lambda}\right) =
\IP\left(\overset{\carr}n_{\lambda}\right) +
     \IP\left(S\in\left\{ n,\ldots,m-1\right\} ,\cG_{\lambda}\neq S\right)
\end{equation}
which directly leads to
\begin{cor}
\label{Cor critical non critical}
The usage of non-critical segments
is smaller than the usage of critical segments, more precisely for
$n\in\left\{ 0,\ldots,N-1\right\} $: \begin{eqnarray}
\IP\left(\overset{\carr}{n}_{\lambda}\right) & = &
\IP\left(\overset{\carr}{\left(\left\lceil n\right\rceil _{\lambda}\right)}_{\lambda}\right)
-\IP\left(S\in\left\{ n,\ldots,\left\lceil n\right\rceil _{\lambda}-1\right\} ,\cG_{\lambda}\neq S\right).\label{critical non critical}
\end{eqnarray}
\end{cor}

To compare the expected length of the largest gap on a wavelength in
the WDM ring with the expected length of the largest gap in the
single wavelength ring, we introduce the enlarged and reduced ring
in Appendix~\ref{AppendixA}. In brief, in the enlarged ring, an
extra node is added on the considered wavelength between the
$\lambda$-neighbors of the source node. This enlargement results in
$(a)$ a set of $\eta+1$ nodes homed on the considered wavelength,
and $(b)$ an enlarged set of active nodes $\cA_{\lambda}^+$
containing the original destination
nodes plus the added extra node (which in a sense represents the
source node on the considered wavelength) for a total of $\ell+ 1$
active nodes. The expected length of the largest gap on this
enlarged wavelength ring with $\ell+1$ active nodes among $\eta+1$
nodes homed on the wavelength (A) is equivalent to $\Lambda$ times
the expected length of the largest gap on a single wavelength ring
with $l = \ell$ destination nodes and one source node among $N$
nodes homed on the ring, and (B) provides an upper bound on the
expected length of the largest gap on the original wavelength ring
(before the enlargement).

In the reduced ring, the left- and right-shifted source node are
merged into one node on the considered wavelength, resulting $(a)$
in a set of $\eta - 1$ nodes homed on the considered wavelength, and
$(b)$ a set $\cA_{\lambda}^-$ of $\ell-1,\ \ell$, or $\ell+1$ active nodes.
The expected length of the largest gap decreases with increasing number of
active nodes, hence we consider the case with $\ell+1$ active nodes
for a lower bound.
The expected length of the
largest gap on the reduced wavelength ring with
$\ell+1$ active nodes among $\eta-1$ nodes homed on the
wavelength (A) is equivalent to $\Lambda$ times the expected length of the
largest gap on a single wavelength ring with $l = \ell$ destination nodes
and one source node among $N$ nodes homed on the ring,
and (B) provides a lower bound on the expected length of the
largest gap on the original wavelength ring (before the reduction).
From these two constructions, which are formally provided in
Appendix~\ref{AppendixA}, we directly obtain:

\begin{prop}
\label{Prop33}
Given that the cardinality of $\cF_{\lambda}$ is $\ell$,
the expected length of the CLG on
wavelength $\lambda$ is bounded by:
\begin{equation}
\Lambda\cdot g\left(\ell,\eta-1\right)\leq
 \IE^{\ell}\left(\left|CLG_{\lambda}\right|\right)\leq
 \Lambda\cdot g\left(\ell,\eta+1\right),
\label{estimates for clg lambda}
\end{equation}
where $g\left(l,N\right)$ denotes the expected length of the CLG
for a single wavelength ring with $N$ nodes, when the active set
is chosen uniformly at random from all subsets of $\left\{ 1,\ldots,N\right\} $
with cardinality $\left(l+1\right)$.
\end{prop}

The expected length of the largest gap $g(l,N)$ \cite{MSHR05} is
given for $l = 0, \ldots, N-1$, by $g(l,N)=\sum_{k=1}^{N}k\cdot q_{l,N}(k)$,
where $q_{l,N}(\cdot)$ denotes the distribution of the length of
the largest gap.
Let $p_{l,N}(k)={{N-k-1 \choose l-1}}/{{N-1 \choose l}}$ denote the
probability that an arbitrary gap has $k$ hops.
Then the distribution $q_{l,N}$ may be computed using the recursion
\begin{eqnarray} \label{Recursion1}
q_{l,N}(k)=p_{l,N}(k)\cdot\sum_{m=1}^{k}q_{l-1,N-k}(m)+
 \sum_{m=1}^{k-1}p_{l,N}(m)\cdot q_{l-1,N-m}(k)
\end{eqnarray}
together with the
initialization $q_{0,N}(k)=\delta_{N,k}$ and $q_{N-1,N}(k)=\delta_{1,k}$,
where $\delta_{N,k}$ denotes the Kronecker Delta.
Whereby, $q_{0,N}(k)=\delta_{N,k}$ means a ring with only one active
node has only one gap of length $N$,
hence the largest gap has length $N$ with probability one.
Similarly, $q_{N-1,N}(k)=\delta_{1,k}$ means a ring with all nodes
active (broadcast case) has $N$ gaps with length one,
hence the largest gap has length $1$ with probability one.
This initialization directly implies $g(0,N) = N$ as well as $g(N-1, N) = 1$.
Obviously, we have to set $g(l, N) = 0$ for $l \geq N$.

\section{Bounds on Segment Utilization for $\lambda\neq\Lambda$}
\label{util_laneqLa:sec}
\subsection{Uniform Traffic}

In the setting of uniform traffic, one has for all $n\in\left\{ -\Lambda+\lambda+1,\ldots,\lambda\right\} $
and $k\in\left\{ 0,\ldots,\eta-1\right\} $, for reasons of symmetry:
\begin{equation}
P_{\alpha}\left(\overset{\carr}{n}_{\lambda}\right)=P_{\alpha}\left(\overset{\carr}{(n+k\Lambda)}_{\lambda}\right).\label{p alpha}
\end{equation}
For $n\in\left\{ -\Lambda+\lambda+1,\ldots,\lambda\right\} $, the
difference between critical and non-critical edges, corresponding
to Corollary \ref{Cor critical non critical}, can be estimated by
\begin{eqnarray}
 0 &\leq& P_{\alpha}\left(S\in\left\{ n,\ldots,\lambda-1\right\},
    \cG_{\lambda}\neq S\right)\nonumber  \\
 &\leq&  P_{\alpha}\left(S\in\left\{ n,\ldots,\lambda-1\right\} \right)
\ \ = \frac{\lambda-n}{N}.
\label{estimating the difference}
\end{eqnarray}

With shortest path routing, on average
$N - E_{\alpha}\left(\left|CLG\right|_{\lambda}\right)$ segments
are traversed on $\lambda$ to serve a uniform traffic packet.
Equivalently, we obtain the expected number of traversed segments
by summing the utilization probabilities of the individual
segments, i.e., as
$\sum_{n=1}^{N}P_{\alpha}\left(\overset{\carr}{n}_{\lambda}\right)
+ \sum_{n=1}^{N}P_{\alpha}\left(\overset{\carl}{n}_{\lambda}\right)$,
which, due to symmetry, equals
$2 \sum_{n=1}^{N}P_{\alpha}\left(\overset{\carr}{n}_{\lambda}\right)$.
Hence,
\begin{eqnarray}
N - E_{\alpha}\left(\left|CLG\right|_{\lambda}\right) & = &
2\sum_{n=1}^{N}P_{\alpha}\left(\overset{\carr}{n}_{\lambda}\right)
\end{eqnarray}
and
\begin{eqnarray}
E_{\alpha}\left(\left|CLG\right|_{\lambda}\right) & = &
N - 2\sum_{n=1}^{N}P_{\alpha}\left(\overset{\carr}{n}_{\lambda}\right) \\
& = & N-2\eta\sum_{k=-\Lambda+\lambda+1}^{\lambda}
P_{\alpha}\left(\overset{\carr}{k}_{\lambda}\right).
\end{eqnarray}
Expressing $P_{\alpha}\left(\overset{\carr}{k}_{\lambda}\right)$
using Corollary \ref{Cor critical non critical}, we obtain
\begin{equation}
E_{\alpha}\left(\left|CLG\right|_{\lambda}\right) =
N - 2 N P_{\alpha}\left(\overset{\carr}{\lambda}_{\lambda}\right)
+ 2\eta\sum_{k=-\Lambda+\lambda+1}^{\lambda}
P_{\alpha}\left(S\in \{ k,\ldots,\lambda -1\}, \cG_{\lambda} \neq S \right).
\label{uniform traffic estimates}
\end{equation}
Solving for $P_{\alpha}\left(\overset{\carr}{\lambda}_{\lambda}\right)$
yields
\begin{equation}
P_{\alpha}\left(\overset{\carr}{\lambda}_{\lambda}\right) =
\frac{1}{2} - \frac{1}{2N} E_{\alpha}\left(\left|CLG\right|_{\lambda}\right)
+ \frac{1}{\Lambda}\sum_{k=-\Lambda+\lambda+1}^{\lambda}
P_{\alpha}\left(S\in \{ k,\ldots,\lambda-1\}, \cG_{\lambda} \neq S \right).
\end{equation}
Hence, the inequalities (\ref{estimating the difference}) lead to
\begin{equation}
\frac{1}{2}-\frac{1}{2N}E_{\alpha}\left(\left|CLG\right|_{\lambda}\right) \leq P_{\alpha}\left(\overset{\carr}{\lambda}_{\lambda}\right)  \leq
\frac{1}{2}-\frac{1}{2N}E_{\alpha}\left(\left|CLG\right|_{\lambda}\right)+\frac{\Lambda-1}{2N}.
\label{pellalphabounds}
\end{equation}
Employing the bounds for $E_{\alpha}\left(\left|CLG\right|_{\lambda}\right)$
from Proposition~\ref{Prop33} gives
\begin{equation}
\frac{1}{2}-\frac{1}{2 \eta} \sum_{\ell=0}^{\eta} g(\ell, \eta + 1)\mu_{\lambda,\ell}  \leq
   P_{\alpha}\left(\overset{\carr}{\lambda}_{\lambda}\right)  \leq
\frac{1}{2}-\frac{1}{2 \eta} \sum_{\ell=0}^{\eta-2} g(\ell, \eta - 1)\mu_{\lambda,\ell} + \frac{\Lambda-1}{2N}.
\label{pellalphabounds}
\end{equation}

\subsection{Hotspot Destination Traffic}
The only difference to uniform traffic is that $N$ cannot be a sender,
since it is already a destination, i.e.,
\begin{equation}
q_{\beta}^{\ell}\left(\overset{\carr}{n}_{\lambda}\right)=
p_{\alpha}^{\ell}\left(\overset{\carr}{n}_{\lambda}\,|\, S\neq N\right).
\end{equation}
Using $p_{\alpha}^{\ell}\left(S=N\right)=\frac{1}{N}$,
we obtain
\begin{eqnarray}
q_{\beta}^{\ell}\left(\overset{\carr}{n}_{\lambda}\right) & = & \frac{N}{N-1}p_{\alpha}^{\ell}\left(\overset{\carr}{n}_{\lambda}\right)-\frac{1}{N-1}p_{\alpha}^{\ell}\left(\overset{\carr}{n}_{\lambda}\,|\, S=N\right) \\
 & = & \frac{N}{N-1}p_{\alpha}^{\ell}
\left(\overset{\carr}{n}_{\lambda}\right)-
\frac{1}{N-1}q_{\gamma}^{\ell}\left(\overset{\carr}{n}_{\lambda}\right).
\label{HSDestsegutil}
\end{eqnarray}
 Due to the factor $\frac{1}{N-1}$, the second term is negligible
in the context of large networks.

\subsection{Hotspot Source Traffic}
Since node $N$ is the sender (and given that there is at least one
destination node on $\lambda$), it sends a packet copy over
segment $\overset{\carr}{u}_{n}$ on wavelength $\lambda$ if the CLG on
$\lambda$ starts at a node with index $n$ or higher.
Hence, the usage probability of
a segment can be computed as
\begin{equation}
q_{\gamma}^{\ell}\left(\overset{\carr}{n}_{\lambda}\right)
=q_{\gamma}^{\ell}\left(\cG_{\lambda}\geq n\right)
\label{gamma usage for ungleich}
\end{equation}
for $n\in\left\{ 1,\ldots,N\right\} $.
We notice immediately that
$q_{\gamma}^{\ell}\left(\overset{\carr}{n}_{\lambda}\right)$ is monotone
decreasing in $n$.
Moreover, for all $n\in\left\{ 1,\ldots,
    \left(\eta-1\right)\Lambda+\lambda\right\} $,
Equation (\ref{critical non critical}) simplifies to
\begin{eqnarray}
q_{\gamma}^{\ell}\left(\overset{\carr}{n}_{\lambda}\right)
& = & q_{\gamma}^{\ell}
 \left(\overset{\carr}{\left(\left\lceil
n\right\rceil _{\lambda}\right)}_{\lambda}\right)
\end{eqnarray}
since the sender is node $N \equiv 0$ and consequently
$\IP\left(S\in\left\{ n,\ldots,\left\lceil n\right\rceil _{\lambda}-1\right\} ,\cG_{\lambda}\neq S\right) = 0$ for
the considered
$n\in\left\{ 1,\ldots,\left(\eta-1\right)\Lambda+\lambda\right\} $.
Since $q_{\gamma}^{\ell}\left(\overset{\carr}{n}_{\lambda}\right)$ is
monotone decreasing in $n$, the maximally used critical segment
on wavelength $\lambda$ is
$\overset{\carr}{u}_{\lambda}$.

With node $N$ being the sender, the CLG on $\lambda$ can only start at the
source node $N \equiv 0$, or at a destination node homed on
$\lambda$. If the CLG does not start at $N \equiv 0$, the
segment $\overset{\carr}{u}_{\lambda}$ leading to the first node
homed on $\lambda$, namely node $\lambda$, is utilized.
Hence,
\begin{equation}
q_{\gamma}^{\ell}\left(\overset{\carr}{\lambda}_{\lambda}\right)
=q_{\gamma}^{\ell}\left(\cG_{\lambda}\neq0\right).
\end{equation}
Observe that
\begin{equation}
q_{\gamma}^{\ell}\left(\cG_{\lambda}=0\right)<q_{\gamma}^{\ell}\left(\cG_{\Lambda-\lambda}=0\right)\textrm{ for }\lambda<\frac{\Lambda}{2},\label{usage halb leer halb voll}\end{equation}
which is exploited in Section~\ref{SummarylaneqLa:sec}.

Enlarging the ring leads to \begin{equation}
q_{\gamma}^{\ell}\left(\cG_{\lambda}=0\right)\leq
q_{\gamma}^{\ell}\left(\cG_{\lambda}^{+}=0\right)=
 \frac{1}{\ell+1},\label{gamma Gap at zero upper}
\end{equation}
since the gaps bordering node $0$ are enlarged whereas the lengths of
all other gaps are unchanged.
A right shifting of $S$ yields the following lower bound:
\begin{eqnarray}
q_{\gamma}^{\ell}\left(\cG_{\lambda}=0\right) & \geq & q_{\gamma}^{\ell}\left(\cG_{\lambda}^{\rightarrow}=0\,|\,\lambda\notin\cF_{\lambda}\right)q_{\gamma}^{\ell}\left(\lambda\notin\cF_{\lambda}\right)\\
 & = & \frac{1}{\ell+1}\left(1-\frac{\ell}{\eta}\right).
\label{gamma Gap at zero lower}
\end{eqnarray}
Thus,
\begin{eqnarray}
1-\frac{1}{\ell+1}\leq
 q_{\gamma}^{\ell}\left(\overset{\carr}{\lambda}_{\lambda}\right)
     & \leq & 1-\frac{1}{\ell+1}\left(1-\frac{\ell}{\eta}\right).
\label{qellgammalambdabounds}
\end{eqnarray}

\subsection{Summary of Segment Utilization Bounds
   and Approximation for
   $\lambda \neq \Lambda$}
\label{SummarylaneqLa:sec}
For $\lambda\neq\Lambda$ we obtain
from (\ref{segment_util:eqn}) and (\ref{HSDestsegutil})
\begin{eqnarray}
\IP\left(\overset{\carr}{n}_{\lambda}\right)
 & = &
\sum_{\ell=0}^{\eta} \left(   p_{\alpha}^{\ell}
             \left(\overset{\carr}{n}_{\lambda}\right)
 \left(\alpha\mu_{\lambda,\ell}+\frac{N}{N-1}\beta\nu_{\lambda,\ell}\right)
+ q_{\gamma}^{\ell}\left(\overset{\carr}{n}_{\lambda}\right)
\left(\gamma\kappa_{\lambda,\ell}-\frac{1}{N-1}\beta\nu_{\lambda,\ell}\right) \right).
\label{p n lambda}
\end{eqnarray}
Using Corollary \ref{Cor critical non critical} for $p_{\alpha}^{\ell}$ and
(\ref{gamma usage for ungleich}) for $q_{\gamma}^{\ell}$ yields
\begin{equation}
\max_{n\in M}\IP\left(\overset{\carr}{n}_{\lambda}\right)=\IP\left(\overset{\carr}{\lambda}_{\lambda}\right),\end{equation}
i.e., the segment number $\lambda$ experiences the
maximum utilization on wavelength $\lambda$.
Moreover, inequality (\ref{usage halb leer halb voll}) yields
\begin{equation}
\max_{\lambda\neq\Lambda}\max_{n\in M}
\IP\left(\overset{\carr}{n}_{\lambda}\right)=
      \IP\left(\overset{\carr}{1}_{1}\right), \label{max lambda neq Lambda}
\end{equation}
i.e., the first segment on wavelength 1, experiences the
maximum utilization among all segments on all wavelengths
$\lambda \neq \Lambda$.

From (\ref{p n lambda}) in conjunction with
(\ref{pellalphabounds}) and (\ref{qellgammalambdabounds})
we obtain
\begin{eqnarray}
\IP\left(\overset{\carr}{1}_{1}\right) & \geq & \frac{1}{2}\left(\alpha+\frac{N}{N-1}\beta\right)-\frac{1}{2\eta}\sum_{\ell=0}^{\eta}g\left(\ell,\eta+1\right)\left(\alpha \mu_{1,\ell}
+\frac{N}{N-1}\beta \nu_{1,\ell}\right)+
\nonumber \\
&  & +\sum_{\ell=0}^{\eta}\frac{\ell}{\ell+1}\left(\gamma \kappa_{1,\ell}-\frac{1}{N-1}\beta \nu_{1,\ell}\right) =: p1l
\label{lower bound 1}
\end{eqnarray}
and
\begin{eqnarray}
\IP\left(\overset{\carr}{1}_{1}\right)  & \leq &
\frac{1}{2}\left(1 + \frac{\Lambda-1}{N}\right)\left(\alpha+\frac{N}{N-1}\beta\right)
%\nonumber \\ & &
-\frac{1}{2\eta}\sum_{\ell=0}^{\eta}g\left(\ell,\eta-1\right)\left(\alpha \mu_{1,\ell}+\frac{N}{N-1}\beta \nu_{1,\ell}\right)
\nonumber \\
&  & +\sum_{\ell=0}^{\eta}\frac{\ell\left(\eta+1\right)}{\left(\ell+1\right)\eta}\left(\gamma\kappa_{1,\ell}-\frac{1}{N-1}\beta \nu_{1,\ell}\right) =: p1u.
\label{upper bound 1}
\end{eqnarray}

We obtain an approximation of the segment utilization by considering
the behavior of these bounds for large $\eta=\frac{N}{\Lambda}$.
Large $\eta$  imply $\frac{\eta+1}{\eta}\sim 1$ as well as
$\frac{N}{N-1} \sim 1 $, and $g(\ell,\eta-1)\sim g(\ell,\eta+1)$.
Intuitively, this last relation means that the expected length of
the largest gap on a ring network with $\ell$ destination nodes
among $\eta - 1$ nodes is approximately equal to the largest gap when there
are $\ell$ destination nodes among $\eta + 1$ nodes.
With these
considerations we can simplify the bounds given above and obtain the
approximation (valid for large $\eta$):
\begin{eqnarray}
\IP\left(\overset{\carr}{1}_{1}\right) & \sim &
\frac{1}{2}(\alpha+\beta)
-\frac{1}{2\eta}\sum_{\ell=0}^{\eta}g(\ell,\eta)
\left(\alpha \mu_{1,\ell}+\beta \nu_{1,\ell}\right)+
\label{sim 1}
%\\&  & + \nonumber
\gamma \sum_{\ell=0}^{\eta}\frac{\ell}{\ell+1}\kappa_{1,\ell}
=:p1a.
\end{eqnarray}

\section{Bounds on Segment Utilization for $\lambda=\Lambda$}
\label{util_laeqLa:sec}
For uniform traffic this case, of course, does not differ from the
case $\lambda\neq\Lambda$.

\subsection{Hotspot Destination Traffic}

Since $N$ is a destination node, by symmetry it is reached by a clockwise
transmission with probability one half, i.e., \begin{equation}
Q_{\beta}\left(\overset{\carr}{N}_{\Lambda}\right)=\frac{1}{2}.\label{p beta N}\end{equation}
 For hotspot destination traffic, node $N$ can not be the sender,
i.e., $Q_{\beta}\left(S=N\right)=0$.
Hence, by Proposition~\ref{Prop31}:
\begin{equation}
Q_{\beta}\left(\overset{\carr}{1}_{\Lambda}\right)
 = \frac{1}{2}-Q_{\beta}\left(\cG_{\Lambda}=0\right).
\label{Qbeta1Lambda1}
\end{equation}
 Moreover, we have from Corollary \ref{Cor critical non critical} with
$n = 1$ and $\lambda = \Lambda$:
\begin{equation}
Q_{\beta}\left(\overset{\carr}{\Lambda}_{\Lambda}\right)=
     Q_{\beta}\left(\overset{\carr}{1}_{\Lambda}\right)
       +Q_{\beta}\left(S\in\left\{ 1,\ldots,\Lambda-1\right\} ,
        \cG_{\Lambda}\neq S\right).
\label{QbetaLambdaLambda1}
\end{equation}

To estimate $Q_{\beta}\left(\cG_{\Lambda}=0\right)$, we introduce,
as before, the left- resp.\ right-shift of $S$, given by \begin{equation}
\left\lfloor S\right\rfloor _{\Lambda}:=\left\lfloor \frac{S}{\Lambda}\right\rfloor \Lambda\textrm{ and }\left\lceil S\right\rceil _{\Lambda}:=\left\lceil \frac{S}{\Lambda}\right\rceil \Lambda.\end{equation}
 Left and right shifting of $S$ leads to the following
bounds for the probability
$q_{\beta}^{\ell}\left(\cG_{\Lambda}=0\right)$, which are proven
in Appendix~\ref{AppendixBB}.
\begin{prop}
\label{Prop in beta} For hotspot destination traffic,
conditioning on the cardinality of $\cF_{\Lambda}$ to be $\ell$,
the probability that the CLG starts
at node $0$ is bounded by:
\begin{equation}
\frac{1}{\ell+1}\left(1-\frac{1}{\ell\eta}\right) \leq
     q_{\beta}^{\ell}\left(\cG_{\Lambda}=0\right)\leq
            \frac{1}{\ell+1}\left(1+\frac{1}{\eta}\right).
\end{equation}
\end{prop}

Inserting the bounds from Proposition \ref{Prop in beta}
and noting that
$0\leq Q_{\beta}\left(S\in\left\{ 1,\ldots,\Lambda-1\right\} ,
        \cG_{\Lambda}\neq S\right) \leq (\Lambda - 1) / (2N)$
leads to
\begin{equation}
Q_{\beta}\left(\overset{\carr}{\Lambda}_{\Lambda}\right)
 \leq\frac{1}{2}-\sum_{\ell=1}^{\eta} \nu_{\Lambda,\ell}
 \frac{1}{\ell+1}\left(1-\frac{1}{\ell\eta}\right)+
  \frac{\Lambda-1}{2N}
\end{equation}
and
\begin{equation}
Q_{\beta}\left(\overset{\carr}{\Lambda}_{\Lambda}\right)\geq
\frac{1}{2}-\sum_{\ell=1}^{N-1} \nu_{\Lambda,\ell}
   \frac{1}{\ell+1}\left( 1 + \frac{1}{\eta}\right).
\end{equation}

\subsection{Hotspot Source Traffic}

Since we know that $N$ is the sender and has drop wavelength $\Lambda$,
we have a symmetric setting on $\cF_{\Lambda}$ and can directly apply
the results of the single wavelength setting~\cite{HeSSR07}.

In particular, we obtain from Section~3.1.3 in~\cite{HeSSR07} for $\ell\in\{0,\ldots,\eta-1\}$
\begin{equation}
q_{\gamma}^{\ell}\left(\overset{\carr}{N}_{\Lambda}\right)=0
\end{equation}
and
\begin{equation}
q_{\gamma}^{\ell}\left(\overset{\carr}{\Lambda}_{\Lambda}\right)=q_{\gamma}^{\ell}\left(\cG_{\Lambda}\neq0\right)=\frac{\ell}{\ell+1}.
\end{equation}

\subsection{Summary of Segment Utilization Bounds
   and Approximation for
   $\lambda = \Lambda$}
\label{SummarylaeqLa:sec}
Inserting the bounds derived in the preceding sections in
(\ref{segment_util:eqn}), we obtain
\begin{eqnarray}
\IP\left(\overset{\carr}{\Lambda}_{\Lambda}\right) & \geq &
\frac{1}{2}\alpha\left(1-\frac{1}{\eta}\sum_{\ell=0}^{\eta}
g\left(\ell,\eta+1\right)\mu_{\Lambda,\ell}\right)+
%\nonumber \\ &  & +
\frac{1}{2}\beta\left(1-\sum_{\ell=1}^{\eta}\frac{2(\eta+1)}{(\ell+1)\eta}
\nu_{\Lambda,\ell}\right)+
\\
 &  & +\gamma\sum_{\ell=0}^{\eta-1}\frac{\ell}{\ell+1}\kappa_{\Lambda,\ell}
=: pLl
\nonumber
\end{eqnarray}
and
\begin{eqnarray}
\IP\left(\overset{\carr}{\Lambda}_{\Lambda}\right) & \leq &
 \frac{1}{2}\alpha\left(1+\frac{\Lambda-1}{N}
 -\frac{1}{\eta}\sum_{\ell=0}^{\eta}
     g\left(\ell,\eta-1\right)\mu_{\Lambda,\ell}\right)+
%\nonumber \\ &  & +
\frac{1}{2}\beta\left(1+\frac{\Lambda-1}{N}-
    \sum_{\ell=1}^{\eta}\frac{2(\ell\eta-1)}{(\ell+1)\ell\eta}
  \nu_{\Lambda,\ell}\right)+
\nonumber \\
 &  & +\gamma\sum_{\ell=0}^{\eta-1}\frac{\ell}{\ell+1}\kappa_{\Lambda,\ell}
  =: pLu,
\end{eqnarray}
whereby $\mu_{\Lambda, \ell}$ is given by setting $\lambda = \Lambda$ in
(\ref{mu_laell:eqn}).
Moreover,
\begin{eqnarray}
\IP\left(\overset{\carr}{N}_{\Lambda}\right) & \geq &
  \frac{1}{2}\alpha\left(1-\frac{1}{\eta}\sum_{\ell=0}^{\eta}
   g\left(\ell,\eta+1\right)\mu_{\Lambda,\ell}\right)+\frac{1}{2}\beta =: pNl
\end{eqnarray}
and
\begin{eqnarray}
\IP\left(\overset{\carr}{N}_{\Lambda}\right) & \leq &
\frac{1}{2}\alpha\left(1+\frac{\Lambda-1}{N}-
\frac{1}{\eta}\sum_{\ell=0}^{\eta}
  g\left(\ell,\eta-1\right)\mu_{\Lambda,\ell}\right)+\frac{1}{2}\beta =: pNu.
\end{eqnarray}

Considering again these bounds for large $\eta$, we obtain the
approximations:
\begin{eqnarray}
\IP\left(\overset{\carr}{\Lambda}_{\Lambda}\right)  & \sim &
\frac{1}{2}(\alpha+\beta)
-\frac{\alpha}{2\eta}\sum_{\ell=0}^{\eta}g(\ell,\eta)
 \mu_{\Lambda,\ell}
\label{sim 2}
%\\ &  & \nonumber
-\beta \sum_{\ell=1}^{\eta}\frac{1}{\ell+1}\nu_{\Lambda,\ell}
+\gamma\sum_{\ell=0}^{\eta-1}\frac{\ell}{\ell+1}\kappa_{\Lambda,\ell}
=: pLa
\end{eqnarray}
as well as
\begin{eqnarray}
\IP\left(\overset{\carr}{N}_{\Lambda}\right) & \sim &
  \frac{1}{2}(\alpha+\beta) -
        \frac{\alpha}{2\eta}\sum_{\ell=0}^{\eta}g(\ell,\eta)
\mu_{\Lambda,\ell} =: pNa.
\label{sim 3}
\end{eqnarray}

\section{Evaluation of Largest Segment Utilization and Selection of
Routing Strategy}
\label{max_segment:sec}

With (\ref{max lambda neq Lambda}) and a detailed
consideration of wavelength $\lambda = \Lambda$,
we prove in Appendix~\ref{AppendixC} the
main theoretical result:

\begin{thm}
\label{Mainthm} The maximum segment utilization probability is
\begin{equation}
\max_{n\in\left\{ 1,\ldots,N\right\} }
 \max_{\lambda\in\left\{ 1,\ldots,\Lambda\right\} }
    \IP\left(\overset{\carr}{n}_{\lambda}\right)=
     \max\left\{
           \IP\left(\overset{\carr}{1}_{1}\right),
           \IP\left(\overset{\carr}{\Lambda}_{\Lambda}\right),
           \IP\left(\overset{\carr}{N}_{\Lambda}
     \right)\right\}. \label{mainthm:eqn}
\end{equation}

\end{thm}

It thus remains to compute the three probabilities on the right hand
side. We have no exact result in the most general setting (it would
be possible to give recursive formulae, but these would be
prohibitively complex).
However, we have given upper and lower bounds
and approximations in Sections~\ref{SummarylaneqLa:sec}
and~\ref{SummarylaeqLa:sec}, which match rather
well in most situations, as demonstrated in the
next section, and have the same asymptotics when $\eta\rightarrow\infty$
while $\Lambda$ remains fixed.

Toward assessing the considered shortest-path routing strategy, we
directly observe, that
$\IP\left(\overset{\carr}{N}_{\Lambda}\right)$ is always less or
equal to $\frac{1}{2}$. On the other hand, the first two usage
probabilities will, for $\gamma$ large enough, become larger than
$\frac{1}{2}$, especially for hotspot source traffic with moderate
to large fanouts. Hence, shortest-path routing will result in a
multicast capacity of less than two for large portions of hotspot
source multi- and broadcast traffic, which may arise in content
distribution, such as for IP TV.

The intuitive explanation for the high utilization of the segments
$\overset{\carr}{1}_{1}$ and $\overset{\carr}{\Lambda}_{\Lambda}$
with shortest-path routing for multi- and broadcast hotspot source
traffic is a follows. Consider the transmission of a given
hotspot source traffic packet with
destinations on wavelength $\Lambda$ homing the hotspot. If the
packet has a single destination uniformly distributed among the
other $\eta-1$ nodes homed on wavelength $\Lambda$, then the CLG is
adjacent and to the left (i.e., in the counter clockwise sense) of the
hotspot with probability one half. Hence, with
probability one half a packet copy is sent in the clockwise
direction, utilizing the segment
$\overset{\carr}{\Lambda}_{\Lambda}$. With an increasing number of
uniformly distributed destination nodes on wavelength $\Lambda$, it
becomes less likely that the CLG is adjacent and to the left of the
hotspot, resulting in increased utilization of segment
$\overset{\carr}{\Lambda}_{\Lambda}$. In the extreme case of a
broadcast destined from the hotspot to all other $\eta-1$ nodes
homed on $\Lambda$, the CLG is adjacent and to the left of the
hotspot with probability $1/\eta$, i.e., segment
$\overset{\carr}{\Lambda}_{\Lambda}$ is utilized with probability $1
- 1/\eta$. With probability $1 -
2/\eta$ the CLG is not adjacent to the hotspot, resulting in two
packet copy transmissions, i.e., a packet copy is sent in each ring
direction.

For wavelength 1, the situation is subtly different due to the
rotational offset of the nodes homed on wavelength 1 from the
hotspot. That is, node 1 has a hop distance of 1 from the hotspot
(in the clockwise direction), whereas the highest indexed node on
wavelength 1, namely node $(\eta-1) \Lambda +1$ has a hop distance
of $\Lambda - 1$ from the hotspot (in the counter clockwise
direction). As for wavelength $\Lambda$, for a given packet with a
single uniformly distributed destination on wavelength 1, the CLG is
adjacent and to the left of the hotspot with probability one half,
and the packet consequently utilizes segment
$\overset{\carr}{1}_{1}$ with probability one half. With increasing
number of destinations, the probability of the CLG being adjacent
and to the left of the hotspot decreases, and the utilization of
segment $\overset{\carr}{1}_{1}$ increases, similar to the case for
wavelength $\Lambda$. For a broadcast destined to all $\eta$ nodes
on wavelength 1, the situation is different from wavelength
$\Lambda$, in that the CLG is never adjacent to the hotspot, i.e.,
the hotspot always sends two packet copies, one in each ring
direction.

\subsection{One-Copy (OC) Routing}
To overcome the high utilization of the segments
$\overset{\carr}{1}_{1}$ and $\overset{\carr}{\Lambda}_{\Lambda}$
due to hotpot source multi- and broadcast traffic, we propose
\textit{one-copy (OC) routing}:
With one-copy routing, uniform traffic and hotspot destination
traffic are still served using shortest path routing.
Hotspot source traffic is served using the following counter-based
policy.
We define the counter $Y_{\lambda}$ to denote the number of
nodes homed on $\lambda$ that would need to be traversed to
reach all destinations on $\lambda$ with one packet transmission in
the clockwise direction (whereby the final reached
destination node counts as a traversed node).
If $Y_{\lambda} < \eta/2$, then one packet copy is sent in the
clockwise direction to reach all destinations.
If $Y_{\lambda} > \eta/2$, then one packet
copy is sent in the counter clockwise direction to reach all destinations.
Ties, i.e., $Y_{\lambda} = \eta/2$, are served in either
clockwise or counter clockwise direction with probability one half.
For hotspot source traffic with arbitrary traffic fanout,
this counter-based one-copy routing ensures a maximum utilization of
one half on any ring segment.
Note that the counter-based policy considers only the nodes
homed on the considered wavelength $\lambda$ to ensure that the
rotational offset between the wavelength $\Lambda$ homing the hotspot and
the considered wavelength $\lambda$ does not affect the routing decisions.

We propose the following
strategy for switching between shortest path (SP) and one-copy (OC) routing.
Shortest path routing is employed if both
(\ref{sim 1}) and (\ref{sim 2}) are less than one half.
If (\ref{sim 1}) or (\ref{sim 2}) exceeds one half, then
one-copy routing is used.
For the practical implementation of this switching strategy, the
hotspot can periodically estimate the current traffic parameters, i.e.,
the traffic portions $\alpha$, $\beta$, and $\gamma$ as well as the
corresponding fanout distributions $\mu_l$, $\nu_l$, and $\kappa_l,\
l = 1, \ldots, N-1$, for instance, through a
combination of traffic measurements
and historic traffic patterns,
similar to~\cite{BiFG05,ElCh05,ElMS03,GeMU03,OkSO02}. From
these traffic parameter estimates, the hotspot can then evaluate
(\ref{sim 1}) and (\ref{sim 2}).

To obtain a more refined criterion for switching between shortest
path routing and one-copy routing we proceed as follows. We
characterize the maximum segment utilization with shortest path
routing more explicitly by inserting (\ref{sim 1}), (\ref{sim 2}),
and (\ref{sim 3}) in (\ref{mainthm:eqn}) to obtain:
\begin{eqnarray}
\nonumber
\max_{n\in \{ 1,\ldots,N \} } \!\!\!\! && \!\!\!\!
\max_{\lambda \in \{ 1,\ldots,\Lambda \} }
\IP\left(\overset{\carr}{n}_{\lambda}\right) =
\frac{1}{2}(\alpha+\beta)-\frac{\alpha}{2\eta}
      \sum_{\ell=0}^{\eta} g(\ell,\eta) \mu_{1,\ell}  \label{mainthm:eqn} \\
 &&+ \max \left\{ 0,\
            -\frac{\beta}{2\eta}\sum_{\ell=0}^{\eta}g(\ell,\eta) \nu_{1,\ell}+
     \gamma \sum_{\ell=0}^{\eta}\frac{\ell}{\ell+1}\kappa_{1,\ell},\
% \right. \nonumber     \\ & & \left. \qquad
- \beta \sum_{\ell=1}^{\eta}\frac{1}{\ell+1}\nu_{\Lambda,\ell}
+  \gamma\sum_{\ell=0}^{\eta-1}\frac{\ell}{\ell+1}
\kappa_{\Lambda,\ell} \right \},
\end{eqnarray}
whereby we noted that the definition of
$\mu_{\lambda,\ell}$ in (\ref{mu_laell:eqn})
directly implies that $\mu_{\lambda,\ell}$  is independent of $\lambda$.
Clearly, the hotspot source traffic does not influence the
maximum segment utilization as long as
\begin{eqnarray}
\gamma \leq \gamma_{th1, 1} :=
\frac{\beta}{2\eta}\ \frac{\sum_{\ell=0}^{\eta}g(\ell,\eta)\nu_{1,\ell}}
      {\sum_{\ell=1}^{\eta}\frac{\ell}{\ell+1}\kappa_{1,\ell}}
\end{eqnarray}
and
\begin{eqnarray}
\gamma \leq \gamma_{th1, \Lambda} :=
\beta \frac{\sum_{\ell=1}^{\eta}\frac{1}{\ell+1}\nu_{\Lambda,\ell}}
{\sum_{\ell=1}^{\eta-1}\frac{\ell}{\ell+1}\kappa_{\Lambda,\ell}}.
\end{eqnarray}
Thus, if $\gamma \leq \gamma_{th1} =
\min(\gamma_{th1, 1},\ \gamma_{th1, \Lambda})$,
then all traffic is served using shortest path routing.

We next note that Theorem~\ref{Mainthm} does not hold for the
one-copy routing strategy.
We therefore bound the maximum segment utilization probability with
one-copy routing by observing
that (\ref{pellalphabounds})
together with Proposition~\ref{Cor critical non critical}
and (\ref{estimating the difference})
implies that asymptotically for all $\lambda\in\{1,\ldots,\Lambda\}$
\begin{equation}
   P_{\alpha}\left(\overset{\carr}{n}_{\lambda}\right)  \sim
\frac{1}{2}-\frac{1}{2 \eta} \sum_{\ell=0}^{\eta-1} g(\ell, \eta)
  \mu_{\lambda,\ell}. \label{Palpha_approx:eqn}
\end{equation}
Hence, $P_{\alpha}\left(\overset{\carr}{n}_{\lambda}\right)$ is
asymptotically constant.
Moreover, similar as in the single wavelength case~\cite{HeSSR07}, we have
\begin{equation}
   P_{\beta}\left(\overset{\carr}{n}_{\lambda}\right) \leq
 P_{\beta}\left(\overset{\carr}{N}_{\Lambda}\right)= \frac{1}{2}.
\end{equation}
Therefore, the maximum segment utilization with one-copy routing is
(approximately) bounded by
\begin{eqnarray}
\max_{n\in\left\{ 1,\ldots,N\right\} }
 \max_{\lambda\in\left\{ 1,\ldots,\Lambda\right\} }
    \IP\left(\overset{\carr}{n}_{\lambda}\right) \leq
   \frac{1}{2} (\alpha + \beta + \gamma)
        -\frac{\alpha}{2 \eta} \sum_{\ell=0}^{\eta-1}
              g(\ell, \eta)\mu_{1,\ell}.
   \label{oc_util:eqn}
\end{eqnarray}

Comparing (\ref{oc_util:eqn}) with (\ref{mainthm:eqn})
we observe that the maximum segment utilization with one-copy routing is
smaller than with shortest path routing
if the following threshold conditions hold:
\begin{itemize}
\item If
$\sum_{\ell=1}^{\eta}\frac{\ell}{\ell+1}\kappa_{1,\ell} > \frac{1}{2}$,
then set
\begin{eqnarray}
\gamma_{th2, 1} =
\frac{\beta}{2\eta}\ \frac{\sum_{\ell=0}^{\eta}g(\ell,\eta) \nu_{1,\ell}}
      {\sum_{\ell=1}^{\eta}\frac{\ell}{\ell+1}\kappa_{1,\ell} - \frac{1}{2}},
\end{eqnarray}
otherwise set $\gamma_{th2, 1} = \infty$.
\item If
$\sum_{\ell=1}^{\eta-1}\frac{\ell}{\ell+1}\kappa_{\Lambda,\ell}
   > \frac{1}{2}$, then set
\begin{eqnarray}
\gamma_{th2, \Lambda} := \beta \frac{
\sum_{\ell=1}^{\eta}\frac{1}{\ell+1}\nu_{\Lambda,\ell}}
{\sum_{\ell=1}^{\eta-1}\frac{\ell}{\ell+1}\kappa_{\Lambda,\ell} - \frac{1}{2}},
\end{eqnarray}
otherwise set $\gamma_{th2, \Lambda} = \infty$.
\end{itemize}
If $\gamma \geq \gamma_{th2} =
\max(\gamma_{th2, 1},\ \gamma_{th2, \Lambda})$,
then one-copy routing is employed.

For $\gamma$ values between $\gamma_{th1}$ and $\gamma_{th2}$,
the hotspot could numerically evaluate the maximum segment utilization
probability of shortest path routing with the derived approximations.
The hotspot could also obtain the segment utilization probabilities with
one-copy routing through discrete event simulations to determine whether
shortest path routing or one-copy routing of the hotspot traffic
is preferable for a given set of traffic parameter estimates.

\section{Numerical and Simulation Results}
\label{num:sec} In this section we present numerical results
obtained from the derived bounds and approximations of the
utilization probabilities as well as verifying simulations.
We initially simulate individual, stochastically independent
packets generated according to
the traffic model of Section~\ref{model:sec} and routed according to the
shortest path routing policy.
We determine estimates of the utilization probabilities of the
three segments $\overset{\carr}{1}_{1}$,
$\overset{\carr}{\Lambda}_{\Lambda}$, and $\overset{\carr}{N}_{\Lambda}$
and denote
these probabilities by $p1s$, $pLs$, and $pNs$.
Each simulation is run until the
99\% confidence intervals of the utilization probability estimates
are less than 1\% of the corresponding sample means.
We consider a networks with $\Lambda = 4$ wavelength channels in
each ring direction.

\subsection{Evaluation of Segment Utilization Probability
Bounds and Approximations for Shortest Path Routing}
We examine the accuracy of the derived bounds and approximations by plotting
the segment utilization probabilities as
a function of the number of network nodes $N = 8, 12, 16, \ldots, 256$
and comparing with the corresponding simulation results.
\begin{figure*}[t]
\centering
\begin{tabular}{ccc}
\includegraphics[width=.33\textwidth,angle=0]{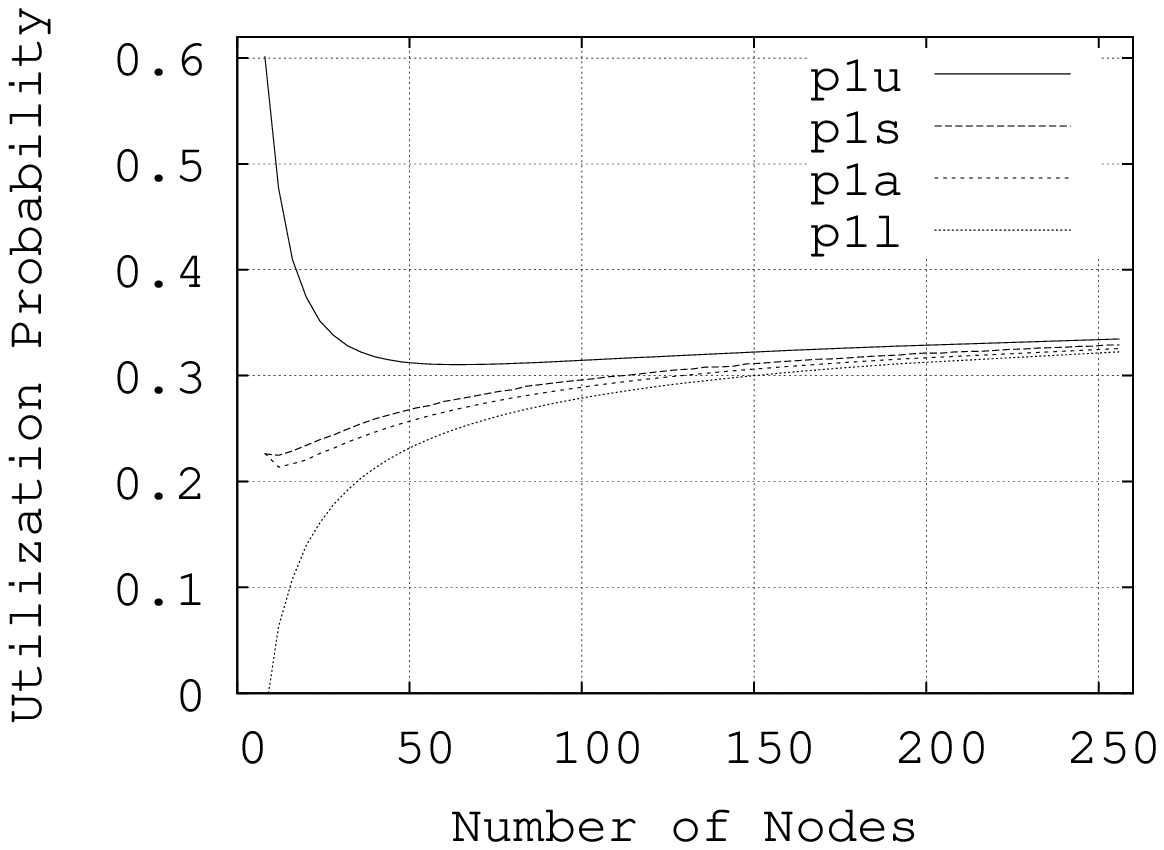}&
\includegraphics[width=.33\textwidth,angle=0]{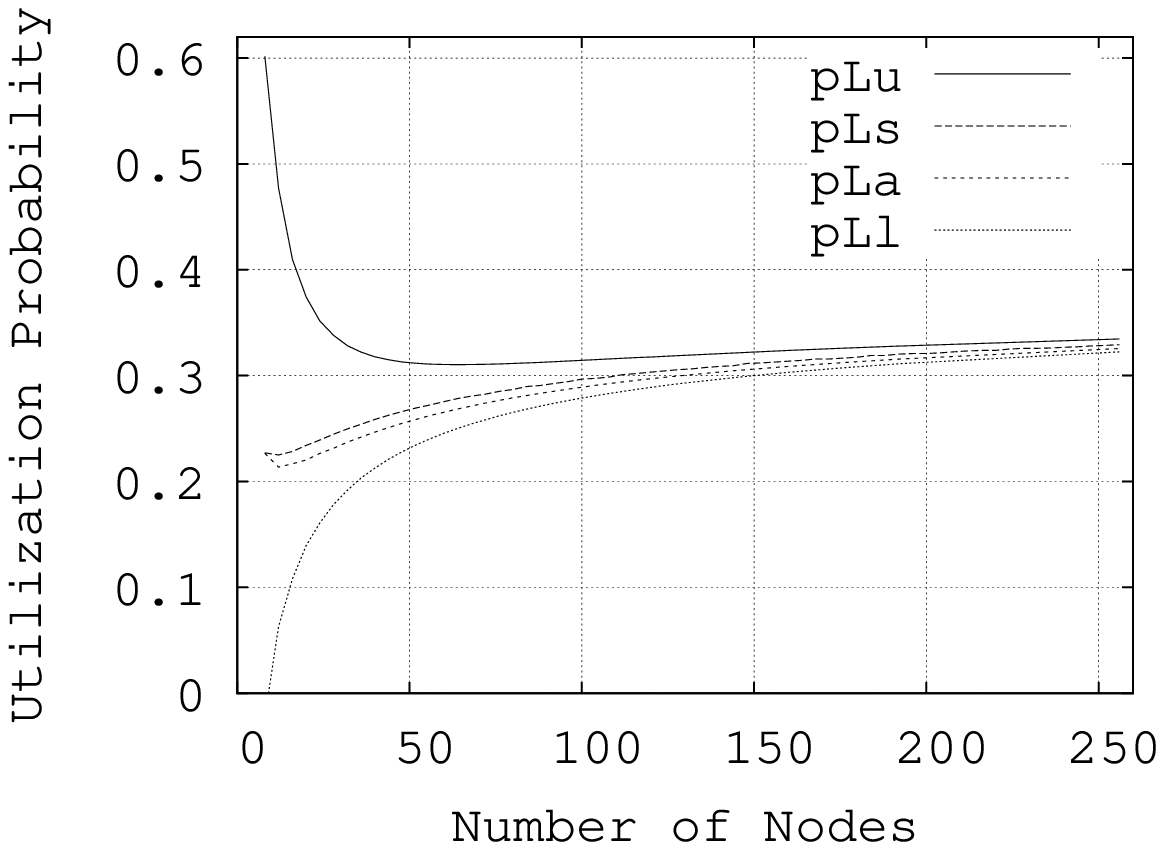}&
\includegraphics[width=.33\textwidth,angle=0]{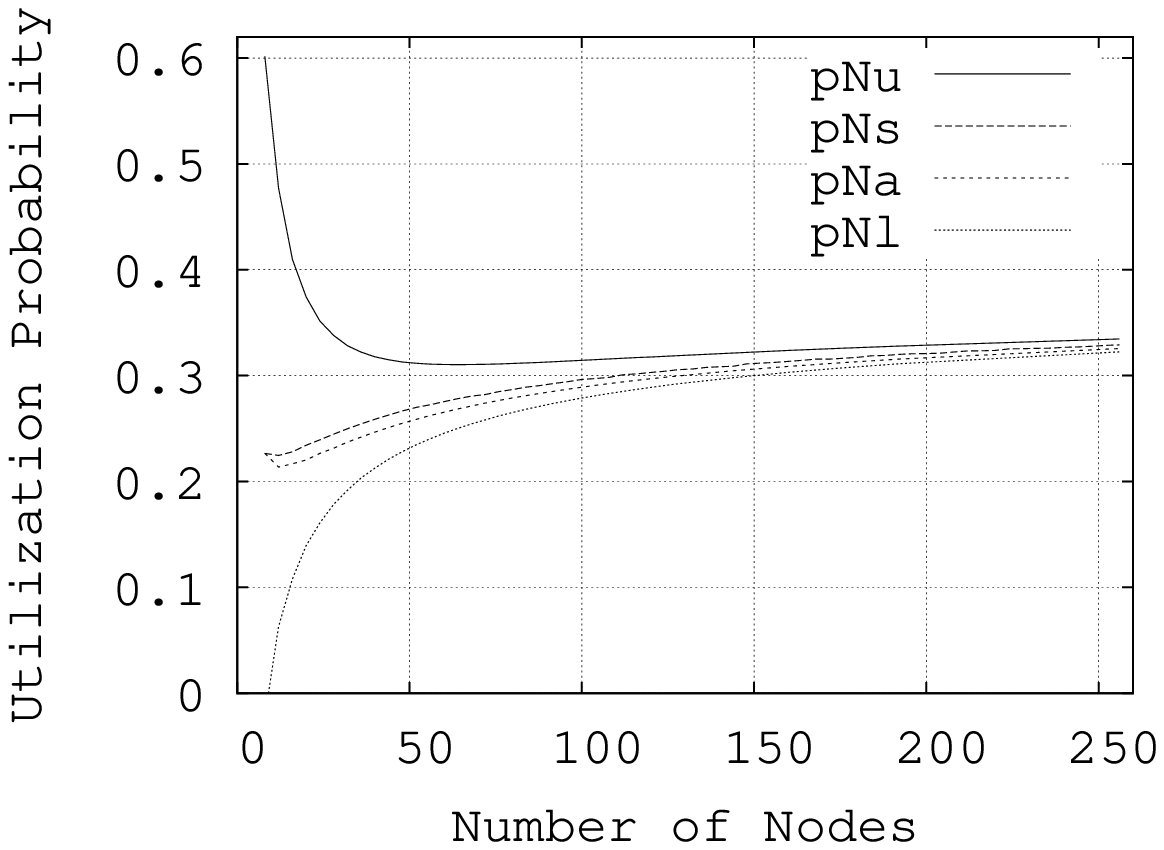}\\
(a) $\IP\left(\overset{\carr}{1}_{1}\right)$   &
(b) $\IP\left(\overset{\carr}{4}_{4}\right)$  &
(c) $\IP\left(\overset{\carr}{64}_{4}\right)$ \\
\end{tabular}
\caption{Segment utilization probability as a function of number of Nodes $N$
for $\alpha = 1$, $\beta = 0$, $\gamma = 0$,
and $\mu_1 = \nu_1 = \kappa_1 = 1/4$ and
$\mu_l = \nu_l = \kappa_l = 3/(4(N-2))$ for $l = 2, \ldots, N-1$.}
\label{Num1a}
\end{figure*}
\begin{figure*}[t]
\centering
\begin{tabular}{ccc}
\includegraphics[width=.33\textwidth,angle=0]{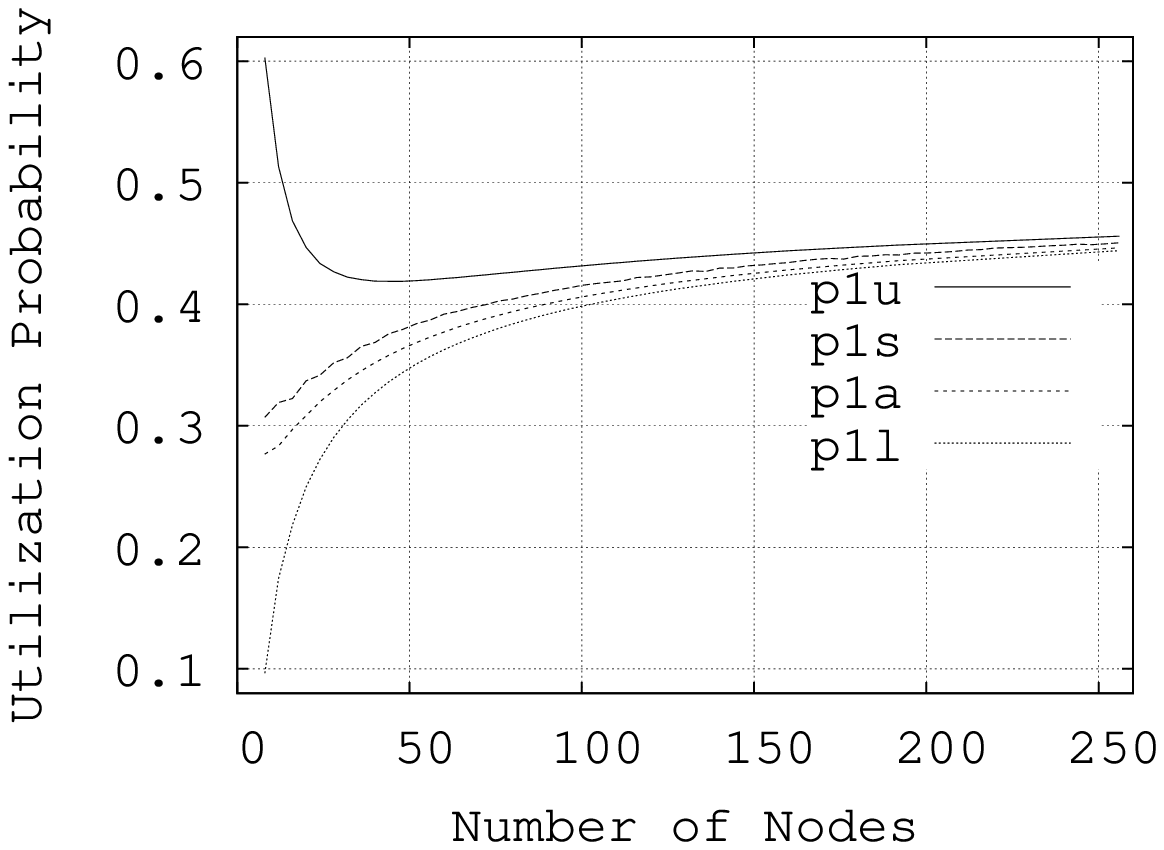}&
\includegraphics[width=.33\textwidth,angle=0]{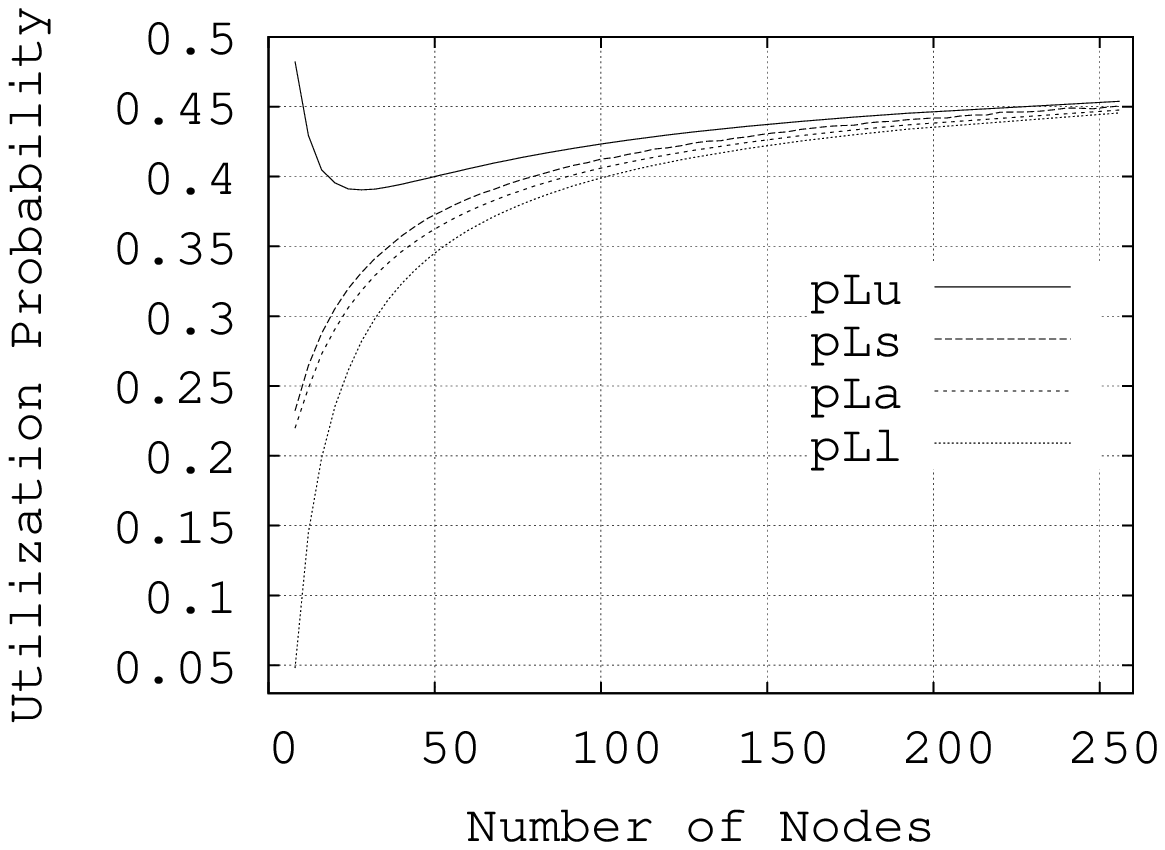}&
\includegraphics[width=.33\textwidth,angle=0]{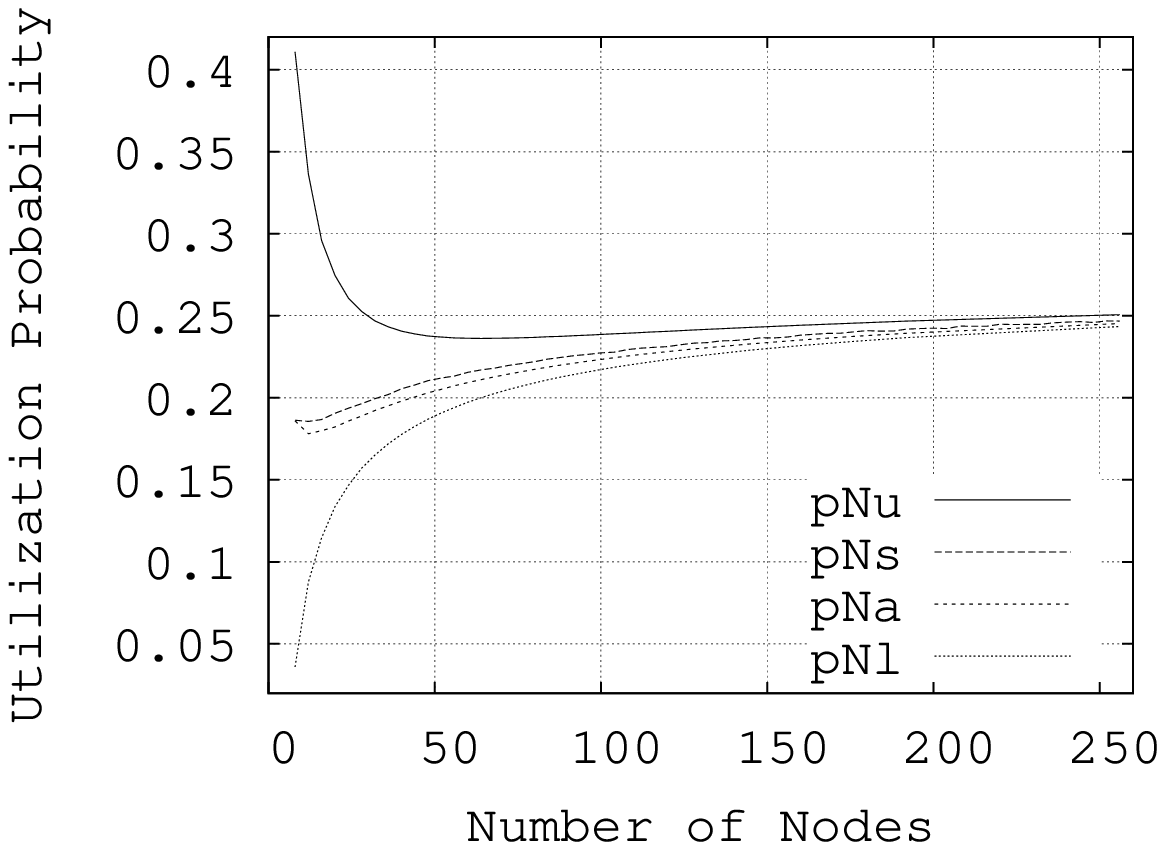}\\
(a) $\IP\left(\overset{\carr}{1}_{1}\right)$   &
(b) $\IP\left(\overset{\carr}{4}_{4}\right)$  &
(c) $\IP\left(\overset{\carr}{64}_{4}\right)$ \\
\end{tabular}
\caption{Segment utilization probability as a function of number of Nodes $N$
for $\alpha = 0.6$, $\beta = 0.1$, $\gamma = 0.3$,
and $\mu_1 = \nu_1 = \kappa_1 = 1/4$ and
$\mu_l = \nu_l = \kappa_l = 3/(4(N-2))$ for $l = 2, \ldots, N-1$.}
\label{Num1b}
\end{figure*}
\begin{figure*}[t]
\centering
\begin{tabular}{ccc}
\includegraphics[width=.33\textwidth,angle=0]{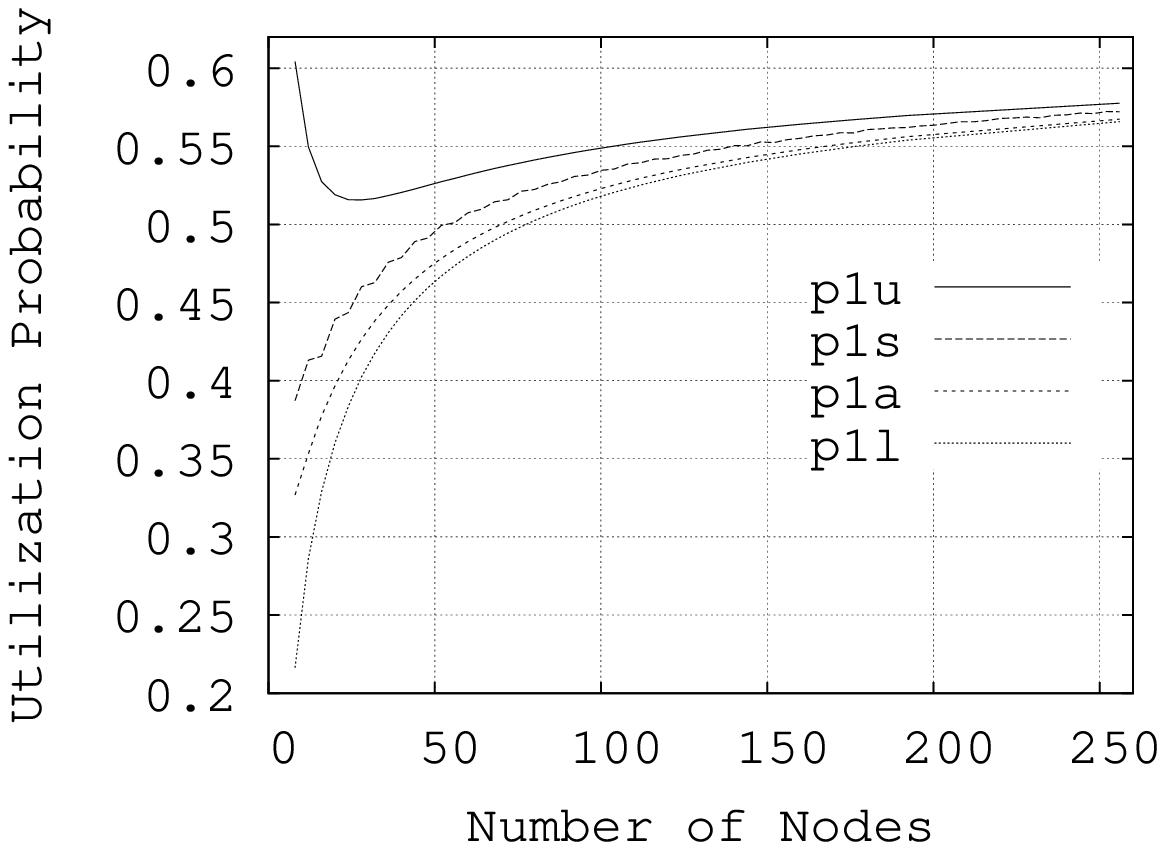}&
\includegraphics[width=.33\textwidth,angle=0]{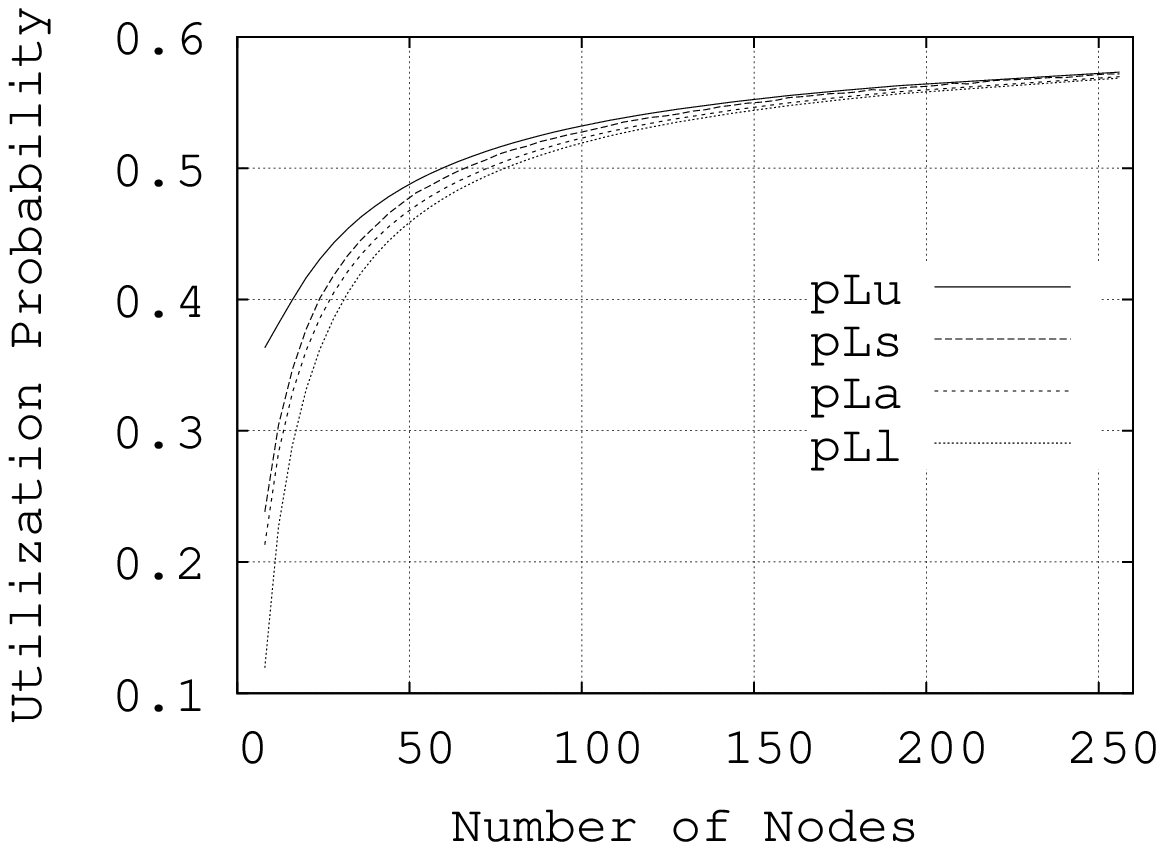}&
\includegraphics[width=.33\textwidth,angle=0]{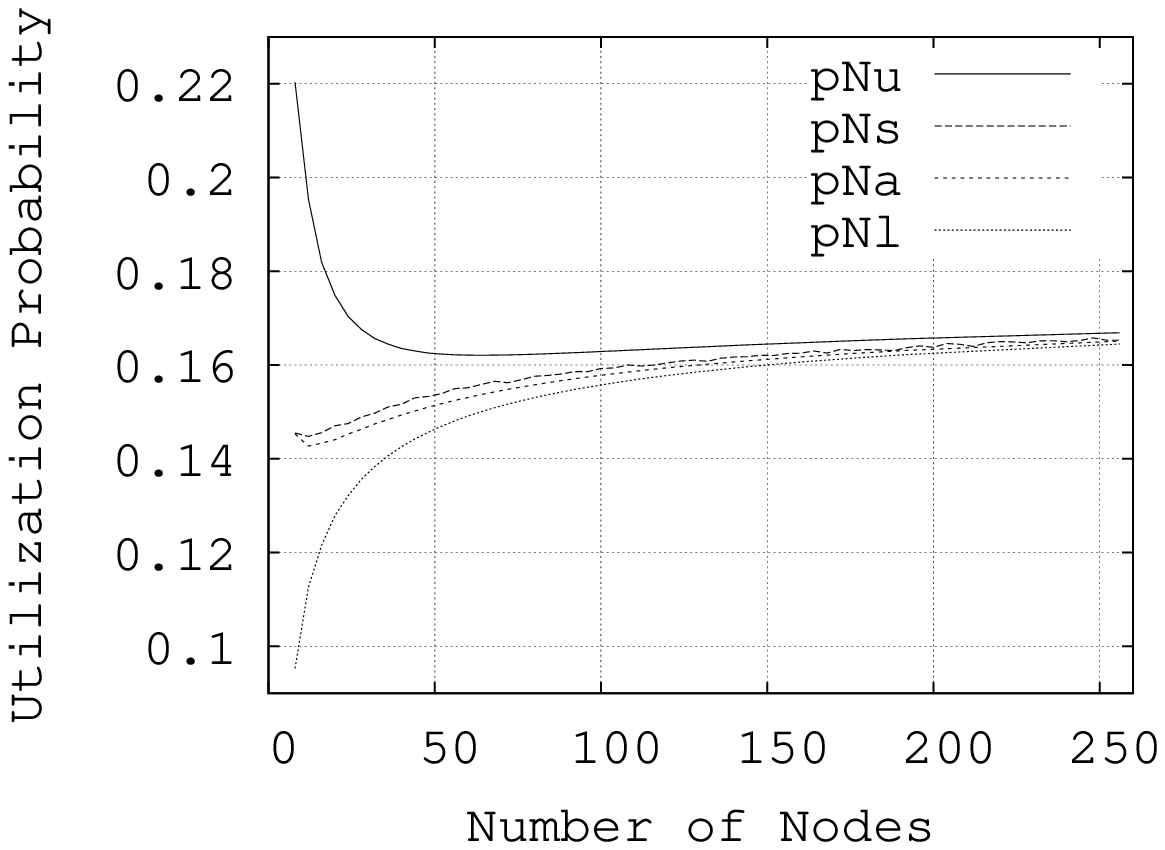}\\
(a) $\IP\left(\overset{\carr}{1}_{1}\right)$   &
(b) $\IP\left(\overset{\carr}{4}_{4}\right)$  &
(c) $\IP\left(\overset{\carr}{64}_{4}\right)$ \\
\end{tabular}
\caption{Segment utilization probability as a function of number of Nodes $N$
for $\alpha = 0.2$, $\beta = 0.2$, $\gamma = 0.6$,
and $\mu_1 = \nu_1 = \kappa_1 = 1/4$ and
$\mu_l = \nu_l = \kappa_l = 3/(4(N-2))$ for $l = 2, \ldots, N-1$.}
\label{Num1c}
\end{figure*}
For the first set of evaluations, we consider multicast traffic with
fixed fanout
$\mu_1 = \nu_1 = \kappa_1 = 1/4$ and
$\mu_l = \nu_l = \kappa_l = 3/(4(N-2))$ for $l = 2, \ldots, N-1$.
We examine increasing portions of hotspot traffic by setting
$\alpha = 1, \beta = \gamma = 0$ for Fig.~\ref{Num1a},
$\alpha = 0.6$, $\beta = 0.1$, and $\gamma = 0.3$ for Fig.~\ref{Num1b},
and
$\alpha = 0.2$, $\beta = 0.2$, and $\gamma = 0.6$ for Fig.~\ref{Num1c}.
We consider these scenarios with
hotspot traffic dominated by hotspot source traffic,
i.e., with $\gamma > \beta$, since many multicast applications involve
traffic distribution by a hotspot, e.g., for IP TV.

\begin{figure*}[t]
\centering
\begin{tabular}{ccc}
\includegraphics[width=.33\textwidth,angle=0]{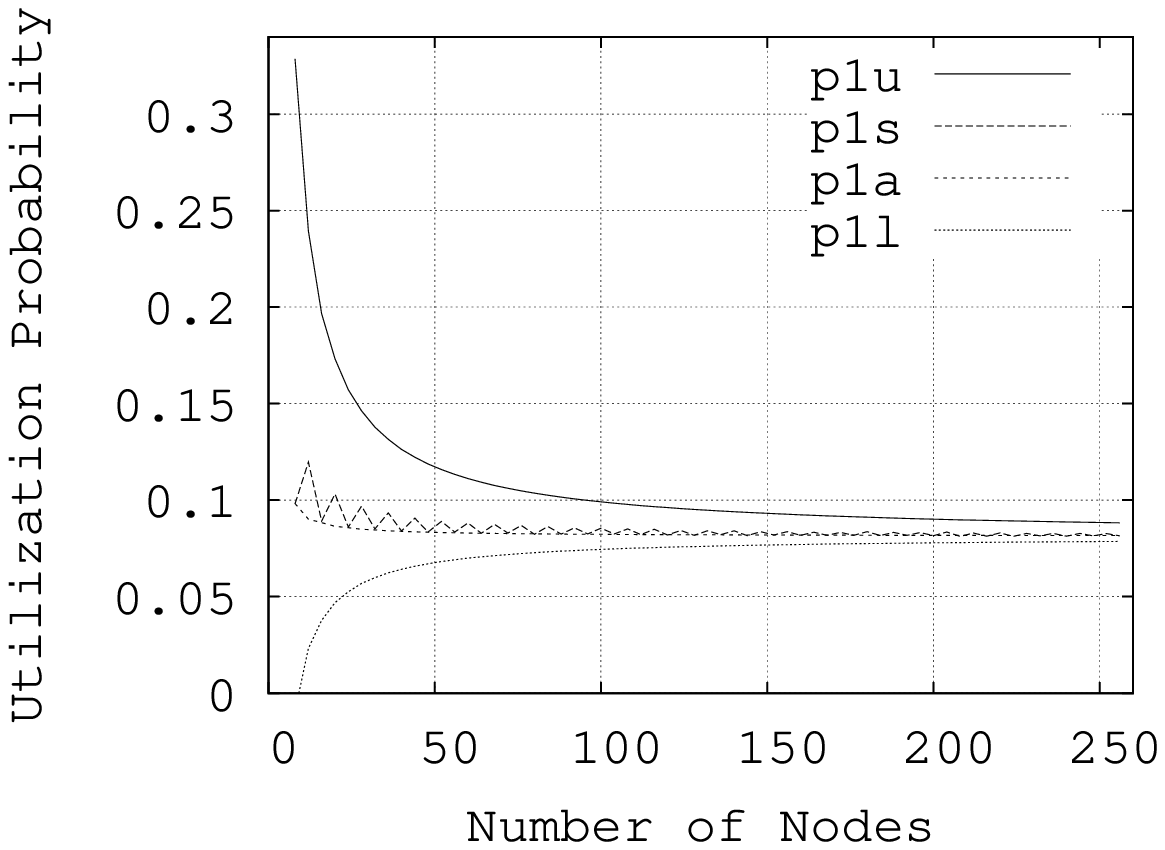}&
\includegraphics[width=.33\textwidth,angle=0]{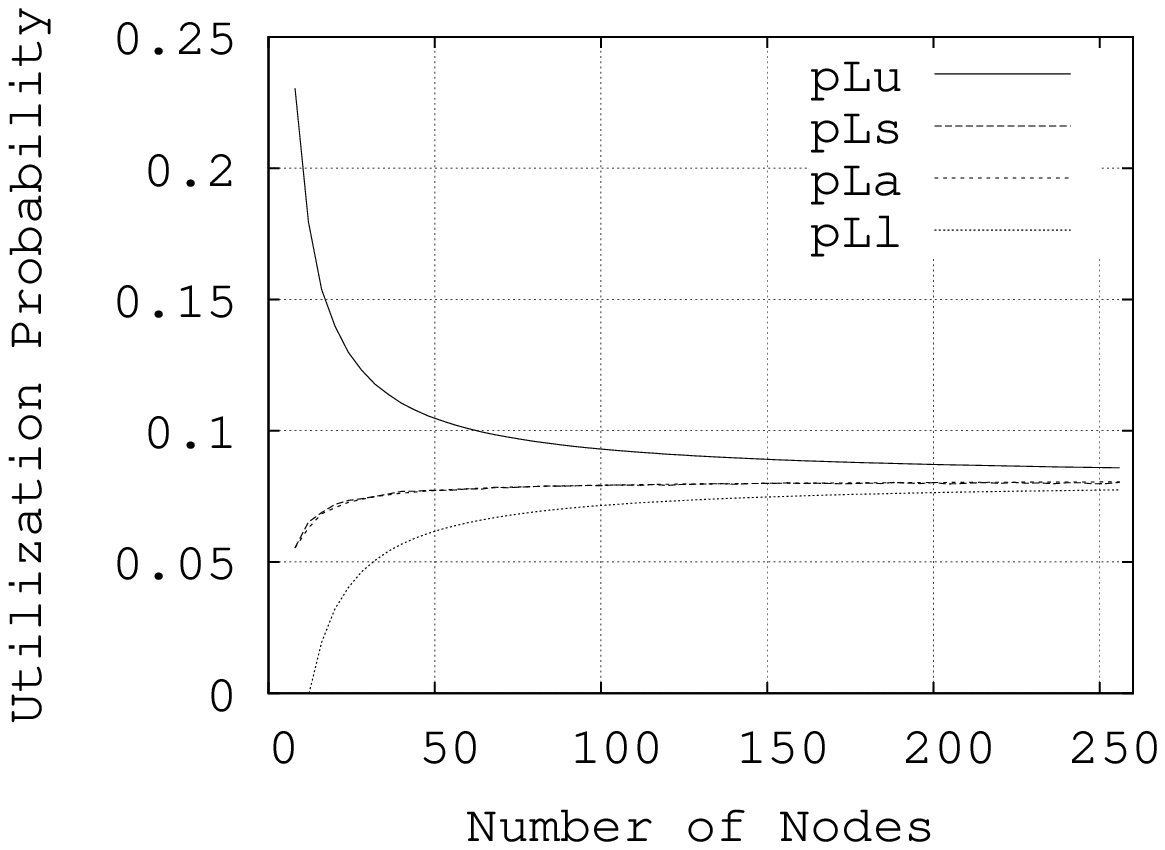}&
\includegraphics[width=.33\textwidth,angle=0]{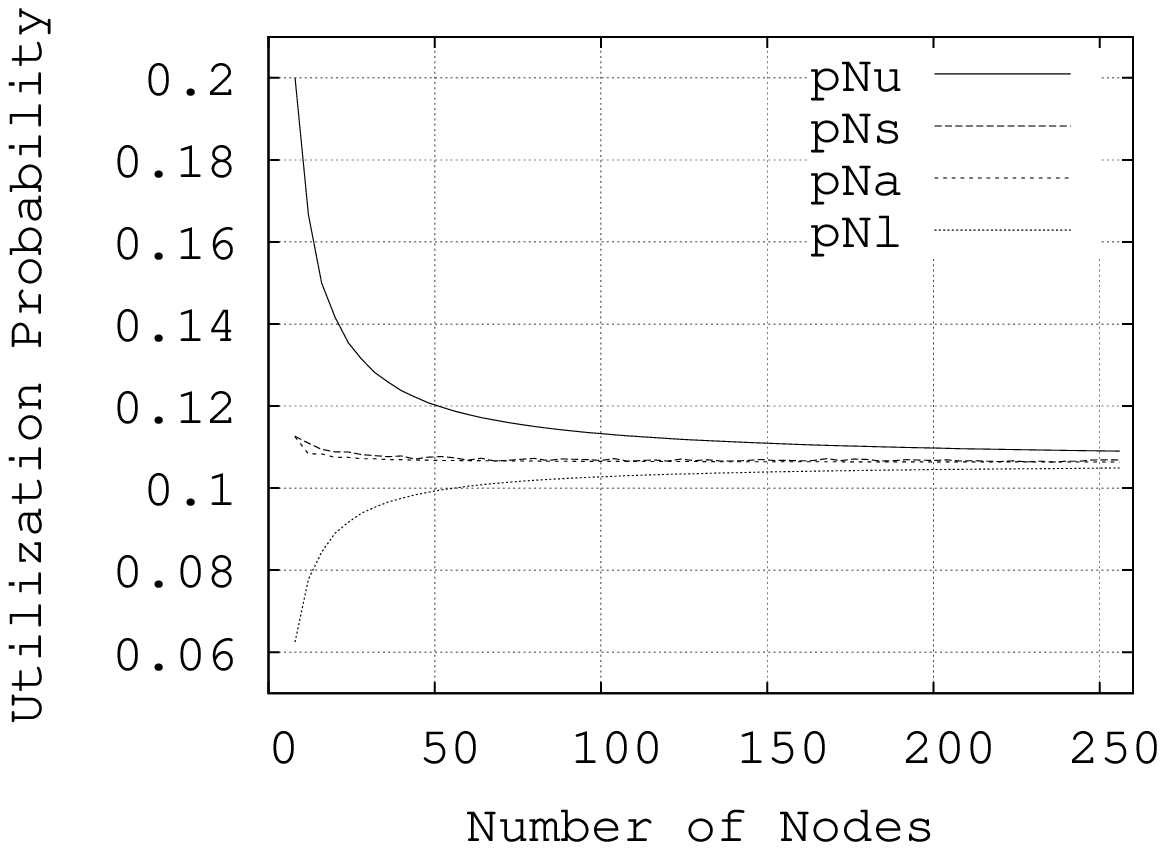}\\
(a) $\IP\left(\overset{\carr}{1}_{1}\right)$   &
(b) $\IP\left(\overset{\carr}{4}_{4}\right)$  &
(c) $\IP\left(\overset{\carr}{64}_{4}\right)$ \\
\end{tabular}
\caption{Segment utilization probability as a function of number of Nodes $N$
for $\alpha = 0.2$, $\beta = 0.2$, $\gamma = 0.6$, and
unicast (UC) traffic with $\mu_1 = \nu_1 = \kappa_1 = 1$.}
\label{Num2a}
\end{figure*}

\begin{figure*}[t]
\centering
\begin{tabular}{ccc}
\includegraphics[width=.33\textwidth,angle=0]{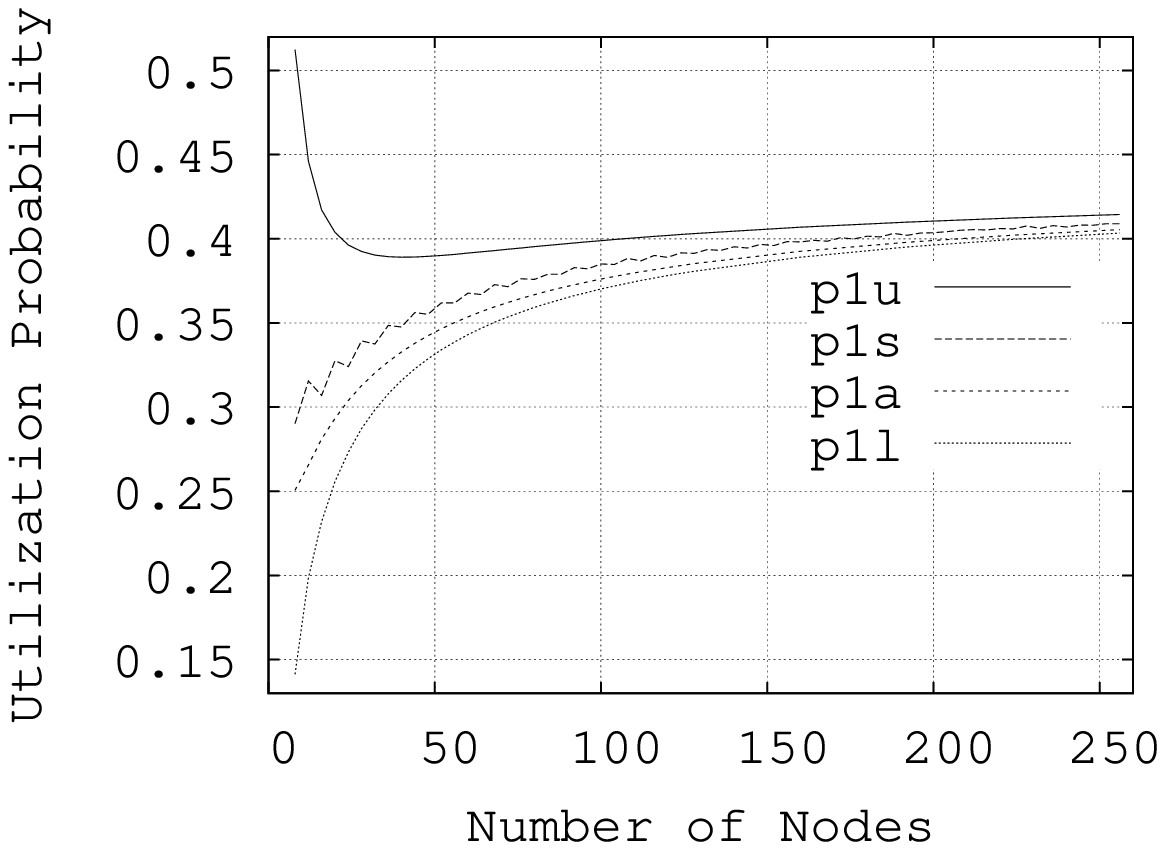}&
\includegraphics[width=.33\textwidth,angle=0]{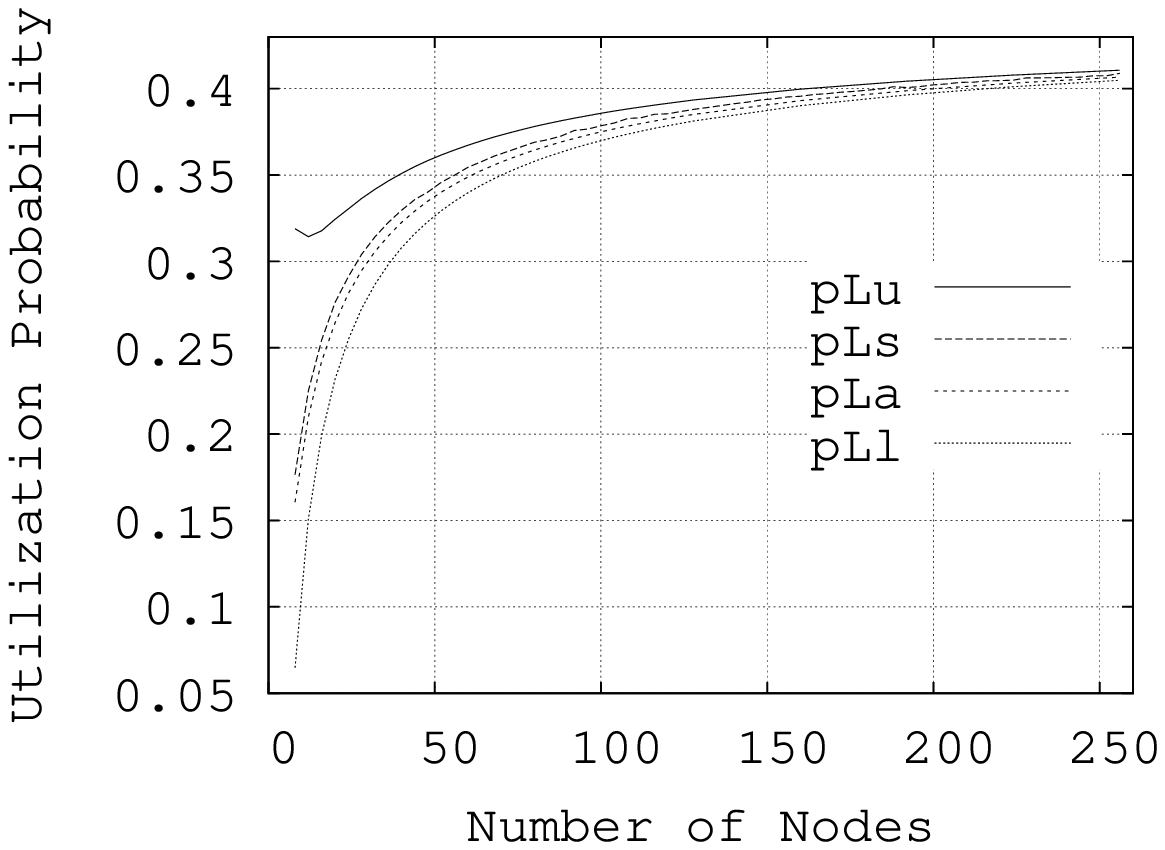}&
\includegraphics[width=.33\textwidth,angle=0]{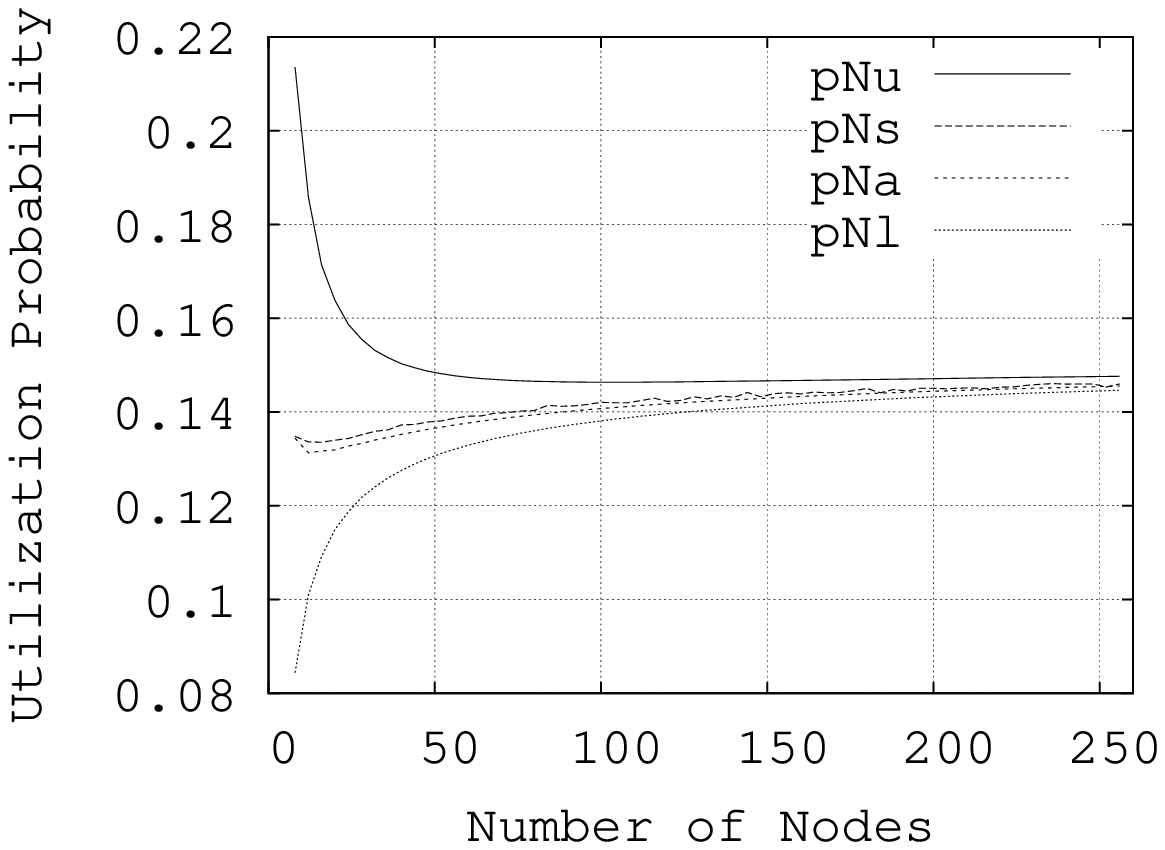}\\
(a) $\IP\left(\overset{\carr}{1}_{1}\right)$   &
(b) $\IP\left(\overset{\carr}{4}_{4}\right)$  &
(c) $\IP\left(\overset{\carr}{64}_{4}\right)$ \\
\end{tabular}
\caption{Segment utilization probability as a function of number of Nodes $N$
for $\alpha = 0.2$, $\beta = 0.2$, $\gamma = 0.6$, for
mixed (MI) traffic with
$\mu_1 = \nu_1 = \kappa_1 = 1/2$ and
$\mu_l = \nu_l = \kappa_l = 1/(2(N-2))$ for $l = 2, \ldots, N-1$.}
\label{Num2b}
\end{figure*}

\begin{figure*}[t]
\centering
\begin{tabular}{ccc}
\includegraphics[width=.33\textwidth,angle=0]{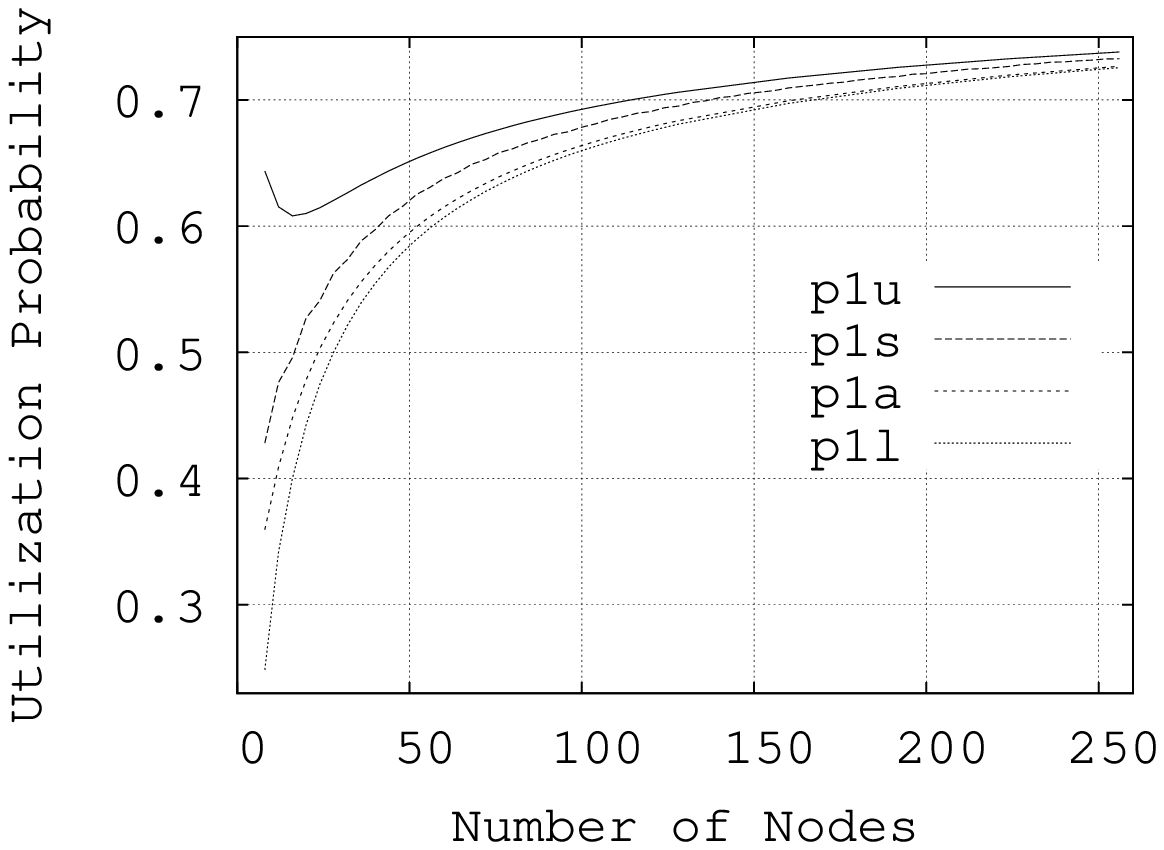}&
\includegraphics[width=.33\textwidth,angle=0]{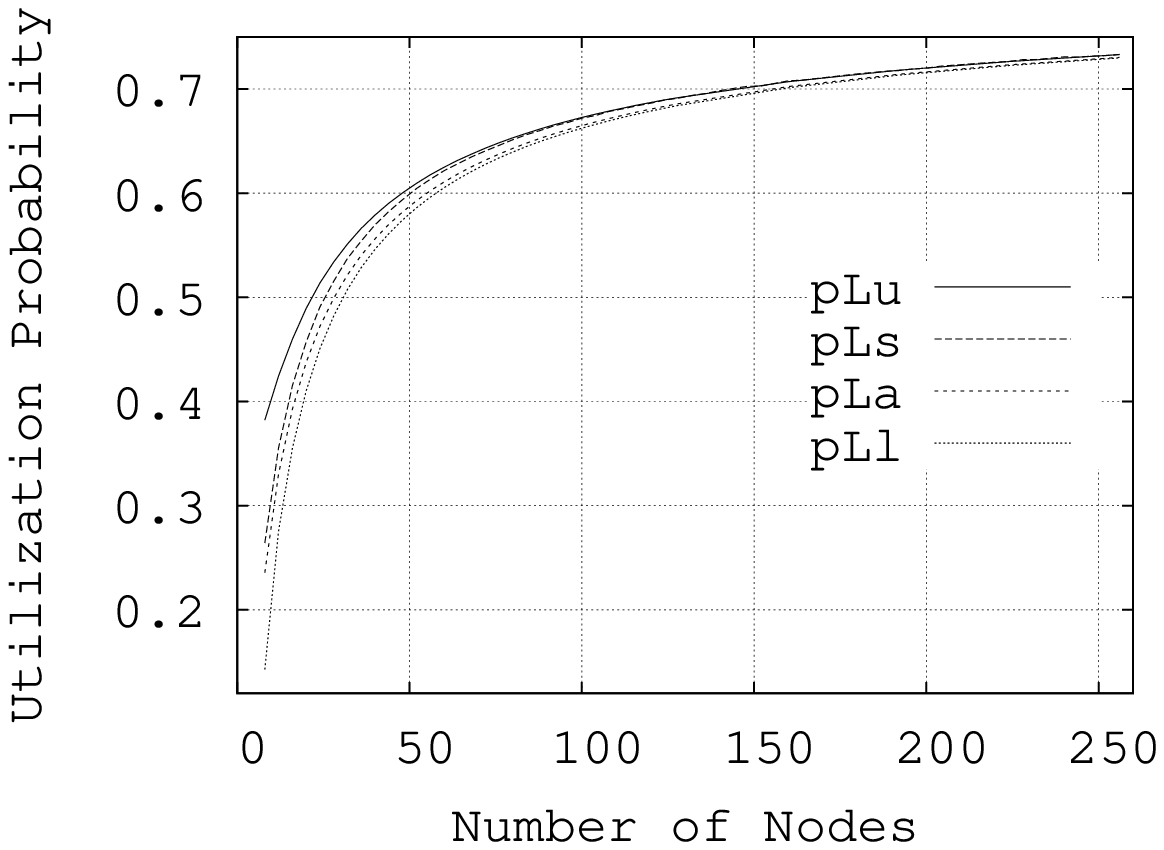}&
\includegraphics[width=.33\textwidth,angle=0]{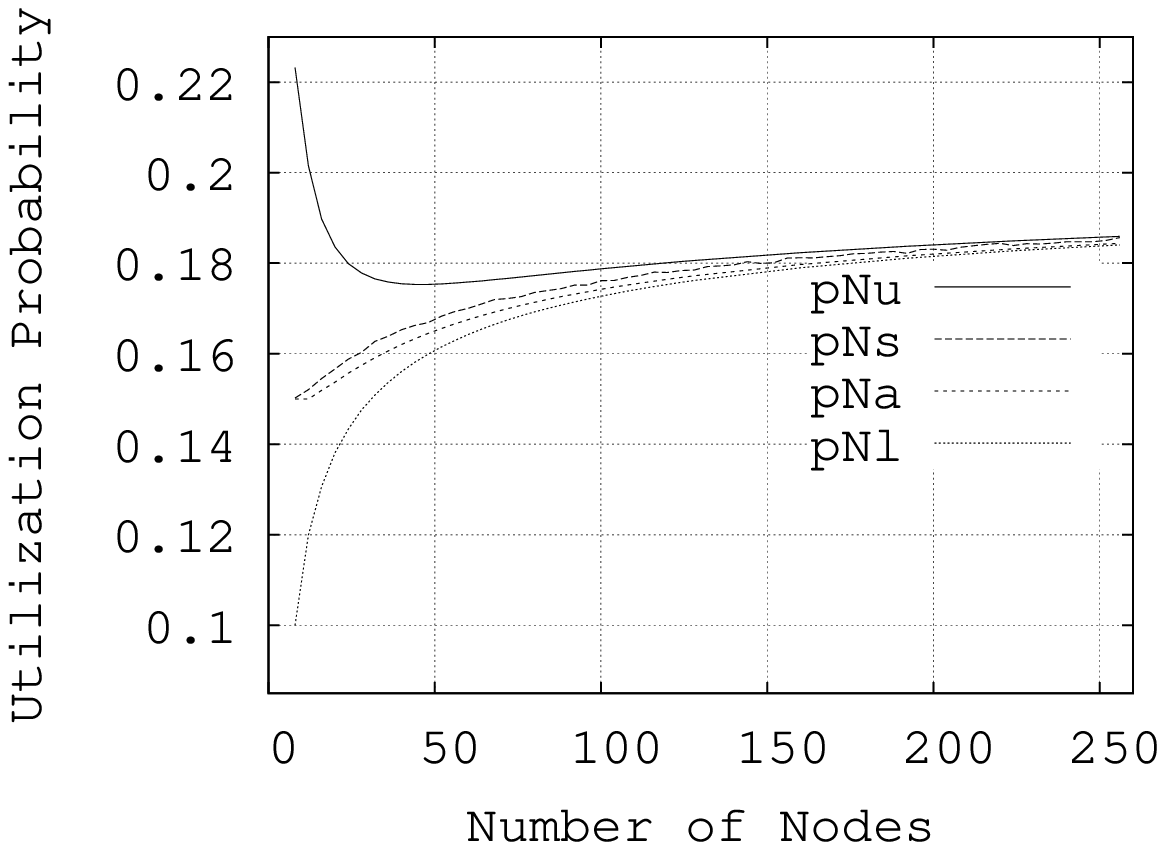}\\
(a) $\IP\left(\overset{\carr}{1}_{1}\right)$   &
(b) $\IP\left(\overset{\carr}{4}_{4}\right)$  &
(c) $\IP\left(\overset{\carr}{64}_{4}\right)$ \\
\end{tabular}
\caption{Segment utilization probability as a function of number of Nodes $N$
for $\alpha = 0.2$, $\beta = 0.2$, $\gamma = 0.6$, for multicast (MC) traffic
with $\mu_l = \nu_l = \kappa_l = 1/(N-1)$ for $l = 1, \ldots, N-1$.}
\label{Num2c}
\end{figure*}

\begin{figure*}[t]
\centering
\begin{tabular}{ccc}
\includegraphics[width=.33\textwidth,angle=0]{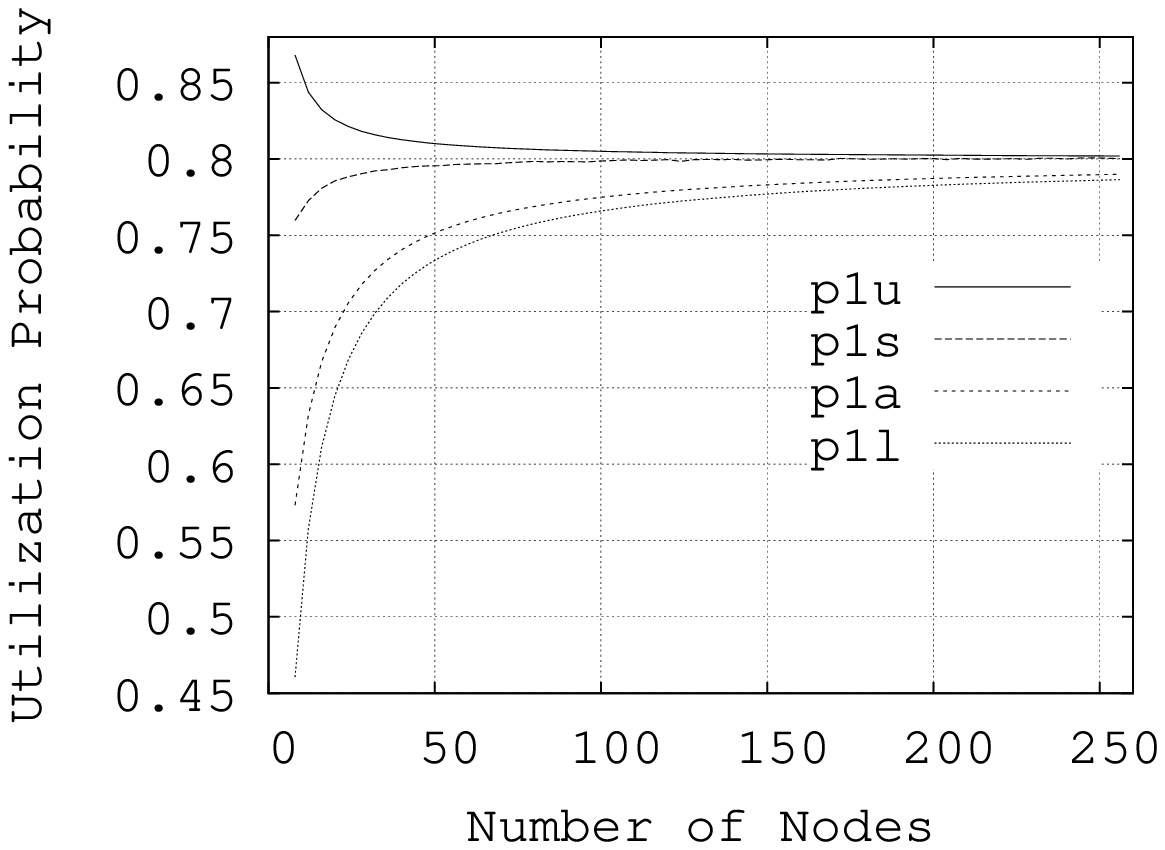}&
\includegraphics[width=.33\textwidth,angle=0]{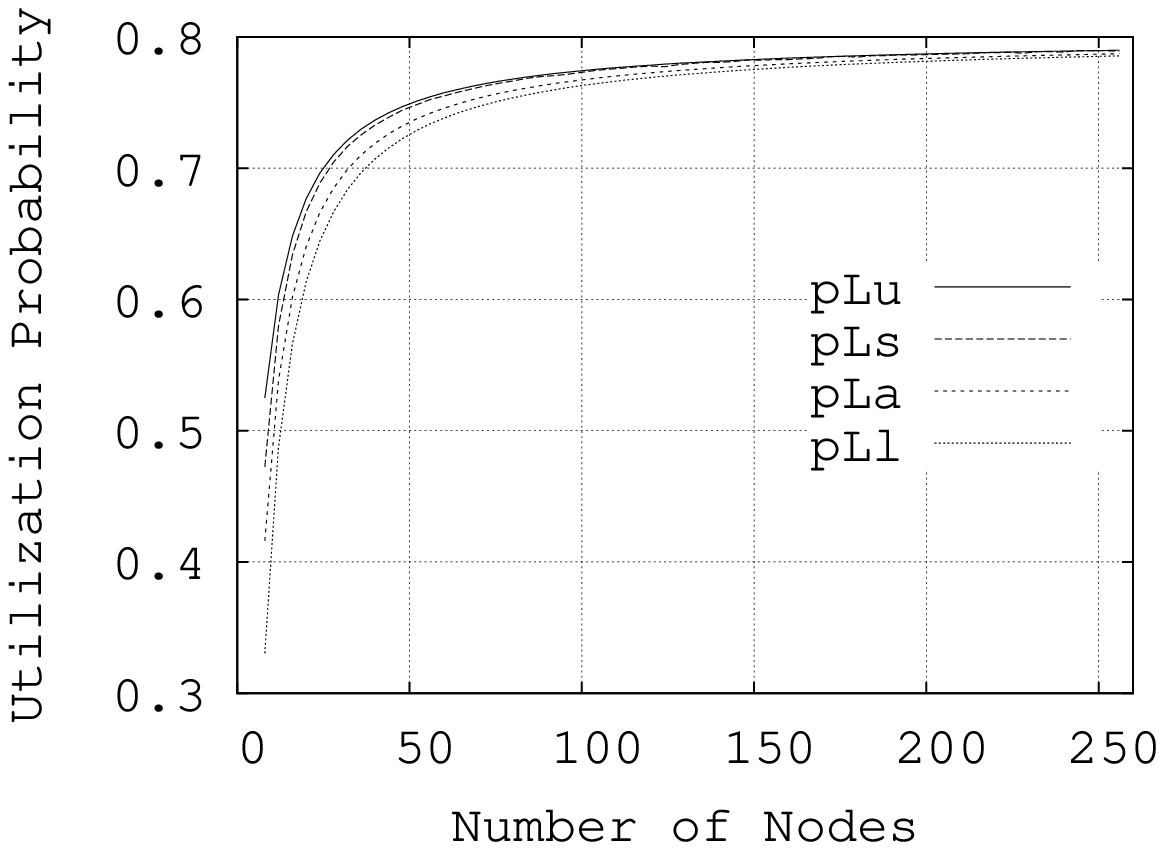}&
\includegraphics[width=.33\textwidth,angle=0]{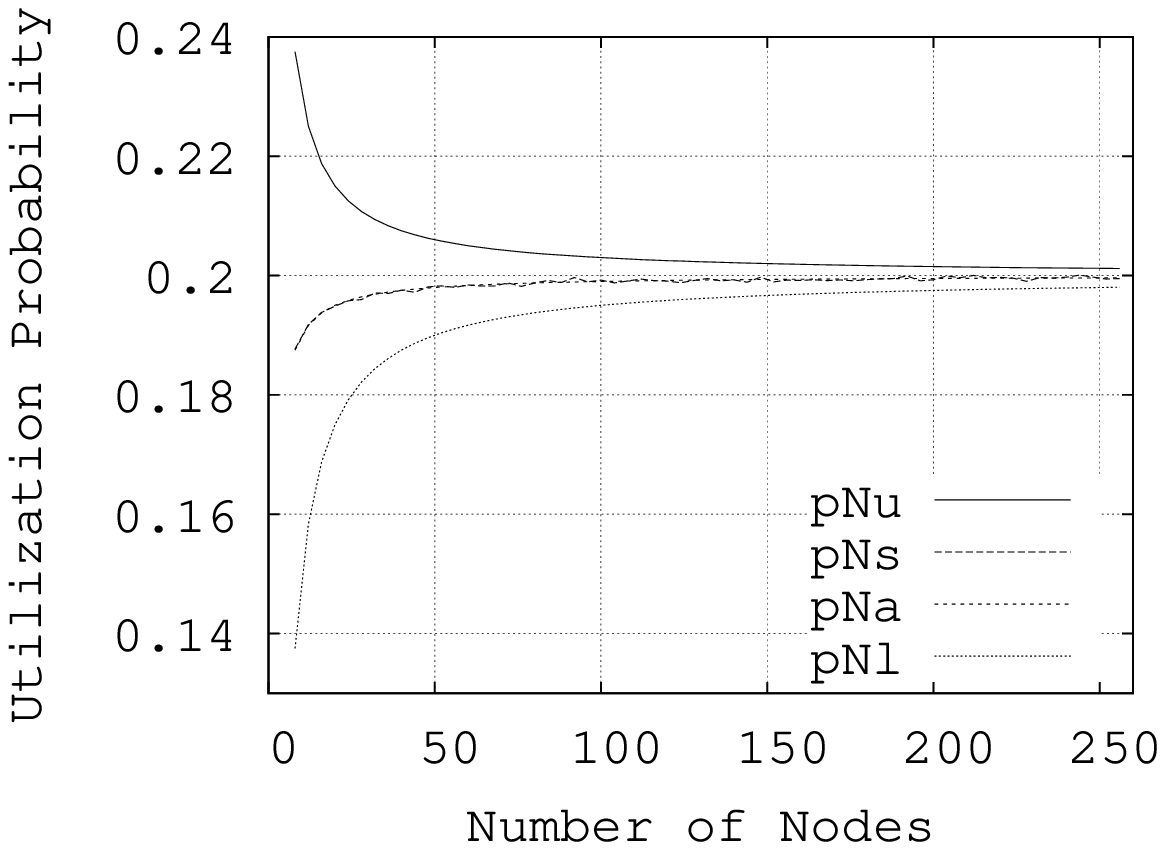}\\
(a) $\IP\left(\overset{\carr}{1}_{1}\right)$   &
(b) $\IP\left(\overset{\carr}{4}_{4}\right)$  &
(c) $\IP\left(\overset{\carr}{64}_{4}\right)$ \\
\end{tabular}
\caption{Segment utilization probability as a function of number of Nodes $N$
for $\alpha = 0.2$, $\beta = 0.2$, $\gamma = 0.6$, for broadcast (BC) traffic
with $\mu_{N-1} = \nu_{N-1} = \kappa_{N-1} = 1$.}
\label{Num2d}
\end{figure*}

We also consider a fixed traffic mix $\alpha = 0.2$,
$\beta = 0.2$, and $\gamma = 0.6$ for increasing fanout.
We consider unicast (UC) traffic with
$\mu_1 = \nu_1 = \kappa_1 = 1$ in Fig.~\ref{Num2a},
mixed traffic (MI) with $\mu_1 = \nu_1 = \kappa_1 = 1/2$ and
$\mu_l = \nu_l = \kappa_l = 1/(2(N-2))$ for $l = 2, \ldots, N-1$
in Fig.~\ref{Num2b},
multicast (MC) traffic with
$\mu_l = \nu_l = \kappa_l = 1/(N-1)$ for $l = 1, \ldots, N-1$
in Fig.~\ref{Num2c},
and broadcast (BC) traffic with
$\mu_{N-1} = \nu_{N-1} = \kappa_{N-1} = 1$
in Fig.~\ref{Num2d}.

We observe from these figures that the bounds get tight for
moderate to large numbers of nodes $N$ and that the
approximations
characterize the actual utilization probabilities fairly accurately
for the full range of $N$.
For instance, for $N = 64$ nodes, the difference between the upper and lower
bound is less than 0.06, for $N = 128$ this difference shrinks to
less than 0.026.
The magnitudes of the differences between the utilization probabilities
obtained with the analytical approximations and
the actual simulated utilization probabilities are less than 0.035 for
$N = 64$ nodes and less than 0.019 for $N = 128$ for the wide range of
scenarios considered in Figs.~\ref{Num1a}--\ref{Num2d}.
(When excluding the broadcast case considered in Fig.~\ref{Num2d},
these magnitude differences shrink to 0.02 for $N$ = 64 nodes
and 0.01 for $N$ = 128 nodes.)

For some scenarios we observe for small number of nodes $N$ 
slight oscillations of the actual utilization
probabilities obtained through simulations,
e.g., in Fig.~\ref{Num2a}(a) and~\ref{Num2b}(a).
More specifically, we observe peaks of the utilization probabilities for
odd $\eta$ and valleys for even $\eta$.
These oscillations are due to the discrete variations in the number of
destination nodes leading to segment traversals.
For instance, for the hotspot source unicast traffic that 
accounts for a $\gamma = 0.6$ portion of the traffic in
Fig.~\ref{Num2a}(a), the utilization of segment $\overset{\carr}{1}_{1}$
is as follows. For even $\eta$, there are $\eta/2$ possible destination
nodes that result in traversal of segment $\overset{\carr}{1}_{1}$, each
of these destination nodes occurs with probability $1/(N-1)$; hence,
segment $\overset{\carr}{1}_{1}$ is traversed with probability
$N/[2 \Lambda(N-1)]$. 
On the other hand, for odd $\eta$, there are $(\eta+1)/2$ possible destination
nodes that result in traversal of segment $\overset{\carr}{1}_{1}$; hence,
segment $\overset{\carr}{1}_{1}$ is traversed with probability
$(N + \Lambda)/[2 \Lambda(N-1)]$. 

Overall, we observe from Fig~\ref{Num1a} that for uniform traffic,
the three segments governing the
maximum utilization probability are evenly loaded.
With increasing fractions of non-uniform traffic (with hotspot source
traffic dominating over hotspot destination traffic), the
segments $\overset{\carr}{1}_{1}$ and $\overset{\carr}{4}_{4}$
experience increasing utilization probabilities
compared to segment $\overset{\carr}{64}_{4}$, as observed in
Figs.~\ref{Num1b} and~\ref{Num1c}.
Similarly, for the non-uniform traffic
scenarios with dominating hotspot source traffic,
we observe from Figs.~\ref{Num2a}--\ref{Num2d}
increasing utilization probabilities for the segments
$\overset{\carr}{1}_{1}$ and $\overset{\carr}{4}_{4}$
compared to segment $\overset{\carr}{64}_{4}$
with increasing fanout.
(In scenarios with dominating hotspot destination traffic, not shown
here due to space constraints, the utilization of
segment $\overset{\carr}{64}_{4}$ increases compared to
segments $\overset{\carr}{1}_{1}$ and $\overset{\carr}{4}_{4}$.)

In Figs.~\ref{Num1c},~\ref{Num2c}, and~\ref{Num2d},
the utilization probabilities
for segments $\overset{\carr}{1}_{1}$ and $\overset{\carr}{4}_{4}$
exceed one half for scenarios with moderate to large numbers of
nodes (and correspondingly large fanouts), indicating the potential
increase in multicast capacity by employing one-copy routing.

\subsection{Comparison of Segment Utilization Probabilities for
SP and OC Routing} 
\begin{figure*}[t]
\centering
\begin{tabular}{cc}
\includegraphics[width=.5\textwidth,angle=0]{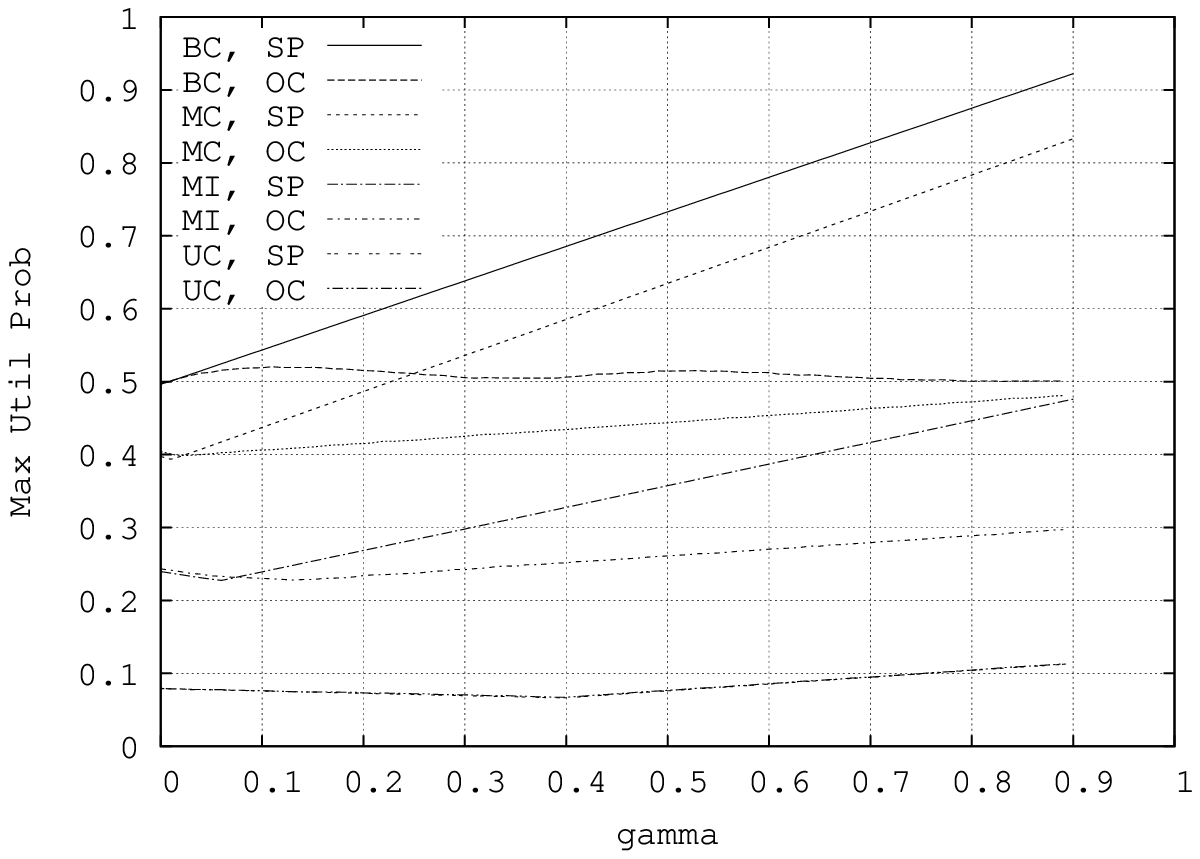} &
   \includegraphics[width=.5\textwidth,angle=0]{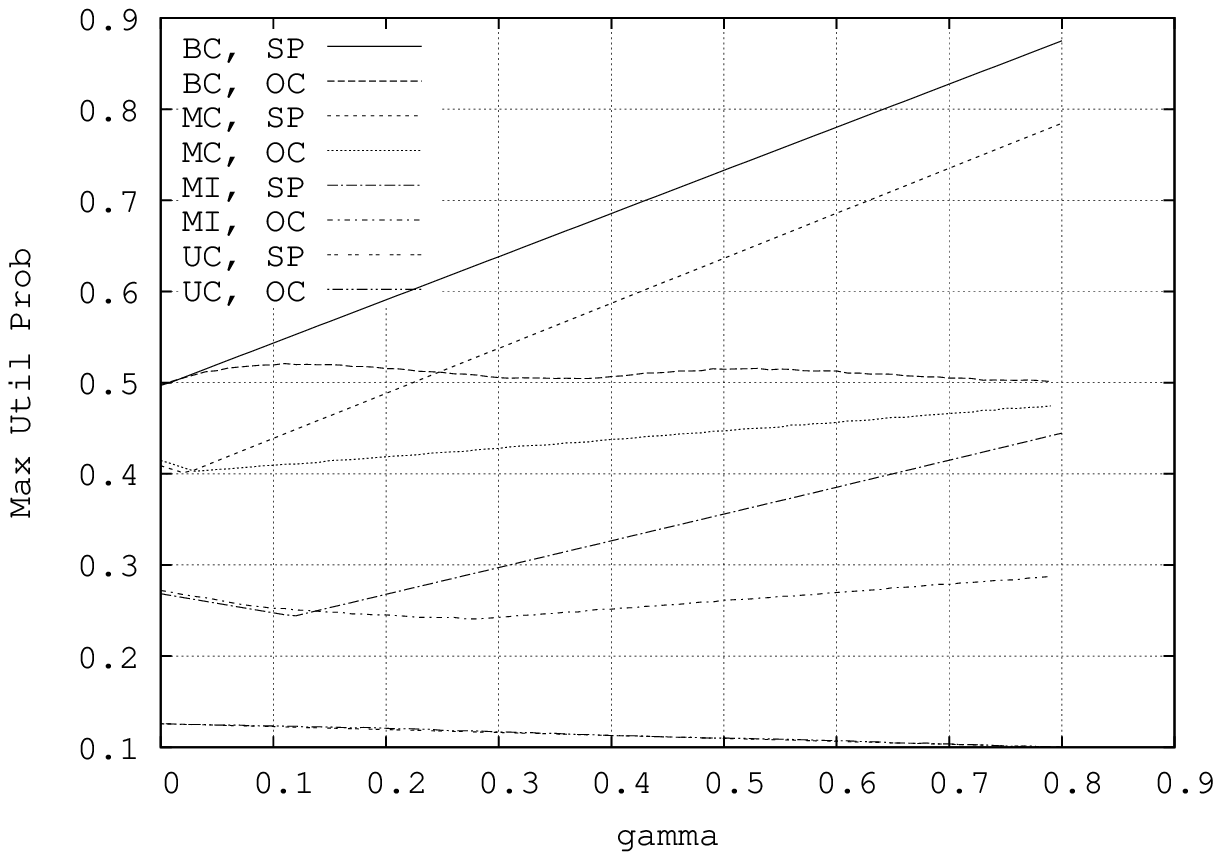}\\
(a) $\beta = 0.1$ & (b) $\beta  =0.2$
\end{tabular}
\caption{Maximum segment utilization probability as a function of
fraction of hotspot source traffic $\gamma$ (with $\alpha = 1 - \beta -\gamma$)
for shortest path (SP) and one-copy routing (OC) for fixed
fraction of hotspot traffic $\beta$ for
unicast (UC) traffic, mixed (MI) traffic, multicast (MC) traffic,
and broadcast (BC) traffic.}
\label{Num4}
\end{figure*}

In Fig.~\ref{Num4} we compare shortest path
routing (SP) with one-copy routing (OC) for unicast (UC) traffic,
mixed (MI) traffic, multicast (MC) traffic, and broadcast (BC)
traffic with the fanout distributions defined above for a network
with $N$ = 128 nodes. The corresponding thresholds $\gamma_{th1}$
and $\gamma_{th2}$ are reported in Table~\ref{gamma_th:tab}. For SP
routing, we plot the maximum segment utilization probability
obtained from the analytical approximations. For OC routing, we
estimate the utilization probabilities of all segments in the
network through simulations and then search for the largest segment
utilization probability.
\begin{table}
\begin{tabular}{|lrr|}\hline
Fanout & $\gamma_{th1}$ &  $\gamma_{th2}$\\
\multicolumn{3}{|c|}{$\beta = 0.1$}\\
UC & 0.397 & $\infty$ \\
MI & 0.059 & 7.32 \\
MC & 0.011 & 0.030 \\
BC & 0.0004 & 0.006 \\ \hline
\multicolumn{3}{|c|}{$\beta = 0.2$}\\
UC & 0.794 & $\infty$ \\
MI & 0.118 & 14.64 \\
MC & 0.022 & 0.061 \\
BC & 0.0008 & 0.013 \\ \hline
\end{tabular}
\caption{Thresholds $\gamma_{th1}$ and $\gamma_{th2}$ for scenarios
considered in Fig.~\ref{Num4}}
\label{gamma_th:tab}
\end{table}
Focusing initially on unicast traffic, we observe that both SP and
OC routing attain the same maximum utilization probabilities. This
is to be expected since the routing behaviors of SP and OC are
identical when there is a single destination on a wavelength. For
$\beta = 0.1$, we observe with increasing portion of hotspot source
traffic $\gamma$ an initial decrease, a minimum value,
and subsequent increase of the
maximum utilization probability. The value of the maximum
utilization probability for $\gamma = 0$ is due to the uniform and
hotspot destination traffic heavily loading segment
$\overset{\carr}{64}_{4}$. With increasing $\gamma$ and consequently
decreasing $\alpha$, the load on segment $\overset{\carr}{64}_{4}$
diminishes, while the load on segments $\overset{\carr}{1}_{1}$ and
$\overset{\carr}{4}_{4}$ increases. For approximately $\gamma =
0.4$, the three segments $\overset{\carr}{1}_{1}$,
$\overset{\carr}{4}_{4}$, and $\overset{\carr}{64}_{4}$ are about
equally loaded. As $\gamma$ increases further, the segments
$\overset{\carr}{1}_{1}$ and $\overset{\carr}{4}_{4}$ experience
roughly the same, increasing load. For $\beta  =0.2$ we observe only
the decrease of the maximum utilization probability, which is due to
the load on segment $\overset{\carr}{64}_{4}$ dominating the maximum
segment utilization. For this larger fraction of hotspot destination
traffic we do not reach the regime where segments
$\overset{\carr}{1}_{1}$ and $\overset{\carr}{4}_{4}$ govern the
maximum segment utilization.

Turning to broadcast traffic, we observe that SP routing
gives higher maximum utilization probabilities
than OC routing for essentially the entire range
of $\gamma$, reaching utilization probabilities around 0.9 for
high proportions of hotspot source traffic.
This is due to the high loading of segments
$\overset{\carr}{1}_{1}$ and $\overset{\carr}{4}_{4}$.
In contrast, with OC routing, the maximum segment utilization stays
close to 0.5, resulting in significantly increased capacity.
The slight excursions of the maximum OC segment utilization probability
above 1/2 are due to uniform traffic.
The segment utilization probability with uniform traffic is
approximated (not bounded) by (\ref{Palpha_approx:eqn}), making
excursions above 1/2 possible even though hotspot destination and
hotspot source traffic result in utilization probabilities less than
(or equal) to 1/2.

For mixed and multicast traffic, we observe for increasing
$\gamma$ an initial decrease, minimum value, and subsequent increase
of the maximum utilization probability for both SP and OC routing.
Similarly to the case of unicast traffic, these dynamics
are caused by initially dominating loading of segment
$\overset{\carr}{64}_{4}$, then a decrease of the loading
of segment $\overset{\carr}{64}_{4}$ while the loads on segments
$\overset{\carr}{1}_{1}$ and $\overset{\carr}{4}_{4}$ increase.
We observe for the mixed and multicast traffic scenarios with
the same fanout for all three traffic types considered
in Fig~\ref{Num4} that SP routing and OC routing give essentially
the same maximum segment utilization for small
$\gamma$ up to a ``knee point'' in the SP curves.
For larger $\gamma$, OC routing gives significantly smaller maximum
segment utilizations.
We observe from Table~\ref{gamma_th:tab} that for
relatively large fanouts (MC and BC), the ranges between
$\gamma_{th1}$ and $\gamma_{th2}$ are relatively small, limiting the
need for resorting to numerical evaluation and simulation for
determining whether to employ SP or OC routing.
For small fanouts (UC and MI), the $\gamma$ thresholds are far apart; further
refined decision criteria for routing with SP or OC are therefore an important
direction for future research.

\begin{figure*}[t]
\centering
\begin{tabular}{cc}
\includegraphics[width=.5\textwidth,angle=0]{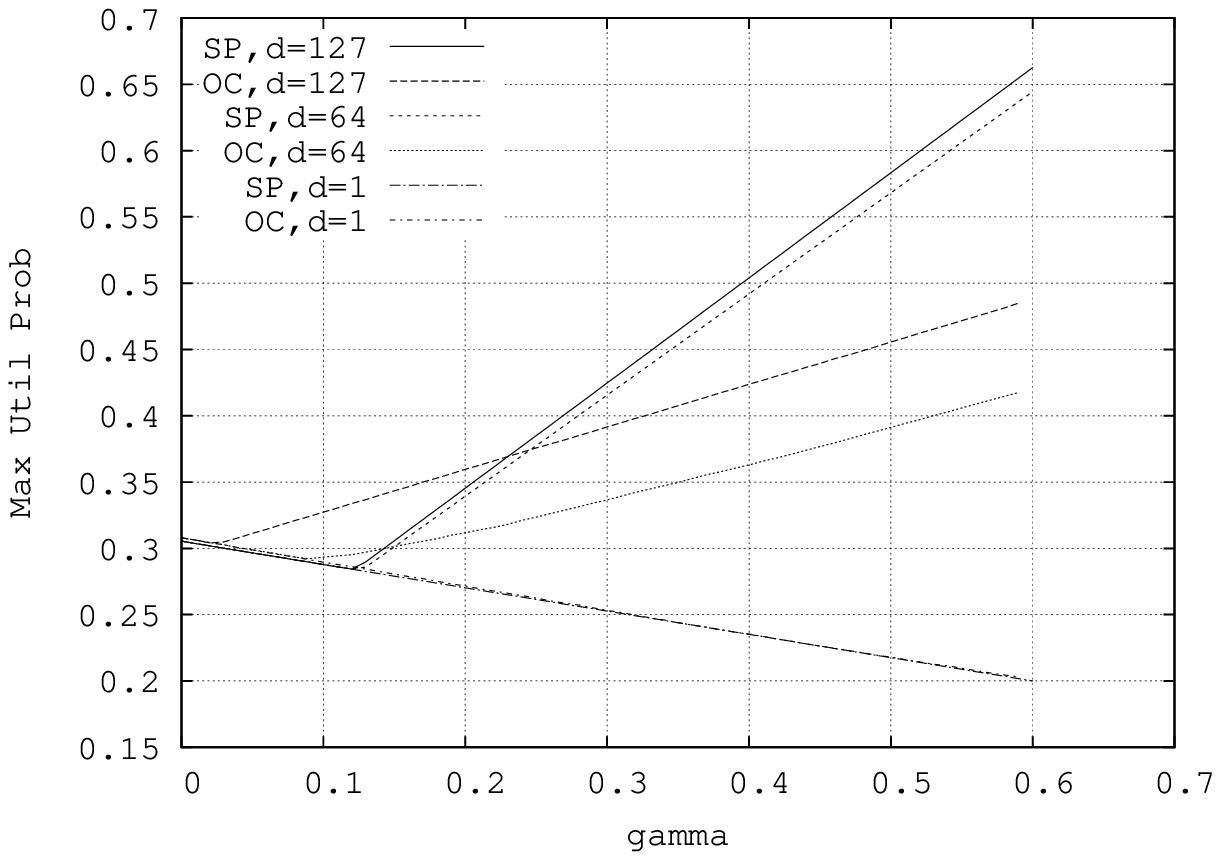} &
   \includegraphics[width=.5\textwidth,angle=0]{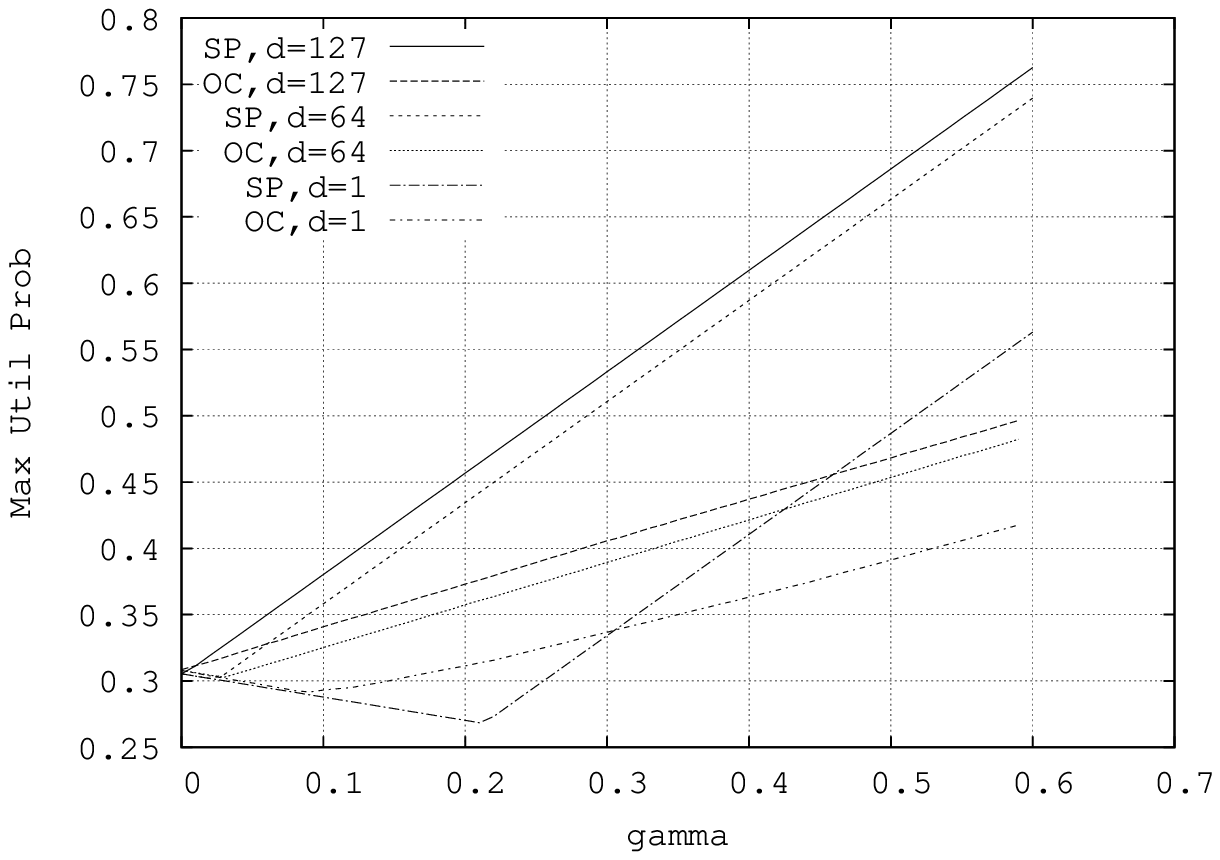}\\
(a) $\nu_8 = 1$, $\kappa_d = 1$ & (b) $\nu_d = 1$, $\kappa_{64} = 1$
\end{tabular}
\caption{Maximum segment utilization probability as a function of
fraction of hotspot source traffic $\gamma$.
Fixed parameters: $N = 128$ nodes, $\beta = 0.4$,
$\mu_l = 1/16$ for $l = 1, \ldots, 16$.}
\label{Num5}
\end{figure*}

We compare shortest path (SP) and one-copy (OC) routing
for scenarios with different fanout distribution for the different
traffic types in Fig.~\ref{Num5} for a ring with $N$ = 128 nodes.
\begin{table}
\begin{tabular}{|lrr|}\hline
Scenario & $\gamma_{th1}$ &  $\gamma_{th2}$\\
\multicolumn{3}{|c|}{$\kappa_d = 1$}\\
$d = 127$ & 0.122 & 0.283\\
$d = 64$ & 0.126 & 0.302 \\
$d = 1$ & 0.972& $\infty$ \\ \hline
\multicolumn{3}{|c|}{$\nu_d = 1$}\\
$d = 127$ & 0.0017 & 0.028 \\
$d = 64$ & 0.025 & 0.073 \\
$d = 1$ & 0.212 & 0.456 \\ \hline
\end{tabular}
\caption{Thresholds $\gamma_{th1}$ and $\gamma_{th2}$ for scenarios
considered in Fig.~\ref{Num5}}
\label{gamma_th_Num5:tab}
\end{table}
We observe from Fig.~\ref{Num5}(a) that for
hotspot source traffic with large fanout, SP routing achieves
significantly smaller maximum segment utilizations than OC routing
for $\gamma$ values up to a cross-over point, which lies
between $\gamma_{th1}$ and $\gamma_{th2}$.
Similarly, we observe from Fig.~\ref{Num5}(b)
that for small $\gamma$,
SP routing achieves significantly smaller maximum segment utilizations
than OC routing for hotspot destination traffic with small fanout.
For example, for unicast hotspot destination traffic
(i.e., $\nu_1 = 1$), for $\gamma = 0.21$, SP routing gives
a multicast capacity of $C_M = 3.72$ compared to
$C_M = 3.19$ with OC routing.
By switching from SP routing to OC routing when
the fraction of hotspot source traffic $\gamma$
exceeds 0.31, the smaller maximum utilization probability, i.e.,
higher multicast capacity can be achieved across the range of
fractions of hotspot source traffic $\gamma$.

\section{Conclusion}
\label{concl:sec}
We have analytically characterized the segment utilization probabilities
in a bi-directional WDM packet ring network with a single hotspot.
We have considered arbitrary mixes of unicast, multicast, and
broadcast traffic in combination with an arbitrary mix of
uniform, hotspot destination, and hotspot source traffic.
For shortest-path routing, we found that there are three segments
that can attain the maximum utilization, which in turn
limits the maximum achievable long-run average multicast packet throughput
(multicast capacity).
Through verifying simulations,
we found that our bounds and approximations of the segment
utilization probabilities, which are exact in the limit for
many nodes in a network with a fixed number of wavelength channels,
are fairly accurate for networks with on the order of ten nodes receiving
on a wavelength.
Importantly, we observed from our segment utilization analysis
that shortest-path routing does
\textit{not} maximize the achievable multicast packet throughput when
there is a significant portion of multi- or broadcast traffic emanating from
the hotspot, as arises with multimedia distribution,
such as IP TV networks.
We proposed a one-copy routing strategy with an achievable
long run average multicast packet throughout of about two simultaneous
packet transmissions for such distribution scenarios.

This study focused on the maximum achievable multicast packet throughput, but
did not consider packet delay.
A thorough study of the packet delay in WDM ring networks with
a hotspot transporting multicast traffic is an important direction
for future research.

\appendix

\vspace{\baselineskip}
\section{Definition of Enlarged and Reduced Ring as well as
 of Left $(\cA_{\lambda}^{\leftarrow})$
and Right Shifting $(\cA_{\lambda}^{\rightarrow})$ of Set of Active Nodes}
\label{AppendixA}

In this appendix, we
first define the enlarging and reducing of the
set of \char`\"{}$\lambda$-active nodes\char`\"{}
$\cA_{\lambda}:=\cF_{\lambda}\cup\left\{ S\right\} $.
Suppose that $\left|\cF_{\lambda}\right|=\ell$.
Depending on the setting,
and with $\cM_{\lambda}$ denoting
the set of nodes homed on a given wavelength $\lambda$,
the set $\cF_{\lambda}$ is chosen uniformly at random among

\begin{itemize}
\item all subsets of $\cM_{\lambda}$ (uniform traffic
and for $\lambda\neq\Lambda$
also hotspot destination and source traffic), or
\item all subsets of $\cM_{\lambda}$ that contain $N$ (hotspot destination
traffic
for $\lambda=\Lambda$ since $N$ is always a destination for
hotspot destination traffic), or
\item all subsets of $\cM_{\lambda}$ that do not contain $N$ (hotspot source
traffic for $\lambda=\Lambda$ since $N$ is always the source for
hotspot source traffic).
\end{itemize}

Assuming $S\notin \cM_{\lambda}$, we define:

\begin{description}
\item [enlarged$\,$ring]
We enlarge the set $\cM_{\lambda}$ by injecting
an extra node homed on $\lambda$
between $\left\lfloor S\right\rfloor _{\lambda}$ and
$\left\lceil S\right\rceil _{\lambda}$
(and correspondingly $\Lambda - 1$ nodes homed on the other wavelengths).
After a re-numeration starting
with $0$ at the new node (which is accordingly homed on wavelength
$\Lambda$ after the re-numeration),
we obtain
$\cM_{\Lambda,\eta+1}:=
\left\{ m\Lambda\,\big|\, m\in\left\{ 0,\ldots,\eta\right\} \right\}$.
We define the enlarged set $\cF_{\lambda}^{+}$ to equal the renumbered
set $\cF_{\lambda}$ united with the new node. This procedure leads
to a random set of active nodes
$\cA_{\lambda}^{+}=\cF_{\lambda}^{+}$
that is uniformly distributed among all subsets of $\cM_{\Lambda,\eta+1}$
with cardinality $\left(\ell+1\right)$ containing node $0$.
Note that the largest gap of the enlarged set is
larger or equal to the largest gap of $\cA_{\lambda}$.

\begin{figure}[h]
\begin{center}
\ineps{.3}{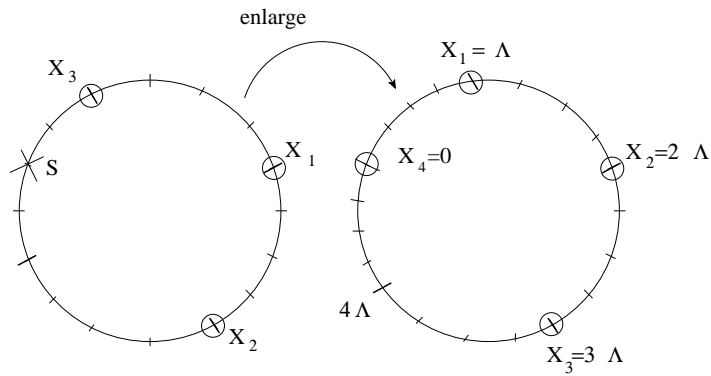} \centering
\end{center}
\caption{Example of enlarging $\cM_3$ for $N=16,\,\Lambda=4$.
The sender homed on wavelength 1 is represented by $S$ in the left
illustration.
The nodes of $\cM_3$ are indicated by longer tick marks
and the nodes of $\cF_3$ are circled.
The enlarged ring has a total of $N + \Lambda = 20$ nodes, with $\eta + 1 = 5$
nodes homed on each wavelength. The added node on wavelength 3 is numbered with
0 and lies between the former $\left\lfloor S\right\rfloor _{\lambda}$ and
$\left\lceil S\right\rceil _{\lambda}$.}
\label{fig2a}
\end{figure}

\item [reduced$\,$ring]
We transform the set $\cM_{\lambda}$ by merging
the nodes $\left\lfloor S\right\rfloor _{\lambda}$ and
$\left\lceil S\right\rceil _{\lambda}$
to a single active node (eliminating the
$\Lambda-1$ nodes inbetween). After re-numeration starting with $0$ at
this merged node, we obtain an active set $\cA_{\lambda}^{-}$ on
$\cM_{\Lambda,\eta-1}$.

Depending on the cardinality of $\cF_{\lambda}\cap\left\{ \left\lfloor S\right\rfloor _{\lambda},\left\lceil S\right\rceil _{\lambda}\right\} $
the new active set $\cA_{\lambda}^{-}$ has $\ell+1$, $\ell$,
or $\ell-1$ elements.
More specifically, if neither the left- nor the right-shifted source node
was a destination node, then $|\cA_{\lambda}^{-}| = \ell + 1$.
If either the left- or the right-shifted source node was a destination node,
then $|\cA_{\lambda}^{-}| = \ell$.
If both the left- and right-shifted source node were destination nodes,
then $|\cA_{\lambda}^{-}| = \ell - 1$.
In each of these cases $\cA_{\lambda}^{-}$ is uniformly distributed
among all subsets of $\cM_{\lambda,\eta-1}$
with cardinality $\left|\cA_{\lambda}^{-}\right|$
that contains node $0$.\\
 Observe that in all cases, the largest gap of $\cA_{\lambda}^{-}$
is smaller or equal to the largest gap of $\cA_{\lambda}$.
\end{description}

\begin{figure}[h]
\begin{center}
\ineps{.3}{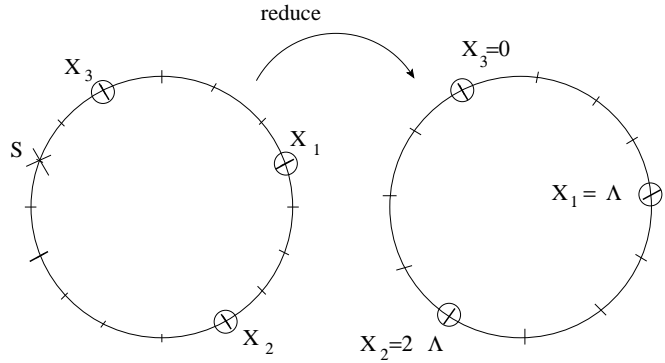} \centering
\end{center}
\caption{Example of reducing for $N=16,\,\Lambda=4$.
The sender is represented by $S$ and the nodes of $\cM_3$
have longer tick marks.
The nodes of $\cF_{3}$ are circled.
The nodes $\left\lfloor S\right\rfloor _{\lambda}$ and
$\left\lceil S\right\rceil _{\lambda}$
(as well as the 3 nodes inbetween) are merged into the node
numbered 0 in the right illustration.}
\label{fig3a}
\end{figure}

We also define the following transformations:

\begin{description}
\item [Left (counter clockwise) shifting]
  Since $S$ is uniformly distributed on $\left\{ 1,\ldots,N\right\}$,
the set
\begin{equation}
\cA_{\lambda}^{\leftarrow}:=\cF_{\lambda}\cup\left\{ \left\lfloor S\right\rfloor _{\lambda}\right\}
\end{equation}
is a random subset of $\cM_{\lambda}$.
We can think of $\cA_{\lambda}^{\leftarrow}$
as being chosen uniformly at random among all subsets of $\cM_{\lambda}$
having cardinality $\left|\cA_{\lambda}^{\leftarrow}\right|$ and
subject to the same conditions as $\cF_{\lambda}$. \\
Notice that
$\left|\cA_{\lambda}^{\leftarrow}\right|= \left|\cF_{\lambda}\right|$
if
$\left\lfloor S\right\rfloor _{\lambda}\in\cF_{\lambda}$
and
$\left|\cA_{\lambda}^{\leftarrow}\right|=\left|\cF_{\lambda}\right|+1$
otherwise.

\begin{figure}[h]
\begin{center}
\ineps{.3}{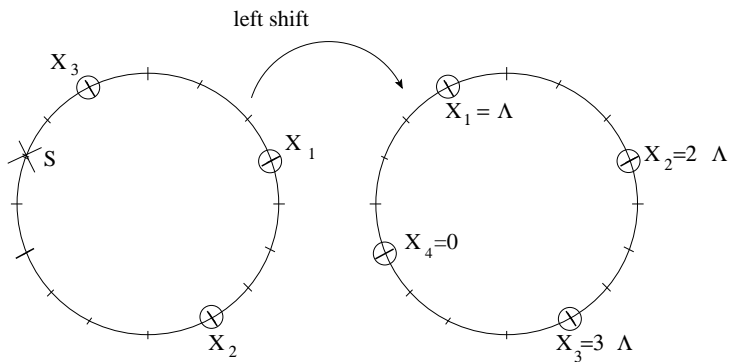} \centering
\end{center}
\caption{Example of left shifting for $N=16,\,\Lambda=4$.
The destination nodes are circled on the left, and the
active nodes are circled on the right.
The nodes are renumbered after the shifting, 
starting with the former sender at 0. Also, the 
active nodes is renumbered, starting with $X_1 >0$, 
the first active node after the former sender. 
The former sender is therefore the last active node, 
i.e., $X_4 = 0$.}
\label{fig4a}
\end{figure}

\item [Right (clockwise) shifting]
Analogously we define
\begin{equation}
\cA_{\lambda}^{\rightarrow}:=\cF_{\lambda}\cup\left\{ \left\lceil S\right\rceil _{\lambda}\right\}.
\end{equation}
 This is a random set chosen uniformly at random among all subsets
of $\cM_{\lambda}$ having cardinality
$\left|\cA_{\lambda}^{\rightarrow}\right|$
and subject to the same conditions as $\cF_{\lambda}$.
\end{description}

\begin{figure}[h]
\begin{center}
\ineps{.3}{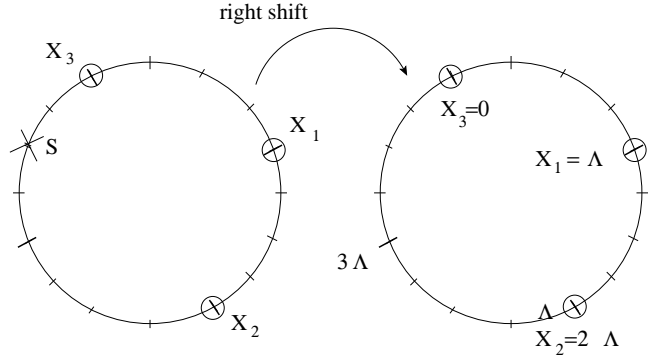} \centering
\end{center}
\caption{Example of right shifting for $N=16,\,\Lambda=4$.
After renumbering, the former sender is $X_3 = 0$.}
\label{fig5a}
\end{figure}

\vspace{\baselineskip}
\section{Proof of Proposition~\ref{Prop in beta} on Bounds for Probability that
CLG starts at Node 0 for Hotspot Destination Traffic for $\lambda = \Lambda$}
\label{AppendixBB}

\begin{proof}
Conditioned on $S\in \cM_{\Lambda}$, we obtain \begin{equation}
q_{\beta}^{\ell}\left(\cG_{\Lambda}=0\,|\, S\in \cM_{\Lambda}\right)=\frac{1}{\ell+1}.\end{equation}
 Hence, we only have to consider the case $S\notin \cM_{\Lambda}$. We
will not explicitly write down this condition.

Consider the right shifting and denote by $\cG_{\Lambda}^{\rightarrow}$
the starting point of the chosen largest gap of $\cA_{\Lambda}^{\rightarrow}$.
Since $N\equiv0$ is the only fixed active node, the first gap, i.e.,
$\left\{ 0,\ldots,X_{\Lambda,1}\right\} $, is the only one that never
shrinks, while the last gap, i.e.,
$\left\{ X_{\Lambda,\ell+1},\ldots,N\right\} $,
is the only one that never grows. Therefore,
\begin{eqnarray}
q_{\beta}^{\ell}\left(\cG_{\Lambda}=0\right) & \leq & q_{\beta}^{\ell}\left(\cG_{\Lambda}^{\rightarrow}=0\right).
\end{eqnarray}
For reasons of symmetry, we have
\begin{eqnarray}
q_{\beta}^{\ell}\left(\cG_{\Lambda}^{\rightarrow}=0\,|\,\left\lceil S\right\rceil _{\Lambda}\notin\cF_{\Lambda}\right) & = & q_{\gamma}^{\ell}\left(\cG_{\Lambda}=0\right)
\nonumber \\
 & = & \frac{1}{\ell+1},
\end{eqnarray}
and
\begin{eqnarray}
q_{\beta}^{\ell}\left(\cG_{\Lambda}^{\rightarrow}=0\,|\,\left\lceil S\right\rceil _{\Lambda}\in\cF_{\Lambda}\right) & = & q_{\gamma}^{\ell-1}\left(\cG_{\Lambda}=0\right)
\nonumber \\
& = & \frac{1}{\ell}.
\end{eqnarray}
The remaining probabilities can be computed as
$q_{\beta}^{\ell}\left(\left\lceil
S\right\rceil _{\Lambda}\in\cF_{\Lambda}\,|\, S\notin \cM_{\Lambda}\right)
=\frac{\ell}{\eta}$,
leading to the desired upper bound.

Analogously, the left shifting yields a lower bound, namely
\begin{eqnarray}
q_{\beta}^{\ell}\left(\cG_{\Lambda}=0\,|\,\left\lfloor S\right\rfloor _{\Lambda}\neq0\right) & \geq & q_{\beta}^{\ell}\left(\cG_{\Lambda}^{\leftarrow}=0\,|\,\left\lfloor S\right\rfloor _{\Lambda}\neq0\right).
\end{eqnarray}
Again for reasons of symmetry, we obtain
\begin{equation}
q_{\beta}^{\ell}\left(\cG_{\Lambda}^{\leftarrow}=0\,|\,\left\lfloor S\right\rfloor _{\Lambda}\notin\cF_{\Lambda}\right)=\frac{1}{\ell+1}
\end{equation}
and
\begin{equation}
q_{\beta}^{\ell}\left(\cG_{\Lambda}^{\leftarrow}=0\,|\,\left\lfloor S\right\rfloor _{\Lambda}\in\cF_{\Lambda}\setminus\left\{ 0\right\} \right)=\frac{1}{\ell}.
\end{equation}
Finally, we have, of course, $q_{\beta}^{\ell}\left(\left\lfloor S\right\rfloor _{\Lambda}\in\cF_{\Lambda}\,|\, S\notin \cM_{\Lambda}\right)=\frac{\ell}{\eta}$
and $q_{\beta}^{\ell}\left(\left\lfloor S\right\rfloor _{\Lambda}\in\cF_{\Lambda}\setminus\left\{ 0\right\} \,|\,
S\notin \cM_{\Lambda}\right)=\ell-1.$
\end{proof}

\vspace{\baselineskip}
\section{Proof of Theorem~\ref{Mainthm} on the Maximal
 Segment Utilization}
\label{AppendixC}

\begin{proof}

Due to Equation (\ref{max lambda neq Lambda}),
we only have to prove the case of drop wavelength $\Lambda$.

Corollary \ref{Cor critical non critical} tell us that it suffices
to consider the critical segments. Let $n\equiv\delta\Lambda$ with
$1\leq\delta<\eta$ be a critical segment for $\Lambda$. Analogously
to the proof of the domination principle in~\cite{HeSSR07},
we reduce the domination principle for hotspot destination traffic
to the statement \begin{equation}
q_{\beta}^{\ell}\left(n\geq\cG_{\Lambda}>n-\Lambda\right)\geq\frac{1}{\eta-\delta}q_{\beta}^{\ell}\left(\cG_{\Lambda}>n-\Lambda\right),\end{equation}
and for hotspot source traffic to:
\begin{equation}
q_{\gamma}^{\ell}\left(\cG_{\Lambda}=n\right)\geq\frac{1}{\eta-\delta}q_{\gamma}^{\ell}\left(\cG_{\Lambda}\geq n\right).
\end{equation}

\begin{figure}[h]
\begin{center}
\ineps{.45}{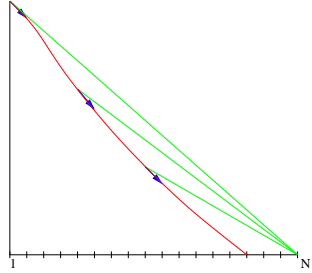} \centering
\end{center}
\caption{Illustration of statement (C.1): the mean slope of a certain period is bigger or equal than the mean slope over all later periods}
\label{fig6a}
\end{figure}

In the $\gamma$ (hotspot source traffic) setting,
we know that $N$ is the sender, and thus
$\cA_{\Lambda}\subset \cM_{\Lambda}.$
Hence, we do not need to consider the nodes on the other drop wavelengths and
the proof is exactly the same as in the single
wavelength case~\cite{HeSSR07}, see also figure C.2.
\begin{figure}[h]
\begin{center}
\ineps{.45}{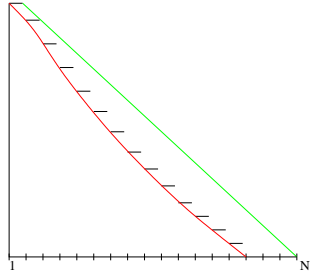} \centering
\end{center}
\caption{Gamma setting: the usage probability stays constant on non critical edges}
\label{fig6}
\end{figure}

We will now use the same
strategy for the more complicated proof in the $\beta$
(hotspot destination traffic) setting.
Let $K_{n}$ denote the number of active nodes finding themselves
between the nodes $N$ and $n$ (clockwise), i.e.,
\begin{equation}
K_{n}:=\left|\cA_{\Lambda}\cap\left\{ 1,\ldots,n-\Lambda\right\} \right|.
\end{equation}
 For $k\in\left\{ 0,\ldots,\left(n-1\right)\wedge\left(\ell-1\right)\right\} $
we denote $q_{\gamma}^{\ell,k}$ for the probability measure $q_{\gamma}^{\ell}$
conditioned on $K_{n}=k.$
We denote again $n\equiv\delta\Lambda$
for $\delta\in\left\{ 1,\ldots,\eta-1\right\} $.
We will show that
\begin{equation}
q_{\beta}^{\ell}\left(n-\Lambda<\cG_{\Lambda}\leq n\right)\geq\frac{1}{\eta-\delta}q_{\beta}^{\ell}\left(\cG_{\Lambda}>n-\Lambda\right).
\end{equation}
\begin{figure}[h]
\begin{center}
\ineps{.45}{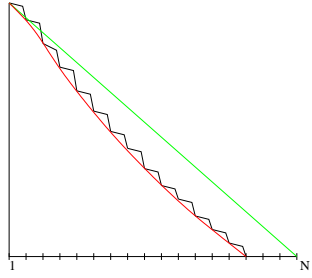} \centering
\end{center}
\caption{Beta setting: the usage probability changes along each segment}
\label{fig7}
\end{figure}

In case that $S\in \cM_{\Lambda}$ we can again use
the proof of the one wavelength scenario.
This is also true if $S\in\left\{1,\ldots,n-\Lambda\right\}$,
since we do not claim anything about these nodes.
Hence, we only have to investigate the
case $S\in\left\{ n-\Lambda+1,\ldots,N\right\} \setminus \cM_{\Lambda}$.
From now on we assume this to be the case.

We decompose the left hand side into two parts, \begin{equation}
q_{\beta}^{\ell}\left(n-\Lambda<\cG_{\Lambda}\leq n\right)=q_{\beta}^{\ell}\left(\cG_{\Lambda}=n\right)+q_{\beta}^{\ell}\left(\cG_{\Lambda}=S,\, n-\Lambda<S<n\right).\label{two parts}\end{equation}
 For the first summand of (\ref{two parts}), we proceed similarly
to the case of a single wavelength, namely \begin{eqnarray}
q_{\beta}^{\ell,k}\left(\cG_{\Lambda}=n\right) & = & q_{\beta}^{\ell,k}\left(\cG_{\Lambda}=n,\,\cG_{\Lambda}\geq n,\,\left\lfloor S\right\rfloor _{\Lambda}\neq n,\, n\in\cF_{\Lambda}\right)\nonumber \\
 & = & q_{\beta}^{\ell,k}\left(\cG_{\Lambda}=n\,\big|\,\cG_{\Lambda}\geq n,\,\left\lfloor S\right\rfloor _{\Lambda}\neq n,\, n\in\cF_{\Lambda}\right)\times\nonumber \\
 &  & \times q_{\beta}^{\ell,k}\left(\cG_{\Lambda}\geq n\,\big|\,\left\lfloor S\right\rfloor _{\Lambda}\neq n,\, n\in\cF_{\Lambda}\right)\nonumber \\
 &  & \times q_{\beta}^{\ell,k}\left(\left\lfloor S\right\rfloor _{\Lambda}\neq n,\, n\in\cF_{\Lambda}\right).\end{eqnarray}
We obtain
\begin{eqnarray}
q_{\beta}^{\ell,k}\left(\cG_{\Lambda}=n\,|\, G\geq n,\,\left\lfloor S\right\rfloor _{\Lambda}\neq n,\, n\in\cF_{\Lambda}\right) & = & q_{\beta,N-n+\Lambda}^{\ell-k}\left(\cG_{\Lambda}=1\,|\,\left\lfloor S\right\rfloor _{\Lambda}\neq1,\,1\in\cF_{\Lambda}\right)\nonumber \\
 & \geq & q_{\beta,N-n+\Lambda}^{\ell-k}\left(\cG_{\Lambda}^{\leftarrow}=1\,|\,\left\lfloor S\right\rfloor _{\Lambda}\neq1,\,1\in\cF_{\Lambda}\right).\end{eqnarray}
This probability can be computed precisely \begin{eqnarray}
 &  & q_{\beta,N-n+\Lambda}^{\ell-k}\left(\cG_{\Lambda}^{\leftarrow}=1\,|\,\left\lfloor S\right\rfloor _{\Lambda}\neq1,\,1\in\cF_{\Lambda}\right)\nonumber \\
 &  & =q_{\gamma,N-n}^{\ell-k-1}\left(\cG_{\Lambda}=0\right)q_{\beta,N-n+\Lambda}^{\ell-k}\left(\left\lfloor S\right\rfloor _{\Lambda}\notin\cF_{\Lambda}\,|\,\left\lfloor S\right\rfloor _{\Lambda}\neq1,\,1\in\cF_{\Lambda}\right)+\nonumber \\
 &  & \quad+q_{\gamma,N-n}^{\ell-k-2}\left(\cG_{\Lambda}=0\right)q_{\beta,N-n+\Lambda}^{\ell-k}\left(\left\lfloor S\right\rfloor _{\Lambda}\in\cF_{\Lambda}\,|\,\left\lfloor S\right\rfloor _{\Lambda}\neq1,\,1\in\cF_{\Lambda}\right)\nonumber \\
 &  & =\frac{1}{\ell-k}\left(1-\frac{\ell-k-1}{\eta-\delta}\right)+\frac{1}{\ell-k-1}\frac{\ell-k-1}{\eta-\delta}\nonumber \\
 &  & =\frac{1}{\ell-k}\left(1+\frac{1}{\eta-\delta}\right).
\end{eqnarray}

We now use the fact that,
conditionally on $S\in\left\{ n-\Lambda+1,\ldots,N\right\}
\setminus \cM_{\Lambda}$,
\begin{equation}
q_{\beta}^{\ell,k}\left(\left\lfloor S\right\rfloor _{\Lambda}\neq n\in\cF_{\Lambda}\right)=q_{\beta}^{\ell,k}\left(\left\lfloor S\right\rfloor _{\Lambda}\neq n\right)q_{\beta}^{\ell,k}\left(n\in\cF_{\Lambda}\right)
\end{equation}
and
\begin{equation}
q_{\beta}^{\ell,k}\left(\cG_{\Lambda}\geq n\,\big|\,\left\lfloor S\right\rfloor _{\Lambda}\neq n\in\cF_{\Lambda}\right)=\frac{q_{\beta}^{\ell,k}\left(\cG_{\Lambda}\geq n\,\big|\, n\in\cF_{\Lambda}\right)}{q_{\beta}^{\ell,k}\left(\left\lfloor S\right\rfloor _{\Lambda}\neq n\right)}.
\end{equation}
Hence, we obtain with $q_{\beta}^{\ell,k}\left(n\in\cF_{\Lambda}\right)=\frac{\ell-k-1}{\eta-\delta}$
that
\begin{eqnarray}
q_{\beta}^{\ell,k}\left(\cG_{\Lambda}=n\right) & \geq & \frac{1}{\eta-\delta}q_{\beta}^{\ell,k}\left(\cG_{\Lambda}\geq n\,\big|\, n\in\cF_{\Lambda}\right)\times\nonumber \\
 &  & \left(1-\frac{1}{\ell-k}\right)\left(1+\frac{1}{\eta-\delta}\right).
\end{eqnarray}
For the second part of (\ref{two parts}), we obtain
\begin{eqnarray}
q_{\beta}^{\ell,k}\left(\cG_{\Lambda}\in I_{\delta}\setminus n\right) & = & q_{\beta}^{\ell,k}\left(\cG_{\Lambda}=S,\,\cG_{\Lambda}\geq S,\,\left\lceil S\right\rceil _{\Lambda}=n,\, n\notin\cF_{\Lambda}\right)\nonumber \\
 & = & q_{\beta}^{\ell,k}\left(\cG_{\Lambda}=S\,\big|\,\cG_{\Lambda}\geq S,\,\left\lceil S\right\rceil _{\Lambda}=n,\, n\notin\cF_{\Lambda}\right)\times\nonumber \\
 &  & \times q_{\beta}^{\ell,k}\left(\cG_{\Lambda}\geq S\,\big|\,\left\lceil S\right\rceil _{\Lambda}=n\notin\cF_{\Lambda}\right)q_{\beta}^{\ell,k}\left(\left\lceil S\right\rceil _{\Lambda}=n,\, n\notin\cF_{\Lambda}\right).
\end{eqnarray}
We have
\begin{eqnarray}
 &  & q_{\beta}^{\ell,k}\left(\cG_{\Lambda}=S\,|\,\cG_{\Lambda}\geq S,\,\left\lceil S\right\rceil _{\Lambda}=n,\, n\notin\cF_{\Lambda}\right)\nonumber \\
 &  & =q_{\beta,N-n+\Lambda}^{\ell-k}\left(\cG_{\Lambda}=S\,|\,\left\lceil S\right\rceil _{\Lambda}=1,\,1\notin\cF_{\Lambda}\right).
\end{eqnarray}
Now, we use that $\left|\cA_{\Lambda}^{\rightarrow}\right|=\left|\cF_{\Lambda}+1\right|$
for $\left\lceil S\right\rceil _{\Lambda}\notin\cF_{\Lambda}$. Hence,
we obtain
\begin{eqnarray}
 &  & q_{\beta,N-n+\Lambda}^{\ell-k}\left(\cG_{\Lambda}=S\,|\,\left\lceil S\right\rceil _{\Lambda}=1,\,1\notin\cF_{\Lambda}\right)\nonumber \\
 &  & \geq q_{\beta,N-n+\Lambda}^{\ell-k}\left(\cG_{\Lambda}^{\rightarrow}=1\,|\,\left\lceil S\right\rceil _{\Lambda}=1,\,1\notin\cF_{\Lambda}\right)\nonumber \\
 &  & =q_{\gamma,N-n}^{\ell-k-1}\left(\cG_{\Lambda}=0\right)\nonumber \\
 &  & =\frac{1}{\ell-k}.
\end{eqnarray}
Note that, conditioned on $S\in\left\{ n-\Lambda+1,\ldots,N\right\}
\setminus \cM_{\Lambda}$,
we have
\begin{equation}
q_{\beta}^{\ell,k}\left(\left\lceil S\right\rceil _{\Lambda}=n\notin\cF_{\Lambda}\right)=q_{\beta}^{\ell,k}\left(\left\lceil S\right\rceil _{\Lambda}=n\right)q_{\beta}^{\ell,k}\left(n\notin\cF_{\Lambda}\right)
\end{equation}
and
\begin{equation}
q_{\beta}^{\ell,k}\left(\cG_{\Lambda}\geq S\,\big|\,\left\lceil S\right\rceil _{\Lambda}=n\notin\cF_{\Lambda}\right)=\frac{q_{\beta}^{\ell,k}\left(\cG_{\Lambda}\geq S\,\big|\,\left\lceil S\right\rceil _{\Lambda}=n\right)}{q_{\beta}^{\ell,k}\left(n\notin\cF_{\Lambda}\right)}.
\end{equation}
Summarizing, we obtain, using $q_{\beta}^{\ell,k}\left(\left\lceil S\right\rceil _{\Lambda}=n\right)=\frac{1}{\eta-\delta}$,
that
\begin{eqnarray}
q_{\beta}^{\ell,k}\left(\cG_{\Lambda}\in I_{\delta}\setminus n\right) & \geq & \frac{1}{\eta-\delta}\frac{1}{\ell-k}q_{\beta}^{\ell,k}\left(\cG_{\Lambda}\geq S\,\big|\,\left\lceil S\right\rceil _{\Lambda}=n\right).
\end{eqnarray}
It remains to show that
\begin{eqnarray}
q_{\beta}^{\ell,k}\left(\cG_{\Lambda}>n-\Lambda\right) & \leq & \left(1-\frac{1}{\ell-k}\right)\left(1+\frac{1}{\eta-\delta}\right)\times\nonumber \\
 &  & \times q_{\beta}^{\ell,k}\left(\cG_{\Lambda}\geq n\,|\, n\in\cF_{\Lambda}\right)+\nonumber \\
 &  & +\frac{1}{\ell-k}\left(1-\frac{\ell-k-1}{\eta-\delta}\right)\times\nonumber \\
 &  & \times q_{\beta}^{\ell,k}\left(\cG_{\Lambda}\geq S\,|\,\left\lceil S\right\rceil _{\Lambda}=n,\, n\notin\cF_{\Lambda}\right).
 \label{nearly finish}
\end{eqnarray}
This can be shown by
\begin{eqnarray}
 &  & q_{\beta}^{\ell,k}\left(\cG_{\Lambda}>n-\Lambda\right)
\nonumber \\
 &  & =\sum_{i=\delta}^{\eta-\left(\ell-k\right)}q_{\beta}^{\ell,k}\left(\cG_{\Lambda}\geq i\Lambda\,|\, X_{k+1}=i\Lambda\right)q_{\beta}^{\ell,k}\left(X_{k+1}=i\Lambda\right)+
\nonumber \\
 &  & \qquad+\sum_{\lambda=1}^{\Lambda-1}q_{\beta}^{\ell,k}\left(\cG_{\Lambda}\geq i\Lambda-\lambda\,|\, X_{k+1}=i\Lambda-\lambda\right)q_{\beta}^{\ell,k}\left(X_{k+1}=i\Lambda-\lambda\right)
\nonumber \\
 &  & \leq\left(1-\frac{1}{\ell-k}\right)q_{\beta}^{\ell,k}\left(\cG_{\Lambda}\geq n\,|\, X_{k+1}=n\right)+\nonumber \\
 &  & \quad+\frac{1}{\ell-k}q_{\beta}^{\ell,k}\left(\cG_{\Lambda}\geq S\,|\,\left\lceil S\right\rceil _{\Lambda}=n\right).\label{looks good}\end{eqnarray}
For the last inequality, we used that for $i\in\left\{ \delta,\ldots,\eta-1\right\} $
and $\lambda\in\left\{ 0,\ldots,\Lambda-1\right\}
$\begin{eqnarray}
 &  & q_{\beta}^{\ell,k}\left(\cG_{\Lambda}\geq i\Lambda-\lambda\,|\, X_{k+1}=i\Lambda-\lambda\right)\nonumber \\
 &  & \leq q_{\beta}^{\ell,k}\left(\cG_{\Lambda}\geq n-\lambda\,|\, X_{k+1}=n-\lambda\right)
\end{eqnarray}
and, for reasons of symmetry,
\begin{equation}
q_{\beta}^{\ell,k}\left(X_{k+1}\in\cF_{\Lambda}\right)=1-\frac{1}{\ell-k}.
\end{equation}
The last step we need is a comparison of (\ref{nearly finish}) and
(\ref{looks good}). The only difference arises, when both of the
events, $\left\{ n\in\cF_{\Lambda}\right\} $ and $\left\{ \left\lceil S\right\rceil _{\Lambda}=n\right\} $,
take place. Then,
\begin{equation}
q_{\beta}^{\ell,k}\left(\cG_{\Lambda}\geq S\,|\,\left\lceil S\right\rceil _{\Lambda}=n\in\cF_{\Lambda}\right)=q_{\beta}^{\ell,k}\left(\cG_{\Lambda}\geq n\,|\,\left\lceil S\right\rceil _{\Lambda}=n\in\cF_{\Lambda}\right).
\end{equation}
This occurs with probability $q_{\beta}^{\ell,k}\left(n\in\cF_{\Lambda}\,|\,\left\lceil S\right\rceil _{\Lambda}=n\right)=\frac{\ell-k-1}{\eta-\delta}$
and explains the additional factor in the decomposition (\ref{nearly finish}).
\end{proof}

\section*{Acknowledgement}
We are grateful to Martin Herzog, formerly of
EMT, INRS, and Ravi Seshachala of Arizona State University for
assistance with the numerical and simulation evaluations.

\bibliographystyle{IEEEtran}
%\bibliography{wdmrings31Aug07}

\end{document}